\documentclass[12pt,a4paper]{article}
\usepackage{latexsym}
\usepackage{graphicx}
\usepackage{verbatim}
\usepackage{amssymb,amsfonts,amscd,amsbsy}

\catcode`\@=11
\def\marginnote#1{}
\newcount\hour
\newcount\minute
\newtoks\amorpm
\hour=\time\divide\hour by60
\minute=\time{\multiply\hour by60 \global\advance\minute by-\hour}
\edef\standardtime{{\ifnum\hour<12 \global\amorpm={am}%
        \else\global\amorpm={pm}\advance\hour by-12 \fi
        \ifnum\hour=0 \hour=12 \fi
        \number\hour:\ifnum\minute<10 0\fi\number\minute\the\amorpm}}
\edef\militarytime{\number\hour:\ifnum\minute<10 0\fi\number\minute}
\def\draftlabel#1{{\@bsphack\if@filesw {\let\thepage\relax
   \xdef\@gtempa{\write\@auxout{\string
      \newlabel{#1}{{\@currentlabel}{\thepage}}}}}\@gtempa
   \if@nobreak \ifvmode\nobreak\fi\fi\fi\@esphack}
        \gdef\@eqnlabel{#1}}
\def\@eqnlabel{}
\def\@vacuum{}
\def\draftmarginnote#1{\marginpar{\raggedright\scriptsize\tt#1}}
\def\draft{\oddsidemargin -.5truein
        \def\@oddfoot{\sl preliminary draft \hfil
        \rm\thepage\hfil\sl\today\quad\militarytime}
        \let\@evenfoot\@oddfoot \overfullrule 3pt
        \let\label=\draftlabel
        \let\marginnote=\draftmarginnote
   \def\@eqnnum{(\theequation)\rlap{\kern\marginparsep\tt\@eqnlabel}%
\global\let\@eqnlabel\@vacuum}  }

\def\preprint{\twocolumn\sloppy\flushbottom\parindent 1em
        \leftmargini 2em\leftmarginv .5em\leftmarginvi .5em
        \oddsidemargin -.5in    \evensidemargin -.5in
        \columnsep 15mm \footheight 0pt
        \textwidth 250mmin      \topmargin  -.4in
        \headheight 12pt \topskip .4in
        \textheight 175mm
        \footskip 0pt
        \def\@oddhead{\thepage\hfil\addtocounter{page}{1}\thepage}
        \let\@evenhead\@oddhead \def\@oddfoot{} \def\@evenfoot{} }

\def\titlepage{\@restonecolfalse\if@twocolumn\@restonecoltrue\onecolumn
     \else \newpage \fi \thispagestyle{empty}\c@page\z@ 
        \def\thefootnote{\fnsymbol{footnote}} }

\def\endtitlepage{\if@restonecol\twocolumn \else  \fi
        \def\thefootnote{\arabic{footnote}}
        \setcounter{footnote}{0}}  

\catcode`@=12
\relax

\def\bea{\begin{array}}
\def\bem{\begin{displaymath}}
\def\beq{\begin{equation}}

\def\eea{\end{array}}
\def\eem{\end{displaymath}}
\def\eeq{\end{equation}}

\def\Im{\mathop{\rm Im}}

\def\ov{\overline}

\def\Re{\mathop{\rm Re}}

\def\s2w{\sin^2 \theta_W}
\def\Tr{\mathop{\rm Tr}}

\relax
\def\dalpha{{\dot\alpha}}

\def\crbig{\\\noalign{\vspace {3mm}}}
\def\bigint{{\displaystyle\int}}

\def\L{{\cal L}}

\def\Fint{\bigint d^2\theta\,}
\def\Fbarint{\bigint d^2\ov\theta\,}
\def\Dint{\bigint d^2\theta d^2\ov\theta\,}

\newcommand{\skipthispart}[1]{}

\relax

\begin{document}
\topmargin-1.1cm
%
%
%
%
\begin{titlepage}
\begin{flushright}
July 28, 2016
\end{flushright}
\vspace{1.3cm}

\begin{center}{\Large\bf
Gauge Coupling Field, Currents, Anomalies and \\
\vspace{5mm}
\boldmath{${\cal N}=1$} Super-Yang-Mills Effective Actions }
\vspace{1.3cm}

{\large\bf Nicola Ambrosetti$^{1}$, Daniel Arnold$^{1}$, \\
\vspace{2mm}Jean-Pierre Derendinger$^1$ and Jelle Hartong$^{2}$}

\vspace{.8cm}
$^{1}$ Albert Einstein Center for Fundamental Physics, \\
Institute for Theoretical Physics, University of Bern, \\
Sidlerstrasse 5, 3012 Bern, Switzerland.



\vspace{.5cm}
$^2$ Physique Th\'eorique et Math\'ematique and International Solvay Institutes,\\
Universit\'e Libre de Bruxelles, C.P. 231, 1050 Brussels, Belgium.\\

\end{center}
\vspace{1.3cm}

\begin{center}
{\large\bf Abstract}
\end{center}
\begin{quote}

Working with a gauge coupling field in a linear superfield, we construct effective Lagrangians for ${\cal N}=1$
super-Yang-Mills theory fully compatible with the expected all-order behaviour or physical quantities. 
Using the one-loop dependence on its ultraviolet cutoff and anomaly matching or cancellation of $R$ and dilatation anomalies, we obtain the Wilsonian effective Lagrangian. With similar anomaly matching or cancellation 
methods, we derive the effective action for gaugino condensates, as a function of the real coupling field. 
Both effective actions lead to a derivation of the NSVZ $\beta$ function from algebraic arguments only. The extension of results to ${\cal N}=2$ theories or to matter systems is briefly considered. 
The main tool for the discussion of anomalies is a generic supercurrent structure
with $16_B+16_F$ operators (the ${\cal S}$ multiplet), which we derive using superspace identities 
and field equations for a fully general gauge theory Lagrangian with the linear gauge coupling superfield, 
and with various $U(1)_R$ currents. As a byproduct, we show under which conditions the ${\cal S}$ multiplet 
can be improved to contain the Callan--Coleman--Jackiw energy-momentum tensor whose trace measures 
the breaking of scale invariance. 

\end{quote}



\end{titlepage}
\setcounter{footnote}{0}
\setcounter{page}{0}
\setlength{\baselineskip}{.6cm}
\setlength{\parskip}{.2cm}

\renewcommand{\theequation}{\thesection.\arabic{equation}}

\newpage
%
%


\section{Introduction}

The approach which identifies coupling constants with background values of fields and superfields has proved, 
following Seiberg \cite{Seiberg}, a useful and powerful tool in the study of perturbative and nonperturbative 
properties of supersymmetric gauge theories. It has been particularly successful for ${\cal N}=2$ theories 
\cite{Seiberg2, SW},  using the factorization ``theorem" of hypermultiplet and vector multiplet scalars and special 
K\"ahler geometry formulated in terms of a holomorphic prepotential. It is also an important ingredient in the study 
of perturbative and nonperturbative moduli spaces of ${\cal N}=1$ theories, described in terms of holomorphic 
invariants \cite{BDFS}.

The situation changes if one introduces a field, the {\it gauge coupling field}, to describe the gauge coupling 
constant in an ${\cal N}=1$ supersymmetric gauge theory, in agreement with the fact that, as shown for instance in 
\cite{SV86, SV, AHM}, the holomorphic dependence on the gauge coupling in $\mathcal{N}=1$ super Yang--Mills 
theory is anomalous. This anomaly is reflected in the discrepancy between the all-order running of the gauge coupling 
and the absence of perturbative corrections to the vacuum angle.

In other words, one cannot in general use a chiral superfield to describe the gauge coupling 
in $\mathcal{N}=1$ super Yang--Mills theory. 
We will show that the correct description is obtained using a real linear superfield which includes in its $4_B+4_F$ 
components a real scalar, the coupling field, and an antisymmetric tensor with gauge invariance. Such a tensor is in 
general dual to a pseudoscalar with axionic symmetry, and the linear superfield to a chiral superfield. 
We will also show that the anomalous dependence on the gauge coupling creates an obstruction to {\it analytically} perform 
the duality transformation, that it provides the adequate information to write all-order effective actions with the linear
superfield and also how the obstruction disappears with extended ${\cal N}=2$ supersymmetry, where holomorphicity 
is relevant. 

When writing effective actions, gauge-invariant operators are needed. The linear superfield introduces, besides the familiar
chiral $\widetilde\Tr{\cal WW}$, a second real, dimension two, operator $\hat L = L -2\Omega$, where $\Omega$ is the 
Chern-Simons superfield. With these two operators, anomaly matching or cancellation of the $R$ and dilatation (rescaling) 
anomalies can be performed. As a tool, we use the appropriate supercurrent superfield equation. 
In the first part of this work, we construct supercurrent structures for supersymmetric gauge 
theories coupled to the linear superfield and study their currents and anomalies. 
These structures naturally involve $16_B+16_F$ fields, as in the ${\cal S}$ structure described by Komargodski and Seiberg, 
\cite{KS} and include both chiral and linear anomaly sources in the supercurrent superfield equation 
$\ov D^\dalpha J_{\alpha\dalpha}=D_\alpha X + \chi_\alpha$. 

This construction of the supercurrent structure for an arbitrary simple gauge group and matter content extends our previous 
work \cite{ADH}. We find again that the supercurrent superfield including the Belinfante energy-momentum tensor 
(obtained when coupling the theory to a background space-time metric) also includes the $U(1)_{\widetilde R}$ current 
with zero $R$ charge for the chiral multiplets. We then derive supercurrent structures with arbitrary $R$ charges for these 
superfields and discuss the corresponding improvement of the energy-momentum tensor. In these supercurrent structures, the 
sources $X$ and $\chi_\alpha$ depend classically on the superfields controlling in the Lagrangian the breaking of $U(1)_R$ in terms of the chiral 
superfield $R$ charges and the breaking of scale invariance with scale dimensions equal to the $R$ charges, as would be required 
by the superconformal algebra. 

In general, the divergence of the dilatation current, which is not present in the supercurrent superfield, is the sum of the divergence
of a virial current and of the trace of the energy-momentum tensor. While the sum is of course unchanged, both contributions are sensitive to improvements of the energy-momentum tensor. In particular, if there exists a Callan-Coleman-Jackiw (CCJ) \cite{CCJ, CJ} 
energy-momentum tensor which cancels the virial current, a scale-invariant theory is also conformal invariant. The CCJ tensor exists
for all renormalisable Lagrangians but many theories have an irreducible virial current: this is the case whenever a linear superfield is 
coupled to chiral and gauge superfields. This has implications for us: supercurrent structures specify the on-shell value of the 
energy-momentum tensor trace only. To get the divergence of the dilatation current, a specific virial current, which we derive, 
is needed, except if the theory would be scale invariant. 

Both source superfields $X$ and $\chi_\alpha$ are supplemented by quantum contributions from
chiral $U(1)_R$ and dilatation anomalies. These quantum corrections use both superfields $\widetilde\Tr{\cal WW}$ and $L-2\Omega$.
The source superfields determine the divergence of the $U(1)_R$ current and the trace ${T^\mu}_\mu$ of the energy-momentum tensor 
in $J_{\alpha\dalpha}$, which is not in general the divergence of the dilatation current, a point which we also carefully discuss.
This is of importance since a non-trivial coupling of the linear superfield always breaks (classically) scale invariance.

We then establish two effective Lagrangians with the gauge coupling field in the linear superfield: the all-order perturbative
Wilsonian Lagrangian for super-Yang-Mills theory and the effective action determining the gaugino condensate. 
In both cases, anomaly matching or compensation is sufficient to derive the all-order renormalisation-group (RG) equation and $\beta$ 
function originally found by Novikov, Shifman, Vainshtein and Zakharov (NSVZ) \cite{NSVZ}. 

The local Wilsonian effective action is obtained from a microscopic theory by integrating short-distance physics up to distance $\mu^{-1}$.
The energy scale $\mu$ which explicitly appears in the (loop-corrected) Wilsonian action acts then as a UV cutoff. When expressed in terms of physical quantities, the Wilsonian action also depends on a second energy scale, $M$, the scale at which quantities like the gauge coupling 
are normalised. Since both $\mu$ and $M$ are arbitrary,\footnote{In general however, $M>\mu$.} two RG equations follow. 
The dependence on the scale $\mu$ is fixed by the fact that the Wilsonian effective action depends holomorphically on $\mu$ and 
therefore runs only to one-loop \cite{SV}. By supersymmetry (and chirality), rescaling $\mu$ is equivalent to an anomalous $U(1)_R$
transformation, or to an anomalous scale transformation. However, there is a residual dilatation anomaly which must be cancelled, by
RG invariance. Since it involves a non-holomorphic dependence on the coupling, it requires the use of the gauge-invariant real 
superfield $\hat L$. The corresponding anomaly counterterm encodes the dependence of the effective action on the physical coupling
$g^2(M)$ identified as the background value of the lowest scalar component $C$ of $\hat L$. While arbitrariness of $\mu$ 
leads to the expected one-loop behaviour of the Wilsonian action, arbitrariness of $M$ leads to the all-order NSVZ $\beta$ function \cite{NSVZ}.
The content of the NSVZ $\beta$ function is thus entirely described by the cancellation of the dilatation anomaly and the one-loop 
$\mu$--dependence of 
the Wilsonian action.\footnote{A supergravity based derivation of the NSVZ $\beta$ function of pure super-Yang-Mills using similar 
anomaly matching arguments has been given long ago \cite{DFKZ2}. }

Similar anomaly matching/cancellation arguments can be used to derive an effective Lagrangian describing gaugino condensates in 
${\cal N}=1$ super-Yang-Mills theory, as a function of the real gauge coupling field $C$.\footnote{Following and extending 
ref.~\cite{BDQQ}.} It actually provides the effective Lagrangian 
version of the derivation performed by NSVZ using instanton methods \cite{NSVZ}. The theory has two superfields, the familiar
chiral $U=\langle\widetilde\Tr{\cal WW}\rangle$ and the real $V=\langle\hat L\rangle$, related by $U=- {1\over2}\ov{DD}V$ as
a consequence of $\widetilde\Tr{\cal WW} = -{1\over2}\ov{DD}\,\hat L$. The effective Lagrangian is again derived by anomaly matching of the
$U(1)_R$ one-loop anomaly by a chiral ($F$-term) counterterm using $U$, and anomaly cancellation of the residual dilatation anomaly
by a real ($D$--term) counterterm using $V$. Since the fundamental condensate field $V$, which also includes the coupling field $C$ as its
lowest component, is real, the effective scalar potential determines the modulus $|\langle\widetilde\Tr\lambda\lambda\rangle|$ of the 
gaugino condensate as a function of $C$ or $g^2(M)$: perturbative anomaly arguments are not able to discretize the $R$-symmetry 
spontaneously broken by the condensate. 
Discretization to $Z_{2N}$ (with $SU(N)$ gauge group) can be easily expressed in a non-perturbative superpotential in $U$ where 
each allowed term can be interpreted as a $k$--instanton contribution. 
Arbitrariness of $M$ in the effective condensate Lagrangian leads again to the all-order NSVZ $\beta$ function \cite{NSVZ}. 

The outline of the paper is as follows. In Section \ref{secGCF}, we define the gauge coupling field as the lowest component $C$ of 
the real linear superfield $L$ and we introduce the gauge-invariant coupling $L$ to the Chern--Simons superfield $\Omega$, 
in the combination $\hat L=L-2\Omega$. The next Section \ref{seclinear} discusses chiral-linear duality in ${\cal N}=1$ 
superspace, repeating for completeness long-known arguments \cite{Siegel}. At this point, the main result is that the dependence 
on the gauge coupling field $C$ of the super-Yang-Mills Lagrangian is not restricted by supersymmetry, that holomorphicity is not relevant
and also that the vacuum angle does not depend on $C$. 
Section \ref{secsupercurrent} presents the supercurrent structures for theories with linear, chiral and gauge superfields. 
We first derive a {\it natural} $16_B+16_F$ structure including the Belinfante improved energy-momentum tensor. 
Tools in the derivation are superfield identities and field equations. 
We then show how to improve this structure to a supercurrent making the scale properties of the theory manifest and consider
the case where the superpotential would be a generic function of the super-Yang-Mills superfield $\widetilde\Tr{\cal WW}$.
Section \ref{secsupercurrent} also provides a 
detailed discussion of scale transformation properties and of the existence (or nonexistence) of the Callan-Coleman-Jackiw (CCJ) 
energy-momentum tensor \cite{CCJ, CJ}. Appendices \ref{AppA}, \ref{AppB} and \ref{AppC} are in support of this Section. 
With this understanding of the supercurrent structures, we next consider the incorporation of anomalies. 
We focus on the source or anomaly superfields $X$ and $\chi_\alpha$ appearing in the on-shell conservation laws of the supercurrent 
multiplet $J_{\alpha\dalpha}$. The study of the $U(1)_R$ and scale perturbative anomalies is the subject of section \ref{secanom}. 

Section \ref{seceff}  discusses the Wilsonian effective Lagrangian for pure ${\cal N}=1$ super-Yang-Mills and the effective Lagrangian for 
gaugino condensates. In both cases, the all-order NSVZ $\beta$ function is derived, using anomaly matching/cancellation only. 
For completeness, it also briefly shows how ${\cal N}=2$ theories escape corrections beyond one-loop.

Finally, we have added a number of appendices. Appendix \ref{AppA} reviews the properties of the supercurrent structure and its improvements 
in component language. Appendices \ref{AppB} and \ref{AppC} give relevant background information on scaling properties of the theory,
on the very particular properties of a certain scale superfield denoted by $\Delta$ and on improvements of the canonical (Noether) 
energy-momentum tensor to the Belinfante and CCJ energy-momentum tensors. Appendix \ref{AppFZ} provides the link between our 
supercurrent structures and the better known Ferrara--Zumino \cite{FZ} structure. Finally, in
Appendix \ref{appLeg}, we collect some useful formulas for the Legendre transformation which appears in linear-chiral 
duality.

\section{The gauge coupling field} \label{secGCF}

Consider the Lagrangian 
\beq
\label{Lagr0}
{\cal L} = {1\over g^2}\, {\cal L}_{\rm{SYM}}, \qquad
{\cal L}_{\rm{SYM}} = -{1\over4} F^a_{\mu\nu}F^{a\,\mu\nu} + {i\over2}\lambda^a\sigma^\mu (D_\mu\ov\lambda)^a
- {i\over2}(D_\mu\lambda)^a\sigma^\mu \ov\lambda^a + {1\over2} D^aD^a,
\eeq
where
\begin{eqnarray}
\label{curv}
F^a_{\mu\nu} & = & \partial_\mu A^a_\nu-\partial_\nu A_\mu^a- {1\over2} f^{abc}A^b_\mu A^c_\nu\,,
\\
\label{covder}
(D_\mu\lambda)^a & = & \partial_\mu\lambda^a- {1\over2}f^{abc}A^b_\mu\lambda^c\,,
\end{eqnarray}
with $f^{abc}$ the structure constants of some simple gauge group with generators $T^a$, i.e.
\begin{equation}
[T^a,T^b]=if^{abc}T^c\,.
\end{equation}
One wants to replace the coupling $g^2$ by a function of a real scalar field $C$,
$$
g^2 \qquad\longrightarrow\qquad h(C),
$$
or simply by a real scalar field $C$. 
It is then easy to see that ${\cal N}=1$ supersymmetry does not provide any information or condition on the 
function $h(C)$. The argument is as follows. Since 
\beq
{\cal L}_{\rm{SYM}} = {1\over4} \, \Fint \widetilde\Tr\,{\cal WW} 
+ {1\over4} \, \Fbarint  \widetilde\Tr\,\ov{\cal WW},
\eeq
where ${\cal W}_\alpha({\cal A}) = -{1\over4}\ov{DD} \, e^{-{\cal A}}D_\alpha e^{\cal A}$ is the chiral superfield 
of gauge curvatures\footnote{To be precise, ${\cal A}$ is the Lie algebra-valued real superfield of gauge 
potentials, ${\cal A} = {\cal A}^aT^a_r$, with generators in some representation $r$ normalized by 
$\Tr(T^a_rT^b_r) = T(r)\delta^{ab}$ and we use the notation
$$
 \widetilde\Tr\,{\cal WW} \equiv T(r)^{-1} \Tr {\cal W}^\alpha {\cal W}_\alpha \,.
$$
For the components of ${\cal A}^a$ in Wess--Zumino gauge we write 
\begin{equation}
{\cal A}^a_{\rm{WZ}}=\theta\sigma^\mu\ov\theta A^a_\mu+i\theta\theta\ov{\theta\lambda}^a-i\ov{\theta\theta}\theta\lambda^a+{1\over2}\theta\theta\ov{\theta\theta}D^a\,.
\end{equation}
We also write $A=A^aT^a_r$, $F_{\mu\nu}=F_{\mu\nu}^aT^a_r$, $D=D^aT^a_r$ 
and $\lambda=\lambda^aT^a_r$.

If needed, the factors $1/2$ in gauge curvatures (\ref{curv}) and covariant derivatives (\ref{covder}) can be
 eliminated by the rescalings ${\cal A}\rightarrow2{\cal A}$ and 
 ${\cal W}_\alpha({\cal A})\rightarrow{1\over2}{\cal W}_\alpha(2{\cal A})$. },
one first observes that there exists a Chern-Simons real superfield $\Omega$ defined
by\footnote{For a detailed study of this superfield, see ref.~\cite{CFV}.}
\beq
\label{CS1}
 \widetilde\Tr\,{\cal WW} = \ov{DD}\,\Omega, \qquad\qquad  \widetilde\Tr\,\ov{\cal WW} = DD\,\Omega
\eeq
such that its gauge variation is linear, $\ov{DD}\,\delta_{\rm gauge}\Omega =0$. One then introduces
a real linear superfield $L$,
\beq
\label{L1}
\ov{DD}\, L = DD\, L =0 ,
\eeq
one postulates that $L$ has gauge variation
\beq
\label{L2}
\delta_{\rm{gauge}} L = 2 \, \delta_{\rm gauge}\Omega
\eeq
and one forms the gauge-invariant real superfield 
\beq
\label{L3}
\hat L = L - 2 \, \Omega.
\eeq
The lowest component of $L$ is a real scalar field $C$ and the gauge-invariant supersymmetric Lagrangian
\beq
\label{Lagr1}
{\cal L} = \Dint {\cal H}(\hat L)
\eeq
includes in its component expansion\footnote{All terms have at most two derivatives.}
\beq
\label{Lagr2}
{\cal L} = {\cal H}_C (C) \, {\cal L}_{\rm SYM} + \ldots , \qquad\qquad
{\cal H}_C (C) = {d\over dC} {\cal H}(C) .
\eeq
Since the function ${\cal H}$ is arbitrary we have a gauge coupling field 
\beq
\label{Lagr3}
{1\over g^2} = {\cal H}_C(C)
\eeq
and ${\cal N}=1$ supersymmetry does not provide information or constraints on the gauge coupling
field. Since\footnote{When dealing with Lagrangians we will sometimes omit total derivative terms when writing equalities.}
\beq
\label{Lagr4}
\Dint \hat L = -{1\over8}\Fint \ov{DD}\hat L + \makebox{h.c.} + \makebox{\rm total deriv.} 
= {\cal L}_{\rm SYM} + \makebox{total deriv.},
\eeq
the linear superfield decouples in a term linear in $\hat L$. 

Hence, since theory (\ref{Lagr1}) does not have a scalar potential, the field equations of the 
linear superfield have a (supersymmetric) solution $\hat L =$ constant, which
allows us to identify this background value of $\hat L$ with the gauge coupling constant.

We will use in this work three components of the superfield $\hat L$:
\beq
\label{L0}
\hat L = C + \theta\sigma^\mu\ov\theta\, \left[ {1\over6} \epsilon_{\mu\nu\rho\sigma}H^{\nu\rho\sigma}
+ \widetilde\Tr \lambda\sigma_\mu\ov\lambda \right] + \theta\theta\ov{\theta\theta} \left[ {1\over4}\Box C +
{\cal L}_{SYM} \right] + \ldots
\eeq
Note the presence of a gaugino axial current besides the tensor field 
\beq
\label{H3is}
H_{\mu\nu\rho} = 3\,\partial_{[\mu} B_{\nu\rho]} - \omega_{\mu\nu\rho},
\eeq
where $\omega_{\mu\nu\rho}$ is the gauge Chern-Simons form, in the $\theta\sigma^\mu\ov\theta$ 
component. The Lagrangian has then kinetic terms $\sim H_{\mu\nu\rho}H^{\mu\nu\rho}$.
This interaction of gauge fields with
an antisymmetric tensor with gauge symmetry is a standard occurence in higher-dimensional global 
and local supersymmetry and in superstring theories. 
It is only in four dimensions that the antisymmetric tensor can be transformed 
into an axion scalar coupled to $\widetilde\Tr F_{\mu\nu}\widetilde F^{\mu\nu}$. It seems then a natural
approach to use $\hat L$, as we do here, to introduce a gauge coupling field since in addition
it does not introduce any dependence on the background value of an axion scalar, {\it i.e.} any explicit 
dependence on the vacuum $\theta$ angle of the Yang-Mills theory.

In the context of four-dimensional effective supergravity descriptions of superstring compactifications, 
the role of the linear supermultiplet as the string loop-counting dilaton field has been originally shown 
by Cecotti, Ferrara and Villasante \cite{CFV}. Its role in anomaly cancellation and in the four-dimensional
Green-Schwarz mechanism \cite{CO} has been displayed in many examples, following the
effective description \cite{DFKZ1} of one-loop gauge threshold corrections in simple orbifolds \cite{DKL}.


\section{The linear superfield and chiral-linear duality}\label{seclinear}
\setcounter{equation}{0}

Like the chiral superfield, the linear superfield \cite{FWZ, Siegel} describes four bosonic and four 
fermionic ($4_B+4_F$) off-shell field components. We use the expansion
\beq
\label{Lexp}
L = C + i\theta\chi_L - i\ov{\theta\chi}_L + {1\over6}\epsilon_{\mu\nu\rho\sigma} \theta\sigma^\mu\ov\theta \,
h^{\nu\rho\sigma} + {1\over2}\,\theta\theta \, \partial_\mu\chi_L\sigma^\mu\ov\theta
+ {1\over2}\ov{\theta\theta}\,\theta\sigma^\mu\partial_\mu\ov\chi_L
+ {1\over4}\theta\theta\ov{\theta\theta}\,\Box C \,,
\eeq
where $h_{\mu\nu\rho} = 3\,\partial_{[\mu} B_{\nu\rho]}$, to solve the linearity 
condition (\ref{L1}). Since $B_{\mu\nu}$ with its gauge invariance
$\delta B_{\mu\nu} =  2\,\partial_{[\mu}\Lambda_{\nu]}$ describes three bosons, the linear superfield 
does not have any scalar auxiliary field and does not generate a specific contribution to the scalar potential 
in a supersymmetric Lagrangian.
When coupled to $\Omega$, as in expression (\ref{L3}), or in conformal supergravity,
the linear superfield $L$ (and its bosonic components $C$ and $B_{\mu\nu}$) has canonical 
scale dimension two. 

In four space-time dimensions, an antisymmetric tensor with gauge invariance, as described in the linear superfield, is dual to a real scalar with axionic shift symmetry. At the Lagrangian level, the supersymmetric version exchanges a chiral and a linear superfield, and this {\it chiral--linear} duality corresponds to the following chain of equalities \cite{Siegel}:
\beq
\label{LS1}
\begin{array}{rcl}
{\cal L} \,\,=\,\,
\Dint {\cal H}(\hat L) &=& \Dint{\cal H}(V)
\crbig&& \displaystyle
+ {1\over8}\Fint S\ov{DD} (V+2\Omega) + {1\over8}\Fbarint \ov S {DD} (V+2\Omega) 
\crbig
&=& \displaystyle \Dint \left[ {\cal H}(V)  - {1\over2} (S+\ov S) V \right] + {\rm derivative}
\crbig \displaystyle
&& \displaystyle + {1\over4}\Fint S \, \widetilde\Tr\,{\cal WW} 
+ {1\over4} \Fbarint \ov S \, \widetilde\Tr\,\ov{\cal WW} 
\crbig
&=& \displaystyle \Dint {\cal K}(S+\ov S) + {\rm derivative}
\crbig
&& \displaystyle  +{1\over4}\Fint S \, \widetilde\Tr\, {\cal WW} 
+ {1\over4} \Fbarint \ov S \, \widetilde\Tr\,\ov{\cal WW} .
\end{array}
\eeq
In the first equality, the Lagrange multiplier chiral superfield $S$ imposes that $V+2\Omega$ is linear.
The third equality (\ref{LS1})
defines the K\"ahler potential of the dual theory in terms of the Legendre transformation
\beq
\label{LS2}
{\cal K}(S+\ov S) = {\cal H}(V) - {1\over2}(S+\ov S)V
\eeq
exchanging variables $V$ and $S+\ov S$, {\it i.e.} with $V$ expressed as a function of $S+\ov S$
by solving the usual relations
\beq
\label{LS2b}
{d{\cal H}\over dV} = {1\over2}(S+\ov S), \qquad\qquad {d{\cal K}\over d(S+\ov S)} = - {1\over2}V.
\eeq
The resulting chiral theory has axionic shift symmetry $\delta S = ia$ ($a$ is a real constant).

Some comments are appropriate.
Firstly, all information on the function ${\cal H}$ goes into the K\"ahler potential ${\cal K}$.
The dual holomorphic gauge kinetic function is always $S$ and the dual gauge coupling constant 
is\,\footnote{We can replace $S$ by a (non constant) function $f(S)$ in equalities (\ref{LS1}).}
\beq
\label{LS3}
{1\over g^2} = \Re s
\eeq
for all functions ${\cal H}$. Secondly, the Legendre transformation exchanges a {\it real}
with a {\it chiral}\/ superfield, with axionic symmetry on $S$ dual to the gauge invariance of 
$B_{\mu\nu}$.
The shift symmetry has an important consequence. Defining the Yang-Mills vacuum angle as
\beq
\label{LS3b}
\langle \Im s \rangle = - {\theta\over8\pi^2} \, ,
\eeq
its contribution to the Lagrangian 
$$
-{\theta\over 32\pi^2}\, \widetilde\Tr [ F_{\mu\nu} \widetilde F^{\mu\nu} -2\,\partial_\mu(\lambda\sigma^\mu\ov\lambda) ]
$$
is a derivative irrespective of ${\cal H}$. Hence, the all-order dependence on the gauge coupling 
and the absence of $\theta$-dependence in perturbation theory are fully compatible with 
supersymmetry. Thirdly, the linear superfield does not have an auxiliary field: $B_{\mu\nu}$ describes 
three off-shell fields and one on-shell helicity zero state. In the dual chiral version, $S$ has a complex
auxiliary field $f_S$ which vanishes in theory (\ref{LS1}). In theories with additional matter chiral superfields
$\Phi$, the auxiliary field $f_S$ is a well-defined linear combination of the auxiliary $f_\Phi$ in $\Phi$. 
Hence, if $S$ is dual to a linear superfield, its auxiliary $f_S$ does not generate an independent 
contribution to the scalar potential and this has clearly implications on the vacuum properties.\footnote{
Reference \cite{DQQ} discusses this point.}

Finally, notice that we may also add a term proportional to ${\cal L}_{SYM}$ (and then independent from 
$L$ or $S$) to theory (\ref{LS1}). Doing this adds a constant term to $g^{-2}$ which is then a 
one-loop correction. Hence, there is no information in the holomorphic coupling $S$, it is naturally 
defined up to a one-loop correction only and its relation to the original coupling field $C$ is fully
included in the Legendre transformation (\ref{LS2}).

Since $\Omega$ has canonical scale dimension two, this is also the case for $L$ and 
$V$ in the equalities (\ref{LS1}). Then, the natural 
canonical dimension of the chiral $S$ is zero. The quantity
$$
\Delta \equiv 2V{\cal H}_V - 2{\cal H} 
$$
measures the violation of scale invariance in the original linear multiplet theory. But, according
to the Legendre transformation (\ref{LS2}) and (\ref{LS2b}), 
$$
\Delta = -2\,{\cal K}
$$
as expected if the scale dimension of $S$ is zero. 
Hence, imposing scale invariance $\Delta=0$ leads to ${\cal H}(\hat L) \propto \hat L$ which is super-Yang-Mills
theory with a constant coupling, {\it i.e.} in which $L$ or $S$ are absent. Clearly, 
this restriction is the obvious statement that there is no scale-invariant propagating gauge coupling field,
in the absence of another dimensionful field. 
Hence, we expect to always find a classical contribution induced by the gauge coupling field to the 
divergence of the dilatation current. 


\section{Supercurrent superfields} \label{secsupercurrent}
\setcounter{equation}{0}

In this Section, we consider a ${\cal N}=1$ theory for chiral superfields $\Phi$ in some representation 
$r$ of the gauge group\footnote{We suppress $i$ indices on $\Phi^i$ and $\bar\Phi_i$.}, gauge superfields ${\cal A}$, ${\cal W}_\alpha$, $\Omega$, 
as defined earlier, and the linear gauge coupling superfield
$L$. These superfields carry linear representations of Poincar\'e supersymmetry, but they 
actually carry representations of the full ${\cal N}=1$ superconformal algebra $SU(2,2|1)$
even if dynamical equations respect in general only Poincar\' e supersymmetry. In other words, fields in the 
theory have well-defined transformation properties under the superconformal algebra, the variation of the
action under these transformations is well-defined, but the invariance of the theory is in general 
generated by the super-Poincar\'e subalgebra only. Since the bosonic subalgebra of $SU(2,2|1)$ is
$$
SU(2,2) \times U(1)_R \supset SO(1,3)_{\rm{Lorentz}} \times SO(1,1)_{\rm{dil}} \times U(1)_R\,,
$$
we may then assign two abelian quantum numbers to all fields, superspace coordinates and superfields, 
a chiral charge $q$ for
$U(1)_R$ transformations, and a scale dimension $w$ for dilatations 
$SO(1,1)_{\rm{dil}}$. As far as the super-Poincar\'e symmetry is concerned, $q$ and $w$ are arbitrary.
But the superconformal algebra introduces further constraints: $w=q$ for chiral superfields\footnote{
In our convention.} and
canonical scale dimensions for gauge superfields.

In addition, unitarity of the quantum theory would introduce further constraints (unitarity bounds) \cite{Mack}.
We are not concerned with them as long as we consider the theory as classical. 

The assigned chiral and scale charges are then as follows:\footnote{In our convention, the
Grassmann coordinates have weights $(q,w)=(3/2,-1/2)$ 
while for gauginos $(q,w)=(3/2,3/2)$.} 
$$
\begin{array}{c}
\Phi: (q,w), \quad\qquad \ov\Phi:(-q,w), \quad\qquad L:(0,2), \quad\qquad {\cal A}:(0,0), 
\crbig
{\cal W}_\alpha: (3/2,3/2), \qquad\qquad \Omega:(0,2).
\end{array}
$$
The charges of $L$ are as required by $\hat L = L-2\Omega$.
If the representation of the chiral superfield is reducible, $r = \oplus_i r_i$, charges
$(q_i,w_i)$ are assigned. The Lagrangian describing the dynamics of these superfields includes 
in general $U(1)_R$ and scale symmetry violating terms. 
In addition, a non--$R$ abelian chiral algebra may act with charge $t$ or $t_i$ on the 
chiral superfields.\footnote{Non-abelian chiral groups will be mostly irrelevant to us. }

Since we will later on be concerned with quantum anomalies in $U(1)_R$ and dilatation 
transformations, the natural setup is to establish a supercurrent structure, {\it i.e.} a supercurrent 
superfield \cite{FZ} $J_{\alpha\dalpha}$, anomaly superfields and the associated supercurrent 
equation. The supercurrent superfield is primarily defined to include the conserved
supercurrent and energy-momentum tensor. It is defined up to improvement transformations.
In this section, our goal is first to construct supercurrent structures for theories with a coupling 
field and then to establish how these transformations encode the relation of the supercurrent 
structure with the assigned chiral and dilatation weights. This will be done for generic 
super-Poincar\'e theories with scale-invariant, conformal or $R$--symmetric theories 
appearing as particular cases. 

We begin with a detailed discussion of the supercurrent structures in a theory with chiral, 
gauge and linear superfields. Some aspects have been studied by Magro, Sachs and 
Wolf \cite{MSW}.\footnote{Our discussion in this section generalizes some of the results of our earlier article
\cite{ADH}, which can be recovered by decoupling the linear superfield. Identical notations are used.} 
Supplementary formulas are provided in Appendix \ref{AppA}. 

\subsection{A superfield identity}

Consider the gauge-invariant real superfield 
\beq
{\cal H} = {\cal H}(\hat L, Y) \qquad\qquad Y = \ov\Phi e^{\cal A} \Phi.
\eeq
In $Y$, the real gauge superfield is Lie algebra-valued, ${\cal A} = {\cal A}^aT_r^a$, 
with generators $T_r^a$ in the 
representation $r$ of the matter chiral superfield $\Phi$. Gauge transformations are
\beq
\label{CSF1}
\Phi\quad\longrightarrow\quad e^\Lambda\, \Phi, \qquad
\ov\Phi\quad\longrightarrow\quad \ov\Phi \,e^{\ov\Lambda}, \qquad
e^{\cal A} \quad\longrightarrow\quad e^{-\ov\Lambda} e^{\cal A} e^{-\Lambda} 
\eeq
with $\Lambda = \Lambda^aT^a_r$ and $\ov D_{\dot\alpha}\Lambda=0$. Gauge-covariant superspace derivatives read
\beq
\label{CSF2}
{\cal D}_\alpha \Phi = e^{-{\cal A}}(D_\alpha e^{\cal A}\Phi) , 
\qquad\qquad
\ov{\cal D}_\dalpha \ov\Phi = (\ov D_\dalpha\ov\Phi e^{\cal A}) e^{-{\cal A}}
\eeq
and 
$$
(\ov{\cal D}_\dalpha\ov\Phi)e^{\cal A}({\cal D}_\alpha\Phi)
= (\ov D_\dalpha\ov\Phi e^{\cal A}) e^{-{\cal A}} (D_\alpha e^{\cal A}\Phi)
$$ 
is gauge invariant.\footnote{In general, the gauge invariant function ${\cal H}$ can depend on
variables $Y_i$ if the representation of the chiral superfields is reducible, $r=\oplus_ir_i$. This
generalization is straightforward. It may also depend on other gauge invariant quantities,
such as holomorphic invariants, which we do not consider here.}

By direct calculation of, for instance,  $\ov{DD}D_\alpha({\cal H}-\hat L{\cal H}_L)$,
the following identity can be derived:
\beq
\label{CSF3}
\begin{array}{l}
2 \ov D^{\dalpha}\Bigl[ (\ov{\cal D}_\dalpha\ov\Phi) {\cal H}_{\Phi\ov\Phi}  ({\cal D}_\alpha\Phi)
- {\cal H}_{LL}(\ov D_\dalpha\hat L)(D_\alpha\hat L) \Bigr]
\crbig \hspace{1.5cm}
= - \hat L \, \ov{DD}D_\alpha{\cal H}_L - (\ov{DD}{\cal H}_\Phi)\, {\cal D}_\alpha\Phi
- \ov{DD}D_\alpha({\cal H}-\hat L{\cal H}_L)
\crbig\hspace{1.9cm}
- 2 \,  \widetilde\Tr\,{\cal WW} \, D_\alpha{\cal H}_L - 4 \, {\cal H}_Y \, \ov\Phi e^{A}{\cal W}_\alpha\Phi ,
\end{array}
\eeq
where subscripts indicate 
derivatives of ${\cal H}$ with respect to either $\Phi$, $\ov\Phi$, $L$ or $Y$. 
We stress that eq.~(\ref{CSF3}) is merely an identity, without any information content. 
The next step is to consider a theory for $\hat L$ and $\Phi$ and to use its field equations 
to rearrange identity (\ref{CSF3}) into a supercurrent equation.

\subsection{The natural (Belinfante) supercurrent structure}

Let us hence consider the theory
\beq
\label{CSF4}
\L=\Dint {\cal H} ( \hat L , Y ) + \Fint W(\Phi) + \Fbarint \ov W(\ov\Phi).
\eeq
Gauge invariance of the holomorphic superpotential $W(\Phi)$, i.e.
\begin{equation}
W_{\Phi^i}(T_r^a)^i{}_j\Phi^j=0\,,
\end{equation}
implies $W_\Phi{\cal D}_\alpha\Phi  = D_\alpha W$. The ${\cal H}$ term in the Lagrangian has in general several chiral symmetries. In particular, 
since $\cal H$ satisfies
\beq
\label{CSF4b}
{\cal H}_\Phi\Phi=\ov\Phi {\cal H}_{\ov\Phi}={\cal H}_YY,
\eeq
it is always invariant under the non-$R$ $U(1)$ symmetry 
rotating all chiral superfields $\Phi$ by the same phase.\footnote{If the representation of the 
matter superfields is reducible, each irreducible component has an associated $U(1)$ global 
symmetry. It extends to $U(n)$ factors if the matter superfields include $n$ 
copies of an irreducible component.} Its chiral symmetries also 
include the $R$ symmetry (that we call $\widetilde R$) which transforms Grassmann coordinates and leaves all superfields in $\hat L$ or $Y$ inert. These chiral symmetries are in general broken 
by the superpotential. 

For completeness, the component expansion of theory (\ref{CSF4}) is as follows:\footnote{Gauge 
invariance of ${\cal H}$ implies ${\cal H}_z Dz = \ov z D{\cal H}_{\ov z}$. }
\beq
\label{bosonicL}
\begin{array}{rcl}
{\cal L} &=& -{1\over2}{\cal H}_{CC} \Bigl[ {1\over2}(\partial_\mu C)(\partial^\mu C) 
+ {1\over12} H_{\mu\nu\rho}H^{\nu\mu\rho}  \Bigr]
+ {\cal H}_{z\ov z} \Bigl[ (D_\mu \ov z)(D^\mu z) + \ov ff \Bigr] 
\crbig
&& + {\cal H}_C \Bigl[ -{1\over4}\widetilde\Tr F_{\mu\nu}F^{\mu\nu} + {1\over2}\widetilde\Tr DD \Bigr]
+ {1\over2}{\cal H}_z D z - W_zf - \ov f \ov W_{\ov z}
\crbig
&& + {i\over12}\epsilon_{\mu\nu\rho\sigma}
H^{\mu\nu\rho} \Bigr[ {\cal H}_{Cz}D^\sigma z - {\cal H}_{C\ov z}D^\sigma\ov z \Bigr]
\crbig
&& \makebox{$+$ fermion terms}\,,
\end{array}
\eeq
where
\begin{eqnarray}
(D_\mu z)^i & = & \partial_\mu z^i+ {i\over2} A^a_\mu(T_r^a)^i{}_jz^j\,,\\
H_{\mu\nu\rho} & = & h_{\mu\nu\rho}-\omega_{\mu\nu\rho}\,,
\end{eqnarray}
in which $\omega$ is the Chern--Simons form normalized such that
\begin{equation}
dH=-\widetilde\Tr F\wedge F\,.
\end{equation}
The kinetic metrics are then ${\cal H}_{z\ov z}$, $-{1\over2}{\cal H}_{CC}$ and ${\cal H}_C$ for the components of superfields $\Phi$, $L$ and ${\cal W}_\alpha$ respectively.

The field equations for theory (\ref{CSF4}) are\footnote{We use the convention 
$\ov {\cal W}_\dalpha = {1\over4}DD e^{\cal A} \ov D_\dalpha e^{-{\cal A}}$, with 
$\ov{\cal W}_\dalpha = - ({\cal W}_\alpha)^\dagger$. }
\beq
\label{CSF5}
\begin{array}{rrcl}
L:& \qquad\qquad \ov{DD}D_\alpha{\cal H}_L &=& 0,
\crbig
\Phi:& \ov{DD}{\cal H}_\Phi &=& 4\,W_\Phi,
\crbig
{\cal A} : &  \ov D^\dalpha \Bigl[ {\cal H}_L \, e^{-{\cal A}} \ov {\cal W}_\dalpha e^{\cal A} \Bigr] 
&=& {\cal W}^\alpha \, D_\alpha{\cal H}_L - T(r)\,{\cal H}_Y\, \Phi \ov\Phi e^{\cal A},
\end{array}
\eeq
with index $\Tr(T^a_rT^b_r) = T(r)\delta^{ab}$.

To derive the field equation for the gauge superfield ${\cal A}$, it is indeed easier to use the dual chiral 
version of the theory,\footnote{To avoid dealing with the complicated non-Abelian 
Chern-Simons superfield \cite{CFV}.}
\beq
\label{CSF6}
\begin{array}{rcl}
{\cal L} &=& \Dint {\cal K}(S+\ov S, Y)
\crbig
&&+ \Fint \left[ W(\Phi) + {1\over4}S \,\widetilde \Tr\,{\cal WW}\right] 
+ \Fbarint \left[ \ov W(\ov\Phi) + {1\over4} \ov{S}\,\widetilde\Tr\,\ov{\cal WW} \right] ,
\end{array}
\eeq
and to apply on the resulting field equation the Legendre transformation into the linear version. 
Variation of eq.~(\ref{CSF6}) and use of the Bianchi identity
\beq
\label{CSF7}
D^\alpha( e^{\cal A} {\cal W}_\alpha e^{-{\cal A}} ) = e^{\cal A} \ov D_\dalpha 
( e^{-{\cal A}} \ov{\cal W}^\dalpha e^{\cal A}) e^{-{\cal A}} 
\eeq
gives then the field equation
\beq
\label{CSF8}
(S+\ov S)\, \ov D_\dalpha( e^{-{\cal A}} \ov {\cal W}^\dalpha e^{\cal A} ) 
= - (D^\alpha S) \, {\cal W}_\alpha - (\ov D_\dalpha\ov S)\, e^{-{\cal A}} \ov {\cal W}^\dalpha e^{\cal A}
+ 2\,T(r)\,{\cal K}_Y\, \Phi \ov\Phi e^{\cal A}.
\eeq
It can be rewritten
\beq
\label{CSF9}
\ov D^\dalpha \Bigl[ (S+\ov S) \, e^{-{\cal A}} \ov {\cal W}_\dalpha e^{\cal A} \Bigr] = 
D^\alpha(S+\ov S) \, {\cal W}_\alpha
- 2\,T(r)\,{\cal K}_Y\, \Phi \ov\Phi e^{\cal A} .
\eeq
Multiplying by ${\cal W}_\beta$ and taking the trace gives
\beq
\label{CSF10}
\ov D^\dalpha \Bigl[ (S+\ov S) \Tr ({\cal W}_\beta e^{-{\cal A}} \ov {\cal W}_\dalpha e^{\cal A}) \Bigr] 
= {1\over2} D_\beta (S+\ov S) \, \Tr{\cal WW} 
+ 2\,T(r)\,{\cal K}_Y\,  \ov\Phi e^{\cal A} {\cal W}_\beta \Phi.
\eeq
The Legendre transformation indicates then that ${\cal K}_Y = {\cal H}_Y$ and 
$S+\ov S = 2{\cal H}_L$, which in turn implies the field equation (\ref{CSF5}) for ${\cal A}$ and the relation
\beq
\label{CSF11}
\ov D^\dalpha \Bigl[ {\cal H}_L \Tr ({\cal W}_\beta e^{-{\cal A}} \ov {\cal W}_\dalpha e^{\cal A}) \Bigr] 
= {1\over2} D_\beta {\cal H}_L \, \Tr{\cal WW} 
+ T(r){\cal H}_Y\,  \ov\Phi e^{\cal A} {\cal W}_\beta \Phi.
\eeq
With field equations (\ref{CSF5}) and relation (\ref{CSF11}), identity (\ref{CSF3}) finally leads to the 
supercurrent structure\footnote{The superfields $J_{\alpha\dalpha}$, $X$ and $\chi_\alpha$
can be calculated directly from the Lagrangian. They are then defined off-shell, but field equations
(\ref{CSF5}) can be used to reformulate them since the superfield 
equation $\ov D^\dalpha J_{\alpha\dalpha} = D_\alpha X + \chi_\alpha$ only holds on-shell.}
\beq
\label{CSF12}
\begin{array}{rcl}
\ov D^\dalpha J_{\alpha\dalpha} &=& D_\alpha X + \chi_\alpha ,
\crbig
J_{\alpha\dalpha} &=&
- 2 \Bigl[ (\ov{\cal D}_\dalpha\ov\Phi) {\cal H}_{\Phi\ov\Phi}  ({\cal D}_\alpha\Phi)
- {\cal H}_{LL}(\ov D_\dalpha\hat L)(D_\alpha\hat L) 
+ 2 \, {\cal H}_L \widetilde\Tr ({\cal W}_\alpha e^{-{\cal A}} \ov {\cal W}_\dalpha e^{\cal A}) \Bigr] ,
\crbig
X &=& 4\, W, 
\crbig
\chi_\alpha &=& \ov{DD}D_\alpha({\cal H}-\hat L{\cal H}_L) .
\end{array}
\eeq
This supercurrent structure can be considered as natural for theory (\ref{CSF4}). 
It actually also applies if ${\cal H}$ is simply a gauge-invariant function of $\hat L$, $\Phi$ and 
$\ov\Phi e^{\cal A}$, instead of a function of $\hat L$ and $Y$.

In the supercurrent structure (\ref{CSF12}), field equations have not been used to generate from
identity (\ref{CSF3}) the source superfield $\chi_\alpha$ and the chiral or linear contributions to the 
supercurrent superfield $J_{\alpha\dalpha}$. Field equations for ${\cal A}$, $\Phi$ and $L$ have been 
respectively used to generate the gauge supercurrent 
term\footnote{Field equations are needed to derive the Yang-Mills Belinfante energy-momentum tensor 
from the canonical tensor.}, the chiral source $X$ and to eliminate the first term in the right-hand
side of identity (\ref{CSF3}).

Using expansion (\ref{appD4}) for the superfield $J_\mu=(\ov\sigma_\mu)^{\dalpha\alpha}J_{\alpha\dalpha}$, 
we find that the supercurrent superfield (\ref{CSF12}) contains the following lowest component:
\beq
\label{CSF13}
j^{\widetilde R}_\mu\equiv  {3\over8}(\ov\sigma_\mu)^{\dalpha\alpha}J_{\alpha\dalpha}\big|_{\theta=0}=-{3\over2}\,{\cal H}_{z\ov z} \,\psi \sigma_\mu\ov\psi + {3\over4}\,{\cal H}_{CC}\, \chi\sigma_\mu\ov\chi 
+ {3\over2}\,{\cal H}_C\, \widetilde\Tr \lambda\sigma_\mu\ov\lambda \, ,
\eeq
where we use the expansions
$$
\hat L = C + i\theta\chi - i\ov\theta\ov\chi + \ldots,
\qquad
\Phi = z + \sqrt2\,\theta\psi-\theta\theta f+\ldots ,
\qquad
{\cal W}_\alpha = -i\lambda_\alpha + \ldots
$$
(and $\ov{\cal W}_\dalpha = -i \ov\lambda_\dalpha+ \ldots$). This is the Noether current of 
$\widetilde R$--transformations with chiral charges $-3/2$, $-3/2$ and $3/2$ for $\chi$, $\psi$ 
and $\lambda$ 
respectively. The chiral charges of superfields $\Phi$, $L$ and ${\cal W}_\alpha$ 
for this $U(1)_{\widetilde R}$ are then $q=0$, $0$, $3/2$ in this supercurrent structure and
$U(1)_{\widetilde R}$ only acts on the Grassmann coordinates. It is an automatic symmetry of
$D$--term Lagrangians and, according to the second eq.~(\ref{appD5}), 
the $\widetilde R$ current is conserved if the superpotential vanishes,
$\partial^\mu j_\mu^{\widetilde R} = -{3\over2}\Im f_X$. 

The supercurrent superfield $J_{\alpha\dalpha}$ of eqs.~(\ref{CSF12}) contains the Belinfante 
improved (symmetric, gauge-invariant) energy-momentum tensor $T_{\mu\nu}$ for theory (\ref{CSF4}). 
Omitting fermions and gauge fields, its expression is 
\beq
\label{CSF13b}
\begin{array}{rcl}
T_{\mu\nu}Ê&=& -{1\over2} {\cal H}_{CC} (\partial_\mu C)(\partial_\nu C)
- {1\over4} {\cal H}_{CC}h_{\mu\rho\sigma}{h_\nu}^{\rho\sigma}
+  {\cal H}_{z\ov z}[(\partial_\mu z)(\partial_\nu\ov z) + (\partial_\nu z)(\partial_\mu\ov z)]
\crbig
&& - \eta_{\mu\nu} \Bigl(  -{1\over4} {\cal H}_{CC} (\partial_\rho C)(\partial^\rho C)
- {1\over24} {\cal H}_{CC}h_{\rho\sigma\lambda}h^{\rho\sigma\lambda}
+ {\cal H}_{z \ov z} [ (\partial_\rho z)(\partial^\rho\ov z) + \ov ff ]  \Bigr)
\crbig
&& + {1\over2} \eta_{\mu\nu} {\cal H}_C \widetilde\Tr(D^2) + {1\over2}\eta_{\mu\nu} \Re f_X ,
\end{array}
\eeq
with auxiliary fields\footnote{The auxiliary field contribution to $T_{\mu\nu}$ is
$\eta_{\mu\nu}V$, where $V$ is the usual scalar potential 
$$
V (C, z, \ov z) = {1\over2} {\cal H}_C \widetilde\Tr D^2 + {\cal H}_{\ov zz}\ov f f.
$$} 
$$f_X = 4W_zf, \qquad \ov f {\cal H}_{\ov zz} = W_z, \qquad
D^a = -{1\over2} {\cal H}_C^{-1} {\cal H}_z T^a_r z
= -{1\over2} {\cal H}_C^{-1} {\cal H}_Y \ov z T^a_rz.
$$
Notice that terms depending on 
${\cal H}_{Cz}$ or ${\cal H}_{C\ov z}$ present in the Lagrangian do not appear
in the Belinfante tensor $T_{\mu\nu}$. If the superpotential vanishes, as we will often 
assume, we have $f=f_X=0$.

\subsection{Scale transformations}\label{secscale}

The supercurrent superfield $J_{\alpha\dalpha}$ includes the $U(1)_{\widetilde R}$ current and the Belinfante
energy-momentum tensor which can then be viewed as partners under Poincar\'e supersymmetry. The superconformal algebra, besides $U(1)_R$ transformations, 
also includes scale transformations, but the dilatation current is not present in $J_{\alpha\dalpha}$.

To discuss the behaviour of the theory under scale transformations, we first use that the source superfield  
$\chi_\alpha$ contributes to the trace of the Belinfante energy-momentum tensor, 
according to the first eq.~(\ref{appD5}). We then define the real superfield
\beq
\label{CSF13c}
\Delta_{(0)} = 2\hat L{\cal H}_{L} - 2{\cal H}, \qquad\qquad\qquad
\chi_\alpha = - {1\over2} \ov{DD}D_\alpha \Delta_{(0)}.
\eeq
Then, using the field equation for $C$, the quantity
\beq
\label{CSF13d}
\delta_{(0)} \equiv \displaystyle
{\partial{\cal L}\over\partial C}2C
+ {\partial{\cal L}\over\partial\partial^\mu C}3\partial^\mu C
+  {\partial{\cal L}\over\partial h^{\mu\nu\rho}}3h^{\mu\nu\rho}
+{\partial{\cal L}\over\partial\partial^\mu z}\partial^\mu z 
+ {\partial{\cal L}\over\partial \partial^\mu\ov z}\partial^\mu\ov z
- 4{\cal L} ,
\eeq
which is the variation of the bosonic Lagrangian under a scale transformation with
scale dimensions $w=2$ for $L$ and $w=0$ for $\Phi$, verifies
\beq
\label{CSF13e}
\begin{array}{rcl}
\delta_{(0)} 
&=& -\partial^\mu[ C{\cal H}_{CC}\partial_\mu C ] + {T^\mu}_\mu
\crbig
&=&
\displaystyle
-{1\over2} \partial^\mu \left[ {\partial\over\partial C}\Delta_{(0)} 
\Bigr|_{\theta=0} \,\partial_\mu C \right] + {T^\mu}_\mu.
\end{array}
\eeq
The expression for $\delta_{(0)}$ is the Lagrangian variation under a dilatation of the fields at fixed coordinates $x$
(or at $x=0$), and this last equation suggests to define a dilatation current
\beq
\label{CSF13f}
j_\mu^{D} = -{1\over2}\left[ {\partial\over\partial C}\Delta_{(0)} \Bigr|_{\theta=0} \,\partial_\mu C \right] 
+ x^\nu T_{\mu\nu}
\eeq
verifying $\partial^\mu j_\mu^{D} = \delta_{(0)}$ as it should. Even if it does not appear in the natural supercurrent 
structure (\ref{CSF12}), this dilatation current is naturally associated by Poincar\'e supersymmetry with the 
$\widetilde R$ current (\ref{CSF13}) present in the supercurrent superfield $J_{\alpha\dalpha}$. 
Both currents correspond to zero $U(1)_{\widetilde R}$ charge $q$ and
scale dimension $w$ for the chiral superfield $\Phi$, the equality $w=q$ following from the superfield supercurrent equations of Poincar\' e supersymmetry and of the underlying superconformal character of the supermultiplets.

If the theory would be scale-invariant, $W=\Delta_{(0)} = 0$ and the anomaly source 
superfields $X=4W$ and $\chi_\alpha= -{1\over2}\ov{DD}D_\alpha \Delta_{(0)}$ would also vanish. An example
is ${\cal H} = \hat L$ which leads to the superconformal super-Yang-Mills Lagrangian. Then,
$\partial^\mu j_\mu^{D}= {T^\mu}_\mu = 0$. 
If however $\Delta_{(0)}\ne 0$, the divergence of the 
dilatation current is not given by the nonzero trace of the Belinfante energy-momentum tensor: with
the linear superfield, there is a virial current.
With scale dimension zero chiral superfields a (two derivative\footnote{The real scale-invariant variable 
$\widetilde\Tr{\cal WW} \,\widetilde\Tr\ov{\cal WW}\hat L^{-3}$ leads to four-derivative terms.}) 
scale-invariant theory is generated by ${\cal H} = \hat L \,{\cal F}(\Phi, \ov\Phi)$.
The first field equation (\ref{CSF5}) for $\hat L$ only makes sense if
${\cal F} = f(\Phi) + \ov f(\ov\Phi)$, in which case
the linear superfield disappears from the dynamical Lagrangian which simply couples the holomorphic
$f(\Phi)$ to $\widetilde\Tr{\cal WW}$.\footnote{Chiral-linear duality as described in Section 
\ref{seclinear} cannot be performed.} 

We now want to generalize this discussion
to the case of a nonzero scale dimension $w$ of the chiral fields, in view of a supersymmetric improvement
of the natural (Belinfante) supercurrent structure.

With respect to a system with chiral and gauge superfields only, the presence of the 
linear superfield introduces some technical subtleties\footnote{See Appendix \ref{AppB}.}
which play a role when discussing the behaviour of the theory under 
scale transformations. Since these subtleties involve scalar fields only,
we omit fer\-mions and gauge fields in this subsection.
Assigning scale dimensions $w$ and two to the superfields $\Phi$
and $\hat L$,
the bosonic quantity which measures the breaking of scale invariance
is
\beq
\label{CSF14}
\begin{array}{rcl}
\delta_{(w)} &=& \displaystyle
{\partial{\cal L}\over\partial C}\, 2C
+{\partial{\cal L}\over\partial z} \, wz
+{\partial{\cal L}\over\partial \ov z} \, w\ov z
+{\partial{\cal L}\over\partial\partial_\mu C} \,3\, \partial_\mu C
+{\partial{\cal L}\over\partial h_{\mu\nu\rho}}\, 3 \, h_{\mu\nu\rho}
\crbig && \displaystyle
+{\partial{\cal L}\over\partial\partial_\mu z}(w+1) \partial_\mu z
+{\partial{\cal L}\over\partial\partial_\mu\ov z}(w+1)\partial_\mu\ov z
- 4{\cal L}.
\end{array}
\eeq 
Using the field equations, it can be written as
\beq
\label{CSF15}
\delta_{(w)} = \partial^\mu{\cal V}_{(w)\mu} + {T^\mu}_\mu
\eeq
in terms of the trace of the Belinfante gauge-invariant energy-momentum tensor $T_{\mu\nu}$ and the virial current
\beq
\label{CSF16}
{\cal V}_{(w)\mu} = -C{\cal H}_{CC} \partial_\mu C
+ wz{\cal H}_{z\ov z}\partial_\mu\ov z + w\ov z{\cal H}_{z\ov z}
\partial_\mu z - {i\over12}w\epsilon_{\mu\nu\rho\sigma}
h^{\nu\rho\sigma}(z{\cal H}_{Cz}-\ov z{\cal H}_{C\ov z}).
\eeq
This in turn indicates that the dilatation current is
\beq
\label{CSF17}
\delta_{(w)} = \partial^\mu j_\mu^{D} 
\qquad\qquad
j_\mu^{D} = {\cal V}_{(w)\mu} + x^\nu T_{\mu\nu}
\eeq
up maybe to a conserved current.
Notice that
\beq
\label{CCJ12}
{\cal V}_{(w)\mu} = {\partial{\cal L}\over\partial\partial^\mu C}2C
+ {\partial{\cal L}\over\partial\partial^\mu z}wz
+ {\partial{\cal L}\over\partial\partial^\mu\ov z}w\ov z
\eeq
is gauge-invariant and does not include a term related to the variation of the antisymmetric 
tensor.\footnote{This result only holds if scale dimension two is assigned to the linear superfield. } 
Notice also that the contribution quadratic in $h_{\mu\nu\rho}$ in the energy-momentum tensor 
(\ref{CSF13b}) would be traceless in six dimensions. This follows from a general result \cite{Pons}: 
in $2(p+1)$ dimensions, the kinetic Lagrangian of a $p$-form field with gauge invariance is scale and 
conformal invariant with canonical dimension $w=p$.

Defining the superfields
\beq
\label{CSF18}
\begin{array}{rcll}
\Delta_{(w)} (L,\Phi, \ov\Phi e^{\cal A}) &=& 2\hat L{\cal H}_L + w{\cal H}_\Phi \Phi
+ w\ov\Phi{\cal H}_{\ov\Phi} - 2{\cal H},
\qquad & \makebox{($\Delta_{(w)}$ real)},
\crbig
\widetilde\Delta_{(w)} (\Phi) &=& {w\over4} \ov{ DD}({\cal H}_\Phi \Phi) - 3W,
\qquad & (\ov D_\dalpha\widetilde\Delta_{(w)} = 0),
\end{array}
\eeq
leads to the relation
\beq
\label{CSF19}
\begin{array}{rcl}
\delta_{(w)} &=& {\Delta_{(w)}}|_{\theta\theta\ov{\theta\theta}}
+ {1\over4}\Box \Delta_{(w)}(C,z,\ov z)
- {1\over2} \partial^\mu \Bigl[ \Delta_{(w)C}(C,z,\ov z)
\partial_\mu C \Bigr]
+ \widetilde\Delta_{(w)}|_{\theta\theta}
+ \ov{\widetilde\Delta}_{(w)}|_{\ov\theta\ov\theta}
\crbig
&=& {1\over2} D_{\Delta_{(w)}} 
- {1\over2} \partial^\mu \Bigl[ \Delta_{(w)C}(C,z,\ov z)\partial_\mu C \Bigr]
- f_{\widetilde\Delta_{(w)}} - \ov f_{\ov{\widetilde\Delta}_{(w)}},
\end{array}
\eeq
with $D_{\Delta_{(w)}}$ as defined in the appendices [eqs.~(\ref{appD3b}) or (\ref{appH4})] and
$\Delta_{(w)C}={\partial\over\partial C}\Delta_{(w)}$. Using equations (\ref{appH1}) and (\ref{appH4}) it can be shown that $\delta_{(w)}$ takes the same functional form as the bosonic Lagrangian (\ref{bosonicL}) but with the substitutions: ${\cal H}$ replaced by $\Delta_{(w)}$ and $W$ replaced by $\widetilde\Delta_{(w)}$. Note the appearance 
of a supplementary derivative term in $\delta_{(w)}$ whenever a linear superfield is present. This equation remains true
in the fully supersymmetric theory with fermion and gauge fields: the supplementary derivative depends 
on scalar fields only. We then have:
\beq
\label{CSF19b}
\begin{array}{rcl}
{\cal V}_{(w)\mu} &=&  -{1\over2}\Delta_{(w)C}\partial_\mu C
+{w\over2}\partial_\mu ( z{\cal H}_z + \ov z{\cal H}_{\ov z})
\crbig
&&- {1\over2}\biggl[ {i\over6}\epsilon_{\mu\nu\rho\sigma}
h^{\nu\rho\sigma}{\partial\over\partial C}
+ (\partial_\mu z){\partial\over\partial z}
- (\partial_\mu\ov z){\partial\over\partial\ov z} \Biggr]
(wz{\cal H}_{z}-w\ov z{\cal H}_{\ov z}).
\end{array}
\eeq
This equality is true for an arbitrary function ${\cal H}(C,z,\ov z)$. Since the choice $w=0$ has been
discussed earlier, we consider now $w\ne0$.

Two cases then exist. Firstly, if the function ${\cal H}$ has a $U(1)$ symmetry with charges proportional to the scale dimension $w$, 
then $wz{\cal H}_{z} = w\ov z{\cal H}_{\ov z}$ and 
\beq 
\label{CSF20}
{\cal V}_{(w)\mu} = -{1\over2}\Delta_{(w)C} \, \partial_\mu C
+ {w\over2}\partial_\mu ( z{\cal H}_z + \ov z{\cal H}_{\ov z}).
\eeq
As shown explicitly in the next subsection, the second term can be eliminated by an improvement to a new
energy-momentum tensor $\Theta_{\mu\nu}$ and to a new virial 
current $\widehat{\cal V}_\mu$ for which, in view of
eqs.~(\ref{CSF15}) and (\ref{CSF19}),
\beq
\label{CSF21}
\partial^\mu j_\mu^{D} = 
\partial^\mu \widehat{\cal V}_\mu + {\Theta^\mu}_\mu
= -{1\over2} \partial^\mu [ \Delta_{(w)C}\partial_\mu C ]
+ {\Theta^\mu}_\mu
\eeq
and 
\beq
\label{CSF22}
{\Theta^\mu}_\mu = {1\over2} D_{\Delta_{(w)}} 
- 2 \Re f_{\widetilde\Delta_{(w)}}.
\eeq
Notice that the $U(1)$ symmetry of ${\cal H}$ does not need to be
an $R$--symmetry.\footnote{This observation extends results stated in ref.~\cite{AB}.}
In this first case, if the theory is scale-invariant, {\it i.e.}~if we have $\Delta_{(w)} = \widetilde\Delta_{(w)}=0$, then it follows that
\beq
\label{CSF22b} 
\widehat{\cal V}_\mu = 
{\Theta^\mu}_\mu = 0
\eeq
and the theory is conformal since the currents
\beq
\label{CSF23}
K_\mu^\alpha = (2x^\alpha x^\nu - \eta^{\alpha\nu}x^2)\,\Theta_{\mu\nu}
\eeq
are conserved, $\partial^\mu K_\mu^\alpha=0$. 
If ${\cal H}$ has a $U(1)$ symmetry but scale invariance is broken,
${\Theta^\mu}_\mu$ is given by the highest components of the superfields
$\Delta_{(w)}$ and $\widetilde\Delta_{(w)}$ which measure the breaking of scale invariance, according to 
eqs.~(\ref{CSF21}) and (\ref{CSF22}). But if $\Delta_{(w)C}\ne0$, the divergence of the dilatation current 
is not given by the trace ${\Theta^\mu}_\mu$. This discussion includes the case $w=0$ considered 
earlier.
Since we restrict ourselves to ${\cal H}(\hat L, Y)$, the $U(1)$ symmetry exists 
and the improvement transformation will be performed at the superfield 
level in the next subsection. 

In the second option, ${\cal H}$ does not have the global $U(1)$ symmetry, 
$wz{\cal H}_{z} \ne w\ov z{\cal H}_{\ov z}$ ($\forall w\ne0$). The
chiral superfield interactions provide then an example of a classical 
theory where scale invariance ($\Delta_{(w)} = \widetilde\Delta_{(w)}=0$) does not imply conformal invariance because the virial current in (\ref{CSF19b}) cannot be transformed away by an improvement transformation. This case is briefly discussed in Appendix \ref{AppC}.

In any case, the message of this subsection is that even when ${\cal H}$ has a $U(1)$ symmetry but the theory is not scale invariant because 
$\Delta_{(w)}\ne 0$ and there is a non-trivial coupling of a linear superfield to chiral superfields such that $\Delta_{(w)C}\partial_\mu C$ is not a 
derivative, one cannot construct an energy-momentum tensor $\Theta_{\mu\nu}$ which is such that $\partial^\mu j_\mu^{D} = {\Theta^\mu}_\mu$. 
Whenever there does exist an energy-momentum tensor $\Theta_{\mu\nu}$ such that $\partial^\mu j_\mu^{D} = {\Theta^\mu}_\mu$ we will refer 
to it as the Callan, Coleman, Jackiw (CCJ) \cite{CCJ, CJ} energy-momentum tensor (see Appendix \ref{AppC}).

\subsection{Improved supercurrent structure: making scale (non-)\newline invariance manifest}\label{secimpsc}

Just like in the previous subsection we assume that the chiral superfields $\Phi$ have an arbitrary scale 
dimension(s) $w$. The canonical value 
is $w=1$, but dimensions can be anomalous.
The dimension of $\hat L$ is always two.\footnote{The 
dimension of $\Omega$ is canonical.
Notice that $L$ contains then a dimension-three vector field
$v_\mu = \epsilon_{\mu\nu\rho\sigma}\partial^\nu b^{\rho\sigma}$ which is conserved or transverse, 
$\partial^\mu v_\mu=0$.} 
In terms of the superfields $\Delta_{(w)}$ and $\widetilde\Delta_{(w)}$ defined in
eqs.~(\ref{CSF18}), the anomaly superfields of the natural supercurrent structure read
\beq
\label{Imp2}
X = -{4\over3} \widetilde\Delta_{(w)} + {4\over3}w \, W_\Phi\Phi, 
\qquad\qquad
\chi_\alpha = -{1\over2}\ov{DD}D_\alpha\Delta_{(w)}
+ {w\over2} \ov{DD}D_\alpha({\cal H}_\Phi \Phi 
+ \ov\Phi{\cal H}_{\ov\Phi}).
\eeq
We may then improve the supercurrent structure using transformation (\ref{appD7}) with
\beq
\label{Imp3}
{\cal G} = -{w\over6}({\cal H}_\Phi\Phi + \ov\Phi{\cal H}_{\ov\Phi})
\eeq
to eliminate the second term in $\chi_\alpha$.
The resulting chiral anomaly superfield is
\beq
\label{Imp5}
\widetilde X = -{4\over3} \widetilde\Delta_{(w)} + {4\over3}w \, W_\Phi\Phi
-{w\over6}\ov{DD}({\cal H}_\Phi\Phi + \ov\Phi{\cal H}_{\ov\Phi}) 
\eeq
and the field equation of $\Phi$ leads then to the supercurrent structure
\beq
\label{Imp4}
\begin{array}{rcl}
\ov D^\dalpha \widetilde J_{\alpha\dalpha} &=& D_\alpha \widetilde X + \widetilde\chi_\alpha ,
\crbig
\widetilde J_{\alpha\dalpha} &=&
- 2 \Bigl[ (\ov{\cal D}_\dalpha\ov\Phi) {\cal H}_{\Phi\ov\Phi}  ({\cal D}_\alpha\Phi)
- {\cal H}_{LL}(\ov D_\dalpha\hat L)(D_\alpha\hat L) 
+ 2 \, {\cal H}_L \widetilde\Tr ({\cal W}_\alpha e^{-{\cal A}} \ov {\cal W}_\dalpha e^{\cal A}) \Bigr] 
\crbig
&& \hspace{2.2cm}
-{w\over3} [D_\alpha,\ov D_\dalpha] ({\cal H}_\Phi\Phi + \ov\Phi{\cal H}_{\ov\Phi}) ,
\crbig
\widetilde X &=& -{4\over3} \widetilde\Delta_{(w)} + {w\over6}\ov{DD}({\cal H}_\Phi\Phi - \ov\Phi{\cal H}_{\ov\Phi}),
\crbig
\widetilde \chi_\alpha &=& -{1\over2}\ov{DD}D_\alpha\Delta_{(w)} .
\end{array}
\eeq
The extension of these formulas to a reducible matter content, with independent scale dimensions 
$w_{(i)}$ for each irreducible component $\Phi_{(i)}$ is straightforward. 

In the case of the canonical Wess-Zumino model, ${\cal H} = \ov\Phi\Phi$, the improved supercurrent 
superfield reduces to 
\beq
\widetilde J_{\alpha\dalpha} = {4\over3}\left[ \left( w-{3\over2}\right) (\ov D_\dalpha\ov\Phi)(D_\alpha\Phi) 
-iw \, (\sigma^\mu)_{\alpha\dalpha}\, \ov \Phi \stackrel{\leftrightarrow}{\partial_\mu} \Phi \right]
\eeq 
with $R$--current
\beq
j_\mu = \left( w-{3\over2}\right)  \psi\sigma_\mu\ov\psi - iw \ov z \stackrel{\leftrightarrow}{\partial_\mu} z ,
\eeq
two results often used in the literature with canonical scale dimension $w=1$.

These formulas hold for a function ${\cal H}(\hat L , \Phi, \ov\Phi e^{\cal A})$. They simplify if 
${\cal H}(\hat L, Y)$, as in our theory (\ref{CSF4}):
\beq
\label{Imp6}
\begin{array}{rcl}
\ov D^\dalpha \widehat J_{\alpha\dalpha} &=& D_\alpha \widehat X + \widehat\chi_\alpha ,
\crbig
\widehat J_{\alpha\dalpha} &=&
- 2 \Bigl[ (\ov{\cal D}_\dalpha\ov\Phi) {\cal H}_{\Phi\ov\Phi}  ({\cal D}_\alpha\Phi)
- {\cal H}_{LL}(\ov D_\dalpha\hat L)(D_\alpha\hat L) 
+ 2 \, {\cal H}_L \widetilde\Tr ({\cal W}_\alpha e^{-{\cal A}} \ov {\cal W}_\dalpha e^{\cal A}) \Bigr] 
\crbig
&& \hspace{2.2cm}
-{2\over3} [D_\alpha,\ov D_\dalpha] (w{\cal H}_Y Y) ,
\crbig
\widehat X &=& -{4\over3} \widetilde\Delta_{(w)} ,
\crbig
\widehat \chi_\alpha &=& -{1\over2}\ov{DD}D_\alpha\Delta_{(w)} .
\end{array}
\eeq
In $\widehat J_{\alpha\dalpha}$, the energy-momentum tensor 
$\Theta_{\mu\nu}$ is related to the
Belinfante tensor by the improvement
\beq
\label{Imp7}
\begin{array}{rcl}
\Theta_{\mu\nu} &=& T_{\mu\nu} - {1\over6} (\partial_\mu\partial_\nu - \eta_{\mu\nu}\Box)
w({\cal H}_zz + \ov z{\cal H}_{\ov z}) \crbig
&=& T_{\mu\nu} - {1\over3} (\partial_\mu\partial_\nu - \eta_{\mu\nu}\Box)
w{\cal H}_yy , \qquad\qquad y=\ov zz
\end{array}
\eeq
and the corresponding improved (scalar) virial current is
\beq
\label{Imp8}
\widehat{\cal V}_\mu = {\cal V}_{(w)\mu} - w\, \partial_\mu
({\cal H}_yy) =
- {1\over2} \Delta_{(w)C}\partial_\mu C
\eeq
as explained in the previous subsection. Using (\ref{appD5}) we find that ${\Theta^\mu}_\mu$ satisfies eq.~(\ref{CSF22}) while the dilatation current verifies eq.~(\ref{CSF21}).

The superfield improvement transformation can also be understood in the following way.
The field equation $\ov{DD}{\cal H}_\Phi= 4 W_\Phi$ implies 
\beq
\label{Imp10a}
{w\over2}\ov{DD} ( {\cal H}_\Phi\Phi + \ov\Phi{\cal H}_{\ov\Phi} ) = 4wW_\Phi\Phi 
- {w\over2}\ov{DD} ( {\cal H}_\Phi\Phi - \ov\Phi{\cal H}_{\ov\Phi} ).
\eeq
The right-hand side vanishes if the theory is invariant under phase rotations of $\Phi$.
In this case,  
\beq
\label{Imp10b}
{\cal Z} = {w\over2} ( {\cal H}_\Phi\Phi + \ov\Phi{\cal H}_{\ov\Phi} ) = -3 \, {\cal G}
\eeq
includes the Noether current of the $U(1)_Z$ symmetry (with charge $w$ on $\Phi$) in its 
$\theta\sigma^\mu\ov\theta$ component and $\ov{DD}\,{\cal Z}=0$ is 
the supersymmetric extension of its conservation equation.
Taking the derivative $D_\alpha$ of eq.~(\ref{Imp10a}), using identity (\ref{appD6}) and the
field equation leads then to the improvement transformation to the supercurrent structure (\ref{Imp4}).
Hence, the supercurrent superfield $\widehat J_{\alpha\dalpha}$ includes in its lowest component the 
current of the $R$ transformation with $R$ charges $0$ and $w$ for $\hat L$ and $\Phi$ respectively. Gauginos, and 
fermions $\psi$ in $\Phi$ and $\chi$ in $L$ have chiral weights $3/2$, $w-3/2$ and $-3/2$ respectively.
Notice that $w$ has been originally introduced as the scale dimension 
of $\Phi$ and it here also plays the role of an $R$ charge. This is reminiscent of the chirality condition
in a superconformal theory, in which the scale dimension and the $U(1)_R$ charge are identified.

We may further improve the structure (\ref{Imp6}) to a Ferrara-Zumino supercurrent with $\chi_\alpha=0$.
This second improvement would lead to a supercurrent depending on the superfield $\Delta_{(w)}$,
\beq
\widehat J_{\alpha\dalpha} \quad\longrightarrow\quad
\widehat J_{\alpha\dalpha} + {1\over3}[D_\alpha,\ov D_\dalpha]\Delta_{(w)}.
\eeq
The content of the supercurrent structure (\ref{Imp6}) is however more intuitive, with the Lagrangian 
superfield ${\cal H}$ defining the supercurrent superfield $\widehat J_{\alpha\dalpha}$ and the 
scale- and $R$-breaking superfields $\Delta_{(w)}$ and $\widetilde\Delta_{(w)}$
defining the source superfields $\widehat X$ and $\widehat\chi_\alpha$.
In the following, we will use the improved supercurrent structure (\ref{Imp6}) as a starting point and we
will be mostly concerned with the case $W(\Phi)=0=\widehat X$. This structure is both $\widetilde R$-- and $R$--symmetric
and is then naturally related to new-minimal supergravity.

If we wish to cancel the virial current completely, we need that ${\cal V}_\mu$ is a derivative, 
and then $\Delta_C$ should be a function of $C$ only. This is the case if
\beq
\label{CL3}
{\cal H}(\hat L,Y) = {\cal F}(\hat L) + {\cal K}(Y) + {\cal I}(\hat L,Y), 
\qquad\qquad
wY{\cal I}_Y + \hat L{\cal I}_L = {\cal I}.
\eeq
Then, 
$$
\Delta_{(w)} = 2\hat L{\cal F}_L -2 {\cal F} +  2wY{\cal K}_Y - 2 {\cal K}, 
\qquad\qquad
\Delta_{(w)LY} = 0.
$$
The second equation (\ref{CL3}) has a very simple significance: 
with dimensions $w$ and two for $\Phi$ and $\hat L$,
the interaction term must be a dimension-two function. Hence, the interaction Lagrangian 
is scale invariant:
\beq
{\cal I}(\hat L, Y) = \hat L \, \widetilde {\cal I}(X), \qquad\qquad
X = Y \hat L^{-w}.
\eeq
Only in this case can we find an energy-momentum tensor such that 
$\partial^\mu j_\mu^{D} = {\Theta^\mu}_\mu$.

For reasons explained in Section \ref{secanom}, it will be natural to write the anomaly source 
superfields as a sum of a classical contribution and an anomalous term as follows
\beq
\label{anom1}
\Delta_{(w)} = \Delta_{\rm{classical}} + \Delta_{\rm{anom.}} , \qquad\qquad  \Delta_{\rm{anom.}} = B \, \hat L,
\eeq
and
\beq
\label{anom2}
\widetilde\Delta_{(w)} = \widetilde\Delta_{\rm{classical}} + \widetilde\Delta_{\rm{anom.}} ,  \qquad\qquad  \widetilde\Delta_{\rm{anom.}}=A \, \widetilde\Tr{\cal WW} ,
\eeq
with some numerical coefficients $A$ and $B$ to be discussed below. An anomalous contribution 
$\Delta_{anom.} = B\hat L$ arises if an anomaly term
\beq
\label{anom3}
{\cal H}_{\rm{anom.}}(\hat L) = {B\over2} ( \hat L\ln\hat L - \hat L)
\eeq
is added to ${\cal H}$. Similarly, an anomalous contribution $\widetilde\Delta_{\rm anom.}$ may be 
obtained if the superpotential is allowed to depend on $\widetilde\Tr{\cal WW}$: this is the subject 
of the next subsection.

\subsection{Adding a dependence to the superpotential on \boldmath{$\widetilde\Tr{\cal WW} $}}\label{secUdep}

The chiral superfield $\widetilde\Tr{\cal WW} $ has a fermionic lowest component.
In principle, the superpotential $W$ could also be a function of $\widetilde\Tr{\cal WW} $,
$W(\Phi,\widetilde\Tr{\cal WW} )$, but this dependence does not play any role in the bosonic 
Lagrangian and for the gauge coupling field, except for a linear term which is already included in
${\cal H}(\hat L,\Phi,\ov\Phi)$ since\footnote{But then ${\cal H}$ is not a function of $Y$ but instead it depends on the (gauge-invariant)  function $f(\Phi)$.}
$$
{1\over2}\Fint f(\Phi)\widetilde\Tr{\cal WW} + {\rm h.c.} =
\Dint [f(\Phi) + {\ov f}({\ov\Phi})] \hat L + \makebox{total deriv.}
$$
It however plays a role in effective Lagrangians like, for instance, in 
the description of gaugino condensates. 
Defining the variable ${\cal U}=\widetilde\Tr{\cal WW} $, the  field equation for the gauge superfield ${\cal A}$
is now
\beq
\label{U1}
\ov D^\dalpha \Bigl[ ({\cal H}_L + 2W_{\cal U} + 2\ov W_{\ov {\cal U}}) 
\, e^{-{\cal A}} \ov {\cal W}_\dalpha e^{\cal A} \Bigr] 
= {\cal W}^\alpha \, D_\alpha({\cal H}_L+ 2W_{\cal U} ) - T(r)\,{\cal H}_Y\, \Phi \ov\Phi e^{\cal A}
\eeq
with $W_{\cal U} = {\partial\over\partial {\cal U}}W(\Phi,{\cal U})$, instead of the third eq.~(\ref{CSF5}). 
Following the same steps, we obtain the ``natural" supercurrent structure
\beq
\label{U2}
\begin{array}{rcl}
\ov D^\dalpha J_{\alpha\dalpha} &=& D_\alpha X + \chi_\alpha ,
\crbig
J_{\alpha\dalpha} &=&
- 2 \Bigl[ (\ov{\cal D}_\dalpha\ov\Phi) {\cal H}_{\Phi\ov\Phi}  ({\cal D}_\alpha\Phi)
- {\cal H}_{LL}(\ov D_\dalpha\hat L)(D_\alpha\hat L) 
\crbig
&& \hspace{3.5cm}
+ 2 \, ({\cal H}_L + 2 W_{\cal U} + 2 \ov W_{\ov {\cal U}})\widetilde\Tr 
({\cal W}_\alpha e^{-{\cal A}} \ov {\cal W}_\dalpha e^{\cal A}) \Bigr] ,
\crbig
X &=& 4\, (W-{\cal U}W_{\cal U}), 
\crbig
\chi_\alpha &=& \ov{DD}D_\alpha({\cal H}-\hat L{\cal H}_L) 
\end{array}
\eeq
instead of expressions (\ref{CSF12}). A violation of scale invariance in the chiral density is measured
by the superfield
\beq
\label{U3}
\widetilde\Delta_{(w)} = wW_\Phi \Phi + 3W_{\cal U}\,{\cal U} - 3W
\eeq
since ${\cal U}=\widetilde\Tr{\cal WW} $ has canonical scale dimension three. 
Relations (\ref{Imp2}) are then unaffected and the same improvement transformation leads  
to the improved supercurent structure $\ov D^\dalpha \widetilde J_{\alpha\dalpha} = D_\alpha\widetilde X
+ \widetilde\chi_\alpha$ with a modified supercurrent superfield
\beq
\label{U4}
\begin{array}{rcl}
\widetilde J_{\alpha\dalpha} &=&
- 2 (\ov{\cal D}_\dalpha\ov\Phi) {\cal H}_{\Phi\ov\Phi}  ({\cal D}_\alpha\Phi)
+ 2{\cal H}_{LL}(\ov D_\dalpha\hat L)(D_\alpha\hat L) 
-{w\over3} [D_\alpha,\ov D_\dalpha] ({\cal H}_\Phi\Phi + \ov\Phi{\cal H}_{\ov\Phi})
\crbig
&& 
- 4 \, ({\cal H}_L + 2W_{\cal U} + 2\ov W_{\ov {\cal U}}) 
\widetilde\Tr ({\cal W}_\alpha e^{-{\cal A}} \ov {\cal W}_\dalpha e^{\cal A})
\end{array}
\eeq
and anomaly superfields as defined in eqs.~(\ref{Imp4}) but with $\widetilde\Delta_{(w)}$ as given in (\ref{U3}). An anomalous contribution 
to $\widetilde\Delta_{(w)}$ as in (\ref{anom2}) follows then, using (\ref{U3}), from a 
Veneziano--Yankielowicz \cite{VY} ``gauge superpotential" 
\beq
\label{U5}
W({\cal U}) = {A\over3}( {\cal U} \ln {\cal U} - {\cal U}).
\eeq

Since $\Dint \hat L = {1\over4}\Fint U + {\rm derivative}$, a theory defined by functions ${\cal H} 
+ (A+\ov A)\hat L$ and $W$ is equivalent to a theory defined by ${\cal H}$ and $W + {1\over2}AU$ 
(A is chiral). All expressions in this section respect this equivalence.

\section{Perturbative anomalies} \label{secanom}
\setcounter{equation}{0}

The improved supercurrent structure (\ref{Imp6}) with scale dimension $w$ for chiral superfields
$\Phi$  includes the Noether current of the $U(1)_R$ acting
with charges $3/2$ on gauginos and $w-3/2$ on chiral fermions in representation $r$.
This $U(1)_R$ group combines the natural $\widetilde R$ transformation described in the natural (Belinfante)
supercurrent structure (\ref{CSF12})\footnote{Obtained with $w=0$ in expressions (\ref{Imp6}).} 
and the non--$R$ 
$U(1)_Z$  acting with charges $w$ on superfields $\Phi$. As explained earlier, the Noether 
current $j_\mu^{(Z)}$ associated with $U(1)_Z$ is in the $\theta\sigma^\mu\ov\theta$ component of superfield
(\ref{Imp10b}),
$$
{\cal Z} = {w\over2} ( {\cal H}_\Phi\Phi + \ov\Phi{\cal H}_{\ov\Phi} ) = w Y {\cal H}_Y.
$$
Using field equations, its superfield conservation equation is of the form 
\beq
\label{an0a}
\ov {DD}{\cal Z} = \Delta_{\cal Z},
\eeq
with a chiral source superfield $\Delta_{\cal Z}$ given in eq.~(\ref{Imp10a}) at the classical level and including in general quantum anomalies.
With identity (\ref{appD6}), this conservation equation can always be turned into an equivalent 
``supercurrent equation"
\beq
\label{an0b}
\begin{array}{rcl}
\ov D^\dalpha J_{\alpha\dalpha} &=& D_\alpha \Delta_{\cal Z} + 3 \ov{DD} D_\alpha {\cal Z},
\crbig
J_{\alpha\dalpha} &=& 2[D_\alpha,\ov D_\dalpha] {\cal Z},
\end{array}
\eeq
where the Noether current $j_\mu^{(Z)}$ is now in $J_{\alpha\dalpha}|_{\theta=0}$. 
In the $\theta\sigma^\mu\ov\theta$ component, the ``energy-momentum" tensor is merely a trivial 
improvement term, according to transformations (\ref{appD9}), its trace is $-3\Box C_{\cal Z}$
and this corresponds to a formal contribution\footnote{Omitting as earlier fermions and gauge fields.}
$$
J_\mu^D = -3\,\partial_\mu C_{\cal Z} = -{3\over2}w\,\partial_\mu ({\cal H}_z z +\ov z{\cal H}_{\ov z})
= -{3\over2} \, \partial_\mu [ \Delta_{(w)}-\Delta_{(0)}]_{\theta=0}
$$
to the dilatation current, in terms of the source superfield (\ref{CSF18}).

\subsection{Mixed ``internal" anomalies}

In this paragraph, we repeatedly use
\beq
\begin{array}{l} \displaystyle
\widetilde\Tr\,{\cal WW}|_{\theta\theta} = - {1\over2}\widetilde\Tr \,F_{\mu\nu}F^{\mu\nu}
- {i\over2}\widetilde\Tr \, F_{\mu\nu}\widetilde F^{\mu\nu} + \ldots
\crbig \hspace{3.5cm} 
\displaystyle
= 2 \, {\cal L}_{SYM} - {i\over2} \, \widetilde\Tr\Bigl[ F_{\mu\nu}\widetilde F^{\mu\nu} 
- 2 \,\partial_\mu(\lambda\sigma^\mu\ov\lambda) \Bigr] ,
\crbig \displaystyle
\ov{DD}{\cal Z}|_{\theta\theta} = -2i \, \partial^\mu j_\mu^{({\cal Z})}  + \ldots ,
\qquad \qquad
\hat L|_{\theta\theta\ov{\theta\theta}} = - {1\over4}\widetilde\Tr \, F_{\mu\nu}F^{\mu\nu} + \ldots
= {\cal L}_{SYM} ,
\end{array}
\eeq
and $\widetilde F_{\mu\nu}={1\over2} \, \epsilon_{\mu\nu\rho\sigma}F^{\rho\sigma}$.

Since a global symmetry $U(1)_Z$ has $U(1)_Z$--gauge--gauge mixed anomaly\footnote{Dots indicate terms generated by supersymmetry.}
\beq
\label{an1}
\partial^\mu j_\mu^{({\cal Z})} = {1\over16\pi^2} \,wT(r)\, \widetilde\Tr \, F^{\mu\nu} \widetilde F_{\mu\nu}
+ \ldots ,
\eeq
the corresponding superfield anomaly equation is
\beq
\label{an2}
\ov{DD}\, {\cal Z} = {1\over4\pi^2} \, wT(r)\, \widetilde\Tr {\cal WW}.
\eeq
At this point, $w$ is the ${\cal Z}$--charge of the superfield $\Phi$ and of its fermionic components.
Identity (\ref{appD6}) with ${\cal G}=-{\cal Z}/3$ leads to
\beq
\label{an3}
-{2\over3} \ov D^\dalpha [ D_\alpha , \ov D_\dalpha ] {\cal Z} = 
-{1\over12\pi^2} \, wT(r)\, D_\alpha \widetilde\Tr{\cal WW}
- \ov{DD} D_\alpha {\cal Z},
\qquad\quad
{\cal Z} = wY{\cal H}_Y.
\eeq
Comparing with the improved supercurrent structure (\ref{Imp6}), the anomaly adds a contribution 
to the chiral source superfield $\widehat X$,
\beq
\label{an4}
\widehat X \quad\longrightarrow\quad \widehat X - {1\over12\pi^2} wT(r) \,\widetilde\Tr{\cal WW},
\eeq
and by supersymmetry a contribution to the energy-momentum trace.

Similarly, the natural $\widetilde R$--symmetry has $U(1)_{\widetilde R}$--gauge--gauge anomaly\footnote{
$C(G) = T(Adj\,G)$ is the quadratic Casimir, $C(G)\delta^{ab} = f^{acd}f^{bcd}$ in terms of structure constants.}
\beq
\label{an5}
\partial^\mu j_\mu^{(\widetilde R)} = {1\over16\pi^2}\, {3\over2}[C(G)-T(r)] \,\widetilde\Tr (F^{\mu\nu} 
\widetilde F_{\mu\nu}) + \ldots
\eeq
According to the second eq.~(\ref{appD5}), this also leads to an anomalous shift of the source superfield 
$\widehat X$:
\beq
\label{an6}
\widehat X \quad\longrightarrow\quad \widehat X - {1\over8\pi^2} \,[ C(G) - T(r) ]\, \,\widetilde\Tr{\cal WW} .
\eeq
Combining both anomalies leads to
\beq
\label{an7}
\widehat X_{(anomaly)} = - {1\over24\pi^2} \, b(w) \, \widetilde\Tr {\cal WW}
\eeq
with coefficient
\beq
\label{an8}
b(w) = b_0 +  2(w-1) T(r),
\qquad\qquad
b_0 = 3C(G)-T(r).
\eeq
This is of course the anomaly of the $R$--symmetry with current described by the lowest component 
of the improved supercurrent (\ref{Imp6}). Writing instead
\beq
\label{an8b}
b(w) = 3C(G) - T(r) (1-\gamma),
\qquad\quad
\gamma = 2(w-1), \qquad\quad  w = 1 + {\gamma\over2},
\eeq
the number $\gamma$ is now the anomalous dimension and $b(w)$ is the numerator of the 
NSVZ $\beta$ function \cite{NSVZ, SV88}.\footnote{While a conserved current has dimension three, see 
for instance ref.~\cite{Gross}, anomalous currents have in general $\gamma>0$.}

Now, according to the first eq.~(\ref{appD5}), the energy-momentum tensor in this supercurrent superfield 
verifies
\beq
\label{an9}
{\Theta^\mu}_\mu = {1\over4} D + {3\over2} \Re f_{\widehat X_{anomaly}} 
= {1\over4} D - {1\over32\pi^2} \,b(w)\, \widetilde\Tr F^{\mu\nu} F_{\mu\nu} + \ldots
\eeq
Since gauginos and chiral fermions have scale dimensions $3/2$ and $w+1/2$ respectively, we expect that
the dilatation current has dilatation--gauge--gauge anomaly
\beq
\label{an10}
\partial^\mu j_\mu^D = - {1\over48\pi^2} \, c(w) \, \widetilde\Tr F^{\mu\nu} F_{\mu\nu}  + \ldots,
\qquad
c(w) = {3\over2} [C(G) + T(r)] + (w-1) T(r).
\eeq
As a consequence, 
\beq 
\label{an11}
D = {1\over4\pi^2} \,d(w)\, \widetilde\Tr F^{\mu\nu}F_{\mu\nu} + \ldots , 
\qquad
d(w) = C(G) - T(r) + {2\over3}(w-1)T(r).
\eeq
This residual dilatation anomaly is introduced in the supercurrent structure by a quantum contribution
\beq
\label{an12}
\chi_{\alpha (anomaly)}= -{1\over4} \ov{DD}D_\alpha U_{(anomaly)},  
\qquad\qquad
U_{(anomaly)} =  - {1\over2\pi^2}\, d(w)\, \hat L
\eeq
added to the source superfield $\widehat\chi_\alpha$.
Hence, with the chiral contribution (\ref{an7}), the improved supercurrent structure including the anomalies is 
\beq
\label{an13}
\begin{array}{rcl}
\ov D^\dalpha \widehat J_{\alpha\dalpha} &=& D_\alpha \widehat X + \widehat\chi_\alpha ,
\crbig
\widehat J_{\alpha\dalpha} &=&
- 2 \Bigl[ (\ov{\cal D}_\dalpha\ov\Phi) {\cal H}_{\Phi\ov\Phi}  ({\cal D}_\alpha\Phi)
- {\cal H}_{LL}(\ov D_\dalpha\hat L)(D_\alpha\hat L) 
+ 2 \, {\cal H}_L \widetilde\Tr ({\cal W}_\alpha e^{-{\cal A}} \ov {\cal W}_\dalpha e^{\cal A}) \Bigr] 
\crbig
&& \hspace{2.2cm}
-{2\over3} [D_\alpha,\ov D_\dalpha] (w{\cal H}_Y Y) ,
\crbig
\widehat X &=& \displaystyle -{4\over3}\left( \widetilde\Delta_{(w)}  
+ {1\over32\pi^2} \,b(w)\, \widetilde\Tr {\cal WW}\right),
\crbig
\widehat \chi_\alpha &=& \displaystyle
-{1\over2}\ov{DD}D_\alpha\left( \Delta_{(w)} - {1\over4\pi^2} \, d(w)\, \hat L \right),
\end{array}
\eeq
with $\widetilde \Delta_{(w)}=0$ if the superpotential vanishes. In the case of pure ${\cal N}=2$ super-Yang-Mills 
theory in which $r=Adj(G)$, $C(G)=T(r)$ and $w=1$ (since both gauginos have same $R$ charge), 
\beq
\label{an14}
b(w) = b_0 = 2C(G), \qquad d(w)=0, \qquad \chi_{\alpha (anomaly)} = 0.
\eeq

\subsection{Matching and cancelling anomalies}  \label{secmatch}

Following the discussion of the previous section, we may use in an effective or phenomenological Lagrangian
local counterterms which, depending on the context, either match an anomaly of the microscopic theory
or compensate an anomaly generated in perturbation theory of the effective theory in order to restore a quantum
symmetry of the underlying theory. 
An exemple of the first situation is the familiar axial current chiral anomaly. 
An example of the second case would be the cancellation of target-space $T$--duality (K\"ahler) anomalies in 
the effective supergravity of string compactifications as originally described in refs.~\cite{CO, DFKZ1}.

Consider
\begin{eqnarray}
\label{Lcor2a}
\Delta{\cal H}_{corr.}(\hat L,Y) &=& \displaystyle
 - {\epsilon\over8\pi^2} \,d(w)\, \hat L (\ln\hat L - 1) ,
\crbig
\label{Lcor2b}
\Delta W_{corr.}(\Phi,{\cal U}) &=& \displaystyle
{\epsilon\over96\pi^2} \,b(w)\, {\cal U}(\ln{\cal U}-1),
\end{eqnarray}
where ${\cal U} = \widetilde\Tr{\cal WW}$ as in Subsection \ref{secUdep} and $\epsilon=\pm1$.
The corresponding scale and $R$--breaking superfields are then
\beq
\label{Lcor3}
\begin{array}{rcl}
\Delta_{corr.} &=& -\displaystyle {\epsilon\over4\pi^2} \, d(w)\, \hat L,
\crbig
\widetilde\Delta_{corr.} &=& \displaystyle {\epsilon\over32\pi^2} \,b(w)\, \widetilde\Tr {\cal WW}.
\end{array}
\eeq
These counterterms are used to obtain effective Lagrangians with ``classical" anomalous behaviour
by modifying the currents in the supercurrent structure.
For $\epsilon=1$, when added as quantum corrections to the function ${\cal H}$ defining an
effective Lagrangian, they would match the microscopic anomaly terms in expressions 
(\ref{an13}). For $\epsilon=-1$, they would cancel or compensate these quantum anomalies to 
describe an exact symmetry, as for instance the renormalization-group does with scale transformations.

If we expand the anomaly counterterm in expression (\ref{Lcor2a}) around a constant background value
\beq
\label{Lcor4}
\hat L \qquad\longrightarrow\qquad  g^2 + \hat L,
\eeq
it can be rewritten
\beq 
\label{Lcor5}
\begin{array}{l} \displaystyle
-\epsilon {d(w)\over8\pi^2} \Dint \left(g^2 +\hat L\right)\left[ \ln g^2 
+ \ln \left(1+{\hat L\over g^2} \right) - 1 \right]
\crbig \displaystyle \hspace{1.2cm} 
= \displaystyle -\epsilon {d(w)\over8\pi^2} \ln g^2 \Dint \hat L + \ldots
= -\epsilon {d(w)\over8\pi^2} \ln g^2 \, {1\over4} \Fint \widetilde\Tr{\cal WW} + {\rm c.c.} + \ldots ,
\end{array}
\eeq
omitting terms of higher orders in $\hat L$. Hence, with a constant coupling, it can be expressed as 
a chiral integral. This is the rescaling anomaly calculated by Arkani-Hamed and Murayama \cite{AHM}\footnote{Their 
eqs.~(2.8) and (2.9) for super-Yang-Mills fields.}. But in terms of the gauge coupling field, it is included in the 
full superspace integral of the real superfield (\ref{Lcor2a}). Actually, ref.~\cite{AHM} evaluates the anomaly induced by the rescaling 
of the gauge superfield ${\cal A}\rightarrow g{\cal A}$ which brings the gauge kinetic terms from 
$-{1\over4g^2}F_{\mu\nu}^a F^{a\mu\nu}$ to the canonical normalization $-{1\over4}F_{\mu\nu}^a F^{a\mu\nu}$.
This rescaling corresponds to $\hat L \rightarrow g^2 \hat L$ in our context. When applied to the 
anomaly-matching term (\ref{Lcor2a}), it produces the correct anomaly variation.

Notice that the chiral anomaly-matching superpotential in expression (\ref{Lcor2b}) generates
\beq
\label{Lcor6}
\begin{array}{l} \displaystyle
{\epsilon\over96\pi^2} \,b(w)\, \Fint {\cal U}(ln\,{\cal U}-1) + {\rm h.c.} =
{\epsilon\over48\pi^2}  \,b(w)\, \ln (\ov uu) {\cal L}_{SYM}
\crbig \hspace{2.5cm}
 \displaystyle
+ {i\,\epsilon\over48\pi^2}  \,b(w)\, \ln (u/\ov u) \left[ -{1\over4} \widetilde\Tr F_{\mu\nu} \widetilde F^{\mu\nu}
+ {1\over2} \partial^\mu (\widetilde\Tr \lambda\sigma_\mu\ov\lambda ) \right] + \ldots
\end{array}
\eeq
which, since $u= - \widetilde\Tr \lambda\lambda$, is a correction to the gauge coupling in
a fermionic background $\langle\widetilde\Tr \lambda\lambda\rangle \ne 0$ only.\footnote{See next Section.}

\section{Effective Lagrangians} \label{seceff}
\setcounter{equation}{0}

We now apply our formalism to two types of effective descriptions of a supersymmetric 
gauge theory, the Wilson effective Lagrangian and the effective action, as defined in
quantum field theory, for the description of gaugino condensates. This section is a
development of refs.~\cite{DFKZ2, BDQQ}.

In this section, it is important to clearly distinguish the scale and the mass dimensions.
As defined earlier, the scale dimension encodes the behaviour under dilatation of 
coordinates and fields (with scale dimensions 
$w_i$). The mass dimension follows from simple dimensional analysis (in energy units) and 
allows for a mass dimension of parameters (which have zero scale dimension).
A Lagrangian has mass dimension four since the action is dimensionless, it does not 
have a well-defined scale dimension in general.\footnote{As a consequence, in a supersymmetric 
theory, the K\"ahler 
potential, the superpotential and $\widetilde\Tr{\cal WW}$ have mass dimensions two, three and three.}
Gauge fields and superfields have identical canonical scale and mass dimensions: this is the case of
superfields ${\cal A}$ ($w=0$), $L$ and $\Omega$ ($w=2$), ${\cal W}$ ($w=3/2$). Chiral superfields have in
general anomalous scale dimensions $w=1+\gamma/2$.
The distinction between scale and mass dimensions disappears if dilatation would be a symmetry: in this 
case, the Lagrangian has scale 
dimension four and all parameters have then vanishing mass 
dimension.

\subsection{Wilson Lagrangian}

The Wilson effective Lagrangian ${\cal L}_{W,\mu}$ explicitly depends on a mass parameter 
$\mu>0$, which acts as an ultraviolet cutoff. Schematically, it is obtained from a fundamental
microscopic quantum field theory by integrating modes with frequencies larger than $\mu$. In perturbation 
theory, the loop expansion in the microscopic theory is matched by the combination of a perturbative 
expansion of the Wilson Lagrangian,
$$
{\cal L}_{W,\mu} = {\cal L}_{W}^{(0)} + \sum_{n>0} {\cal L}_{W,\mu}^{(n)} 
$$
($n$ is the loop order in the microscopic theory) and loops generated from ${\cal L}_{W,\mu}$, with cutoff $\mu$.
If the microscopic theory includes only fields with masses lighter than $\mu$, the classical 
${\cal L}_W^{(0)}$ coincides with the microscopic quantum field theory Lagrangian. If the microscopic
theory includes fields with masses heavier than $\mu$, ${\cal L}_W^{(0)}$ also includes higher-dimensional operators controlled by these mass parameters.
The Wilson Lagrangian is local and the scale $\mu$ is arbitrary. 
Its dependence on $\mu$ is then dictated by a specific renormalization-group (RG)
equation.\footnote{We assume here that $\mu$ is sufficiently far from particle thresholds, to avoid 
a detailed treatment of these thresholds.} 

We wish to consider the Wilson effective Lagrangian of a microscopic $N=1$ gauge theory with zero superpotential:
\beq
\label{Wils1}
{\cal L}_{micro.} = \sum_i Z_i \Dint \ov\Phi_i e^{{\cal A}_i} \Phi_i 
+ {1\over4\ov g^2}\Fint \widetilde\Tr{\cal WW} + {\rm h.c.}
\eeq
where $\ov g$ is the bare coupling of the gauge group assumed simple and the 
sum is over irreducible representations. The wave function renormalisation matrix $Z$ is  
diagonal with zero superpotential. As we will see later on, it is not always wise to assume that 
$Z_i\rightarrow1$ in the limit $\ov g\rightarrow 0$. The Lagrangian is classically scale invariant 
with canonical scale dimensions
$w=1$ and $w=3/2$ for $\Phi$ and ${\cal W}$ respectively, and $\ov g$ has mass dimension zero.

We are interested in the Wilson effective Lagrangian expressed with a supersymmetrized 
background field $C$ for the gauge coupling. Hence, the background value $\langle C\rangle$ 
will be identified with
the physical gauge coupling $g^2(M)$ at a reference energy scale $M$. This scale can be viewed as
defining the renormalization scheme in the microscopic theory, or as the scale used to normalize quantities in
the renormalized theory. For instance, there exists in general subtraction schemes 
(in the microscopic theory) where $g^2(M) = \ov g^2$ (for a given $M$). It can also be regarded as a physical quantity like
a unification scale. 
The Wilson Lagrangian depends on the reference scale $M$ implicitly via $g^2(M)$ or
$C$ and explicitly via the ratio $\mu/M$ and RG equations 
reflect the arbitrariness of these mass parameters.

Our goal in this section is to algebraically derive some of the all-order results of Novikov, Shifman, Vainshtein
and Zakharov (NSVZ) \cite{NSVZ}\footnote{And Jones for super-Yang-Mills theory \cite{Jones}.} with the
gauge coupling background or propagating field $C$ which actually plays a central role in the undersanding of 
the higher order 
contributions to the $\beta$ function. In spirit, our
discussion is very close or identical to the interpretation of Shifman and Vainshtein and to the anomaly
argument of Arkani-Hamed and Murayama for constant coupling parameters \cite{AHM}.
Using then the formulation presented in Section \ref{secsupercurrent}, it immediately follows that the
presence in the $\beta$ function of these higher-order contributions is fully compatible with the 
supercurrent superfield structure expected in ${\cal N}=1$ theories.

\subsubsection{\boldmath{${\cal N}=1$} super-Yang-Mills theory}\label{seceffSYM}

At tree-level, or in the microscopic theory, we would certainly use
\beq
\label{SYM1}
{\cal H}_{(0)} = m^2\ln (\hat L/m^2)
\eeq
with a mass parameter $m$ to keep track of the mass dimensions of the function ${\cal H}$ and of $C$. 
The bosonic Lagrangian is then\footnote{Omitting a derivative.}
\beq
\label{SYM2}
\begin{array}{rcl}
{\cal L}_{(0)} &=& \Dint {\cal H}_{(0)}
\crbig
&=& \displaystyle
{m^2\over C}\left[ -{1\over4} \widetilde\Tr F_{\mu\nu}F^{\mu\nu} + {1\over2}\widetilde\Tr DD \right]
+ {1\over4} {m^2\over C^2} \left[ (\partial^\mu C)(\partial_\mu C) +
{1\over6} H^{\mu\nu\rho} H_{\mu\nu\rho} \right].
\end{array}
\eeq
With the identification $C=m^2 g^2$ of the tree-level gauge coupling field and 
\beq
\label{Wils8b}
C = m^2 \, g^2(M)
\eeq
in general, the quantity $m$ does not play any role in the gauge Lagrangian. It appears in the kinetic Lagrangian 
of the linear superfield where it naturally keeps track of the violation of scale invariance unavoidable with the 
gauge coupling field. Actually,
\beq
\label{Wils8d}
\Delta_{(0)} = 2\hat L{\partial\over\partial\hat L}{\cal H}_{(0)} - 2 {\cal H}_{(0)} = - 2 {\cal H}_{(0)} + 2m^2 
= - m{d\over dm}{\cal H}_{(0)} 
\eeq
indicates that the logarithmic choice (\ref{SYM1}) appropriate for the tree-level Yang-Mills Lagrangian in
expression (\ref{SYM2}) is a function ${\cal H}_{(0)}$ with scale dimension zero.\footnote{The constant terms 
$2m^2$ in $\Delta_{(0)}$ is irrelevant.} The last equality indicates that since $m$ is used to obtain 
the appropriate mass dimensions, a scale transformation of $C$ can be compensated by a rescaling of 
$m$:\footnote{Which is not a scale transformation.} for any  ${\cal H} = m^2{\cal F}(\hat L/m^2)$,
\beq
2\hat L {\cal H}_{\hat L} - 2{\cal H} = -m{d\over dm}{\cal H}.
\eeq

We wish to write a loop-corrected Wilson Lagrangian
\beq
\label{Wils5}
{\cal L}_{W,\mu} = {1\over g_{W,\mu}^2} {\cal L}_{SYM} + \dots 
=  -{1\over 4g_{W,\mu}^2} \widetilde\Tr F_{\mu\nu}F^{\mu\nu} + \ldots,
\eeq
where the Wilson gauge coupling $g_{W,\mu}^2$ is expressed as a function of $C/m^2$ identified with the 
ordinary observable gauge coupling constant $g^2(M)$ at an arbitrary normalisation scale 
$M$,\footnote{Strictly speaking, we always work at a finite nonzero value of $g^2$ and we are not 
concerned with the definition of or the relation with a perturbative renormalization scheme. This question
is discussed for instance in refs.~\cite{JJN} (relation with the DR scheme) or in refs.~\cite{KataevS} 
(higher-derivative regularization). } 
as in eq.~(\ref{Wils8b}).\footnote{And ${\cal L}_{SYM} = {1\over4}\int d^2\theta\, \widetilde\Tr{\cal WW} 
+ {\rm h.c.}$ is defined in eqs.~(\ref{Lagr0}).} 
Without matter superfield, we certainly have
\beq
\label{Wils2}
{\cal L}_{W,\mu} = \Dint {\cal H}( \hat L/m^2, \mu/M)
\eeq
and the Lagrangian has necessarily (classical) $\widetilde R$ symmetry rotating Grassmann coordinates and
fermions. This implies that the corresponding natural supercurrent structure (\ref{CSF12}) including 
the Belinfante energy-momentum tensor has vanishing chiral source superfield $X$:
\beq
\label{Wils12}
\begin{array}{rcl}
\ov D^\dalpha J_{(W)\alpha\dalpha} &=& \chi_{(W)\alpha} ,
\crbig
J_{(W)\alpha\dalpha} &=&
2 \, {\cal H}_{LL}(\ov D_\dalpha\hat L)(D_\alpha\hat L) 
- 4 \, {\cal H}_L \widetilde\Tr ({\cal W}_\alpha e^{-{\cal A}} \ov {\cal W}_\dalpha e^{\cal A}), 
\crbig
\chi_{(W)\alpha} &=& \ov{DD}D_\alpha({\cal H}-\hat L{\cal H}_L).
\end{array}
\eeq
The lowest component of ${3\over8}J_{(W)\alpha\dalpha}$ is the current of $\widetilde R$ symmetry:
\beq
\label{Wils13b}
 j_\mu^{\widetilde R} = {3\over2}\,{\cal H}_C\,\widetilde\Tr\, \lambda\sigma^\mu\ov\lambda
+ {3\over4} {\cal H}_{CC} \chi\sigma_\mu\ov\chi
=  {1\over g_{W,\mu}^2} \, q_\lambda \,\widetilde\Tr\, \lambda\sigma^\mu\ov\lambda
- {{\cal H}_{CC}\over2} \, q_\chi \, \chi\sigma_\mu\ov\chi
\eeq
with $\widetilde R$ charges $q_\lambda = 3/2$ and $q_\chi=-3/2$ as in eq.~(\ref{CSF13}). 
Quantum corrections to the effective Lagrangian appear in the metric factors ${\cal H}_C$ and 
$-{\cal H}_{CC}/2$. But the one-loop chiral anomaly of the $R$--symmetry current generated by massless
gauginos leads formally to\footnote{Eq.~(\ref{an7}).} a chiral source superfield 
\beq
\label{Wils14}
X_{(W),anomaly} = -{C(G)\over8\pi^2} \, \widetilde\Tr{\cal WW} ,
\eeq
as in the anomaly-corrected supercurrent structure (\ref{an13}). 

Two different renormalization-group equations follow. Firstly,
since the perturbative dependence on $\mu$ is restricted to one-loop \cite{SV},
\beq
\label{Wils3}
\mu{d\over d\mu} {\cal L}_{W,\mu} = {b_0\over32\pi^2} \Fint\widetilde\Tr{\cal WW} + {\rm h.c.},
\qquad\qquad
b_0=3C(G),
\eeq
we infer that\footnote{Since $\Dint\hat L = {1\over4}\Fint\widetilde\Tr{\cal WW} + 
\makebox{h.c. $+$ derivative}$.}
\beq
\label{Wils4}
\begin{array}{rcl}
{\cal L}_{W,\mu} &=& \displaystyle  \Dint \hat{\cal H}(\hat L) 
+ {b_0\over32\pi^2} \ln{\mu\over M} \Fint \widetilde\Tr{\cal WW} + {\rm h.c.},
\crbig
{\cal H} (\hat L, \mu/M) &=& \displaystyle \hat{\cal H}(\hat L) + {b_0\over8\pi^2} \ln\left({\mu\over M}\right)
\hat L.
\end{array}
\eeq
The one-loop correction is scale invariant: it will not appear in the divergence 
of the dilatation current: $\Delta_{(1-loop)} = \Delta_{(0)}$. But it is not invariant under the rescalings of the parameters $M$ or $\mu$.
Since the Wilson coupling $g_{W,\mu}$,  which is not a physically significant quantity, is
\beq
\label{Wils5b}
{1\over g^2_{W,\mu}} = \hat{\cal H}_C(C) + {b_0\over8\pi^2}\ln{\mu\over M},
\eeq
a rescaling of $\mu$ in ${\cal L}_{W,\mu}$ is controlled by 
\beq
\beta_W(g_{W,\mu}^2) \equiv \mu{d \over d\mu} g_{W,\mu}^2 = -{b_0\over8\pi^2} \, g_{W,\mu}^{\,4}
\eeq
which is exhausted at one-loop.

Secondly, since $M$ is arbitrary, the RG implies that
\beq
\label{Wils6}
0 = M{d\over dM} \Bigl[ \hat{\cal H}_C(C) + {b_0\over8\pi^2}\ln{\mu\over M}\Bigr]
\eeq
and, with\footnote{We always define the $\beta$ function 
as $\beta \equiv M{d\over dM} g^2$.}
\beq
\label{Wils7}
M{d\over dM} C = \beta(C) = m^2\,\beta(g^2)
\eeq
since we identify $C/m^2$ with the physical gauge coupling $g^2(M)$,
\beq
\label{Wils8}
\beta(C) = {1\over8\pi^2} {b_0\over \hat{\cal H}_{CC}}.
\eeq
The $\beta$ function is then proportional to the inverse of the linear gauge coupling superfield 
kinetic metric $-{1\over2}{\cal H}_{CC}$,  which is positive. With identifications (\ref{Wils8b}) and (\ref{Wils7}),
\beq
\label{Wils8c}
\beta(g^2(M)) = {1\over8\pi^2m^2}\, {b_0\over \hat{\cal H}_{CC}} = - {m^2 \over C^2\hat{\cal H}_{CC}} 
\beta_{1-loop}.
\eeq
The tree-level ${\cal H}_{(0)}$, eq.~(\ref{SYM1}), leads to $\beta(g^2)=\beta_{1-loop}$, corrections
to ${\cal H}_{(0)}$ generate higher order contributions.\footnote{At one-loop only, the equality of the 
$\beta$ functions implies ${\cal H} = m^2\ln\hat L + b \hat L$, with an arbitrary constant $b$ which can be eliminated by a redefinition of the scale $\mu$.}

Under a rescaling $\mu\rightarrow e^\lambda \mu$ of the Wilson scale,
\beq
\delta {\cal L}_{W,\mu} = \lambda {b_0\over32\pi^2} \Fint\widetilde\Tr{\cal WW} + {\rm c.c.}
= - \lambda {b_0\over32\pi^2} \widetilde\Tr F_{\mu\nu}F^{\mu\nu} + \ldots
\eeq
This variation is the supersymmetry partner of the anomalous variation induced by the $\widetilde R$ symmetry 
rotating the gaugino: under $\lambda_\beta \rightarrow e^{{3\over2}i\alpha} \lambda_\beta$,
\beq
\delta {\cal L}_{W,\mu} = -i\alpha {b_0\over32\pi^2} \Fint\widetilde\Tr{\cal WW} + {\rm c.c.}
= - \alpha {b_0\over32\pi^2} \widetilde\Tr F_{\mu\nu}\widetilde F^{\mu\nu} + \ldots ,
\eeq
a variation which can be deduced from the anomaly-matching term (\ref{Lcor6}).
In this sense, the one-loop term in the Wilson Lagrangian can be understood as a matching term for 
the anomaly of the $\widetilde R$--symmetry. 

Following section \ref{secmatch}, we should then cancel the residual scaling anomaly (\ref{an12})
with coefficient $d(w)=C(G)$ by adding to the tree-level Lagrangian function ${\cal H}_{(0)}$ the 
contribution (\ref{Lcor2a}) with $\epsilon=-1$. This countertem removes all dependence on the physical 
scale $M$ and defines the $\beta$ function. The resulting function ${\cal H}$ is 
\beq
\label{Wils10}
{\cal H}(\hat L) = m^2 \ln{\hat L\over m^2} + {C(G)\over8\pi^2}
\left[ \hat L\ln {\hat L\over m^2} - \hat L \right] + {b_0\over8\pi^2} \ln{\mu\over M}  \hat L,
\eeq
which in turn leads to the Wilson gauge coupling
\beq
\label{Wils10a}
\begin{array}{rcl} \displaystyle
{1\over g^2_{W,\mu}} &=& {\cal H}_C \,\,=\,\, \displaystyle {m^2\over C} + {C(G)\over8\pi^2} \ln {C\over m^2}
+ {b_0\over8\pi^2}\ln {\mu\over M}
\crbig
&=& \displaystyle {1\over g^2(M)} + {C(G)\over8\pi^2} \ln g^2(M) 
+ {b_0\over8\pi^2}\ln {\mu\over M}.
\end{array}
\eeq
Arbitrariness of $M$ in this expression, or directly formula (\ref{Wils8c}), leads to the beta function
\beq
\label{Wils9} 
\beta(g^2) = -{g^4\over8\pi^2} \, {3\,C(G) \over 1- {C(G)\over8\pi^2} g^2} 
\eeq
which is the all-order NSVZ beta function \cite{NSVZ, SV88, Jones}.

In the function ${\cal H}(\hat L)$, the first term is the classical, tree-level contribution, the second term encodes all perturbative contributions
beyond one-loop and the third, $\mu$-dependent term, is the one-loop correction. 
Hence, the NSVZ beta function can be derived from algebraic and anomaly arguments only, 
including its denominator, in the formalism with the gauge coupling field which introduces a second, 
real, anomaly-matching (or cancelling) superfield $\hat L$.\footnote{This 
Lagrangian has been obtained long ago, using similar 
arguments and somewhat obscure conformal supergravity methods, in ref.~\cite{DFKZ2}.}
Choosing $M=\mu$ leads to the relation \cite{SV}
\beq
\label{Wils11}
{\cal H}_C =
{1\over g_{W,\mu}^2} = {m^2\over C} + {C(G)\over8\pi^2} \ln \left({C\over m^2}\right) = {1\over g^2(\mu)}
+ {C(G)\over 8\pi^2}\ln g^2(\mu)
\eeq
and in the chiral version $g_{W,\mu}^{-2} = \Re s$. We note however that with the higher-order terms, 
the Legendre transformation (\ref{LS2}) leading to the chiral version of the theory cannot be solved 
analytically: the Wilson gauge coupling field $\Re s$ is well-defined (and physically meaningless)
but its K\"ahler potential ${\cal K}(S+\ov S)$ cannot be obtained in a closed form. In this sense, the linear 
theory (\ref{Wils10}) contains more information than the dual chiral version and its symmetry or anomaly 
behaviour is explicit.

The loop-corrected Wilson Lagrangian for pure super-Yang Mills is then 
\beq
\label{Wils10b}
\begin{array}{rcl}
{\cal L}_{W,\mu} &=& \displaystyle
\Dint \left( m^2 \ln{\hat L\over m^2} + {C(G)\over8\pi^2}
\left[ \hat L\ln{\hat L\over m^2} - \hat L \right] \right)
\crbig
&& \displaystyle + {b_0\over32\pi^2} \ln{\mu\over M} \Fint \widetilde\Tr{\cal WW} + {\rm h.c.}
\end{array}
\eeq
Notice that this Lagrangian does not have a potential since the auxiliary $D$ vanishes and the linear superfield 
does not have an auxiliary field. The value of the coupling constant remains arbitrary in ${\cal L}_{W,\mu}$.

In the supercurrent structure (\ref{Wils12}), the source superfield  $\chi_{(W)\alpha}$ dictates
the behaviour of the Wilson Lagrangian under scale transformations and includes then the anomaly
contribution (\ref{an12}):\footnote{It is an off-shell expression.}
\beq
\label{Wils13}
\begin{array}{rcl}
\chi_{(W)\alpha} &=& \displaystyle -{1\over2} \ov{DD}D_\alpha \Delta_{W,\mu}
\crbig
\Delta_{W,\mu} &=& \displaystyle
2[\hat L{\cal H}_{\hat L}- {\cal H}] \,\,=\,\, 2m^2 \Bigl[1-\ln{\hat L\over m^2}\Bigr] + {C(G)\over4\pi^2} \hat L 
\,\,=\,\, \Delta_{(0)} + {C(G)\over4\pi^2} \hat L .
\end{array}
\eeq
Again, the first term is due to the classical scale breaking with the gauge coupling field, as induced by the
scale dimension $w=2$ of $C$, while the second term is due to the anomaly-cancelling counterterm which
encodes the corrections beyond one-loop. The source superfield $\chi_{(W)\alpha}$ generates the trace of the Belinfante energy-momentum tensor using the on-shell equality ${T^\mu}_\mu = D/4$.  Off-shell, omitting fermions,
\beq
\label{Wils21}
\begin{array}{rcl} \displaystyle
{D\over4} &=& \displaystyle - {m^2\over C}\left[1-{C(G)\over8\pi^2}{C\over m^2}\right]
( \Box C + 2 {\cal L}_{SYM}) 
+ {m^2\over 2C^2} \left[ (\partial_\mu C)(\partial^\mu C) - {1\over6}H_{\mu\nu\rho}H^{\mu\nu\rho} \right] 
\crbig
&=& \displaystyle
\delta - \partial^\mu {\cal V}_\mu, 
\end{array}
\eeq
where $\delta$ is the scale variation of the bosonic Lagrangian and ${\cal V}_\mu$ is the virial current,
\beq
\label{Wils22}
{\cal V}_\mu = {m^2\over C} \left[ 1- {C(G)\over8\pi^2}{C\over m^2}\right]\partial_\mu C 
= \partial_\mu\left[ m^2\ln {C\over m^2}- {C(G)\over8\pi^2} C \right],
\eeq
according to expressions (\ref{CSF14}) and (\ref{CSF20}).\footnote{Notice that $D=2D_\Delta$,
as defined in eq. (\ref{CSF19}).}
Concentrating on the super-Yang-Mills part ${\cal H}_C {\cal L}_{SYM}$ of the Wilson Lagrangian 
(\ref{Wils10b}), or equivalently working in a constant background $C = m^2g^2(M)$,
we have firstly
\beq
\label{Wils23}
\mu{d\over d\mu} {\cal H}_C {\cal L}_{SYM} = M{dC\over dM}{\cal H}_{CC}{\cal L}_{SYM}
= {b_0\over 8\pi^2} {\cal L}_{SYM}
\eeq
since $M$ is arbitrary. This expresses the one-loop dependence of the Wilson coupling on the Wilson scale $\mu$.
Secondly 
\beq
\label{Wils24}
\delta = {D\over4} = - {2\over g^2(M)}\left[1-{C(G)\over8\pi^2} g^2(M)\right] {\cal L}_{SYM} 
= 2C{\cal H}_{CC} {\cal L}_{SYM},
\eeq
so that
\beq
\label{Wils25}
M{d\over dM} \delta = M{dC\over dM} \left[{2\over C^2}{\cal L}_{SYM} \right]
= 2\,{\beta(g^2)\over g^4} {\cal L}_{SYM}.
\eeq
Since $\delta = {T^\mu}_\mu$ on-shell, this result is a version of the trace anomaly formula \cite{Trace}.

The Wilson Lagrangian defined by the function (\ref{Wils10}) also describes the dynamics of the three-index tensor
$H_{\mu\nu\rho} = 3 \, \partial_{[\mu}B_{\nu\rho]} - \omega_{\mu\nu\rho}$ with a simple quadratic Lagrangian:
\beq
\begin{array}{rcl}
{\cal L}_B &=& \displaystyle - {1\over24} {\cal H}_{CC}\,H_{\mu\nu\rho}H^{\mu\nu\rho} - {1\over24}
\epsilon^{\mu\nu\rho\sigma}H_{\mu\nu\rho} \, {\cal J}_\sigma
\crbig
&=& \displaystyle {m^2\over C^2}\Bigl[ 1 - {C(G)\over8\pi^2}{C\over m^2} \Bigr] 
{1\over24} H_{\mu\nu\rho}H^{\mu\nu\rho} 
- {1\over24} \epsilon^{\mu\nu\rho\sigma}H_{\mu\nu\rho}\,{\cal J}_\sigma
\crbig
&=& \displaystyle {1\over g^4 m^2} \Bigl[ 1 - {C(G)\over8\pi^2}g^2 \Bigr] 
{1\over24} H_{\mu\nu\rho}H^{\mu\nu\rho} 
- {1\over24} \epsilon^{\mu\nu\rho\sigma}H_{\mu\nu\rho}\,{\cal J}_\sigma ,
\end{array}
\eeq
where the current $J_\sigma$ is
\beq
J_\sigma = {\cal H}_{CCC} \, \hat\chi\sigma^\mu\ov{\hat\chi} 
= {2m^2\over C^3}\Bigl[ 1 - {C(G)\over16\pi^2}{C\over m^2} \Bigr] \, \hat\chi\sigma^\mu\ov{\hat\chi} \,,
\eeq
in terms of the gauge invariant spinor $\hat\chi=\chi-{1\over2}\sigma^\mu \widetilde\Tr\, \ov\lambda a_\mu$.
The antisymmetric tensor with gauge invariance is equivalent to a pseudoscalar with shift symmetry. 
The duality transformation is performed by first considering $H_{\mu\nu\rho}$ as an unconstrained three-form field
with Bianchi identity 
\beq
\epsilon^{\mu\nu\rho\sigma} \partial_\mu H_{\nu\rho\sigma} = -3 \, \widetilde 
\Tr \Bigl[ F_{\mu\nu}\widetilde F^{\mu\nu}
- 2 \, \partial_\mu(\lambda\sigma^\mu\ov\lambda) \Bigr]
\eeq
imposed by a Lagrange multiplier scalar $a$. Eliminating $H_{\mu\nu\rho}$ with
\beq
H_{\mu\nu\rho} = -2 \, {\cal H}_{CC}^{-1}\, \epsilon_{\mu\nu\rho\sigma}[\partial^\sigma a 
+ {1\over4}{\cal J}^\sigma], 
\eeq
the dual theory is
\beq
{\cal L}_{axion} = - {2\over{\cal H}_{CC}} \, {1\over2} \, \Bigl[\partial^\mu a +{1\over4}{\cal J}^\mu \Bigr]
\Bigl[ \partial_\mu a +{1\over4}{\cal J}_\mu \Bigr]
- {a\over2} \, \widetilde\Tr \Bigl[ F_{\mu\nu}\widetilde F^{\mu\nu} 
- 2 \, \partial_\mu(\lambda\sigma^\mu\ov\lambda) \Bigr]
\eeq
and $a$ is an axion field with standard coupling to $\widetilde\Tr F_{\mu\nu}\widetilde F^{\mu\nu}$ and 
a kinetic metric inverse of the gauge coupling field metric. Since
\beq
 - {2\over{\cal H}_{CC}} = 2\, m^2 g^4 \Bigl[ 1- {C(G)\over8\pi^2}\Bigr]^{-1}
 = - m^2 \, {16\pi^2\over 3C(G)} \beta,
\eeq
the canonically normalized axion field $ma$ couples with scale $m^{-1}$ to 
$\widetilde\Tr F_{\mu\nu}\widetilde F^{\mu\nu}$.

Notice that it is legitimate to use $a$ instead of $h_{\mu\nu\rho}$ as supersymmetry partner of the gauge coupling field $C$.
Simply, while $C$ and $h_{\mu\nu\rho}$ belong to an off-shell linear representation of supersymmetry,
$a$ and $C$ have nonlinear supersymmetry variations depending on the Lagrangian function ${\cal H}$,
as prescribed by the duality transformation. 

\subsubsection{\boldmath{${\cal N}=2$} super-Yang-Mills}

The ${\cal N}=2$ super-Yang-Mills theory adds a chiral $X$ in the adjoint representation to the 
gauge superfield ${\cal W}_\alpha$. The chiral superfield
$$
\widetilde\Tr{\cal WW} - {1\over2} \ov {DD} (\ov X e^{\cal A} X)
$$
transforms with a derivative under the second supersymmetry and a superpotential for $X$ is not permitted.\footnote{
A Fayet-Iliopoulos term linear in $X$ would be allowed in a $U(1)$ theory.}  
The classical Lagrangian can be written in various equivalent forms:\footnote{Even though written in ${\cal N}=1$ superspace, 
this Lagrangian has ${\cal N}=2$ off-shell supersymmetry. This is not the case for theories with hypermultiplets.} 
\beq
\label{N2clas}
\begin{array}{rcl}
{\cal L}_{cl.} &=& \displaystyle
 {1\over g^2} \Fint \left[{1\over4}  \widetilde\Tr{\cal WW} - {1\over8} \ov {DD} (\ov X e^{\cal A} X) \right] 
 + {\rm h.c.}
\crbig
&=& \displaystyle {1\over g^2}\Dint \ov X e^{\cal A} X +
{1\over4g^2}\Fint \widetilde\Tr{\cal WW} + {\rm h.c.} + {\rm derivative}
\crbig
&=& \displaystyle {1\over g^2}\Dint \Bigl[ \hat L + \ov X e^{\cal A} X  \Bigr] + {\rm derivative}.
\end{array}
\eeq
The supersymmetry variations are of course independent from the gauge coupling constant $g$.
In the last expression, the linear superfield $L$ would be non-dynamical: its contribution to the Lagrangian 
is a derivative.

The traditional introduction of renormalized quantities in a theory with chiral matter superfields amounts
to writing
\beq
\label{Lmatter}
{\cal L} =  {1\over 4g^2} \Fint \widetilde\Tr{\cal WW} + {\rm h.c.}
+ \sum_i \Dint Z_i \ov\Phi_{r_i} e^{{\cal A}_{r_i}}\Phi_{r_i},
\eeq
where the sum is over irreducible components $r_i$ of the matter representation $r$. Then, in ${\cal N}=2$
super-Yang-Mills theory,
\beq
\label{N2Z}
Z_X = {1\over g^2}
\eeq
and the corresponding anomalous dimension is
\beq
\label{gammais}
\gamma_X = - M{d\over dM} \ln Z_X = {1\over g^2}\beta(g^2)
\eeq
in terms of the renormalisation scale $M$ used to normalize quantities. In ${\cal N}=2$, the beta function
is purely one-loop and, with $b_0=2\,C(G)$ for super-Yang-Mills theory, 
\beq
\label{N2}
\label{N=20}
\beta(g^2) = - {g^4\over4\pi^2} C(G), \qquad\qquad
\gamma_X(g^2) = - {g^2\over4\pi^2} C(G) .
\eeq
The last result provides a derivation of the gauge contribution to the scheme-inde\-pen\-dent one-loop
anomalous dimension of a chiral superfield in irreducible representation $r$: since the anomalous dimension
follows from the two-point function of this superfield, the relevant group quantity is 
\beq
\label{Cris}
\sum_{a,j}{(T_r^a)^i}_j{(T_r^a)^j}_k 
\equiv C(r) \, \delta^i_k
\eeq
instead of $T(r)$ in $\beta$ functions. But $C(Adj G) = T(Adj G) = C(G)$ and then 
\beq
\label{gammar}
\gamma_{r, gauge} = - {g^2\over4\pi^2} C(r) .
\eeq
Inserting the values of $\gamma_X$ and $b_0$ in the NSVZ formula \cite{NSVZ}
\beq
\label{betaNSVZ}
\beta_{\rm NSVZ} (g^2) = -{g^4\over8\pi^2} {b_0 + \sum_i\gamma_{r_i}T(r_i) \over 1 - {C(G)\over8\pi^2}g^2},
\qquad\qquad \gamma_{r_i} = -M{d\over dM} \ln Z_i,
\eeq
the denominator simplifies and the one-loop $\beta$ function (\ref{N=20}) is obtained.

An alternative formulation is to redefine the renormalization constant $Z_X$ as
\beq
Z_X = {1\over g^2} \widehat Z_X
\eeq
and to reexpress the ${\cal N}=2$ super-Yang-Mills $\beta_{\rm NSVZ}$ as
\beq
\label{betaNSVZ2}
\beta_{\rm NSVZ} (g^2) = -{g^4\over8\pi^2} {b_0 + \widehat\gamma_XT(r_X) \over 1 - {g^2\over8\pi^2}
[C(G)-T(r_X)]},
\qquad\qquad
\widehat\gamma_X = - M{d\over dM} \ln\widehat Z_X.
\eeq
Since $T(r_X)=C(G)$ the denominator disappears and $\widehat\gamma_X=0$, see eq.~(\ref{N2Z}).

With the gauge coupling field $C$, the natural ${\cal N}=2$ extension of the super-Yang-Mills 
Wilson Lagrangian (\ref{Wils10b}) is clearly
\beq
\begin{array}{rcl}
{\cal L}_{W,\mu} &=&
m^2 \Dint \ln \Bigl[ \hat L + \ov X e^{\cal A} X \Bigr]
\crbig && \displaystyle
+ {b_0\over 32\pi^2} \ln{\mu\over M}\Fint \Bigl[ \widetilde\Tr{\cal WW} 
- {1\over2} \ov {DD} (\ov X e^{\cal A} X) \Bigr] + {\rm h.c.}
\end{array}
\eeq
Since we now have scalar fields in $X$, we will use this Lagrangian for zero background value of 
$X|_{\theta=0}$, {\it i.e.} in the phase with unbroken gauge symmetry.
Expand:
\beq
\begin{array}{rcl}
{\cal L}_{W,\mu} &=& \displaystyle
\Dint \Bigl[ m^2\ln\hat L + \left( {m^2\over\hat L} + {b_0\over 8\pi^2} \ln{\mu\over M} \right)
\ov X e^{\cal A} X \Bigr] + \ldots
\crbig
&& \displaystyle
+ {b_0\over 32\pi^2} \ln{\mu\over M}\Fint \widetilde\Tr{\cal WW} + {\rm h.c.}
\crbig
&=& \displaystyle
{1 \over g^2_{W,\mu}} \left[ {1\over4}\Fint \widetilde\Tr{\cal WW} + {\rm h.c.} + 
\Dint \ov X e^{\cal A} X \right] + \ldots
\end{array}
\eeq
with $b_0 = 2C(G)$ and
\beq
{1 \over g^2_{W,\mu}}  = {m^2\over C} + {b_0\over 8\pi^2} \ln{\mu\over M} .
\eeq
The Wilson wave-function renormalization constant for $X$ is then
\beq
Z_{W,X} = {1 \over g^2_{W,\mu}}  ={m^2\over C} + {b_0\over 8\pi^2} \ln{\mu\over M} = {1\over g^2(M)} 
+ {b_0\over 8\pi^2} \ln{\mu\over M}
\eeq
with anomalous dimension
\beq
\gamma_{W,X} = - \mu {d\over d\mu} \ln Z_{W,X} = - {C(G)\over 4\pi^2} g_{W,\mu}^2,
\eeq
as expected. Notice that, as earlier, the scale $M$ is arbitrary in the Wilson Lagrangian which only changes
if the Wilson scale $\mu$ is varied. 

We should maybe remark here that the anomalous dimension of the superfield $X$, as defined in 
eqs.~(\ref{Lmatter}) and (\ref{betaNSVZ}), does not vanish\footnote{As occasionally stated, see for instance 
ref.~\cite{LS}.}: its purely one-loop (and anyway scheme-independent)
value is needed to cancel the higher-order terms in the NSVZ $\beta$ function (\ref{betaNSVZ}).
The point is that the second supersymmetry correlates $Z_X$ and $\gamma_X$ with the inverse
gauge coupling and the $\beta$ function. After rescaling to canonical gauge kinetic terms,
all fields in the super-Yang-Mills multiplet have the canonical scale dimension required by gauge invariance. 
As observed in ref.~\cite{AHM}, the rescaling is not anomalous: in our formulation, this is the absence of the 
contribution (\ref{Lcor2a}).

The introduction of ${\cal N}=2$ hypermultiplets implies the presence of a superpotential. It should be 
$g$--independent to be compatible with the real coupling field. Using chiral superfields ${\cal H}_i$ and
$\widetilde{\cal H}^i$ in representations $r_{\cal H}$ and $\ov r_{\cal H}$ to describe the hypermultiplets,
the appropriate ${\cal N}=2$ Lagrangian reads 
\beq
\begin{array}{rcl}
{\cal L} &=& \displaystyle
{1\over g^2} \Fint \left[{1\over4}  \widetilde\Tr{\cal WW} - {1\over8} \ov {DD} (\ov X e^{\cal A} X) \right]
+ {\rm h.c.}
\crbig
&& \displaystyle + \Dint \Bigl[  \ov{\widetilde{\cal H}}e^{-{\cal A}_{\cal H}} \widetilde{\cal H}
+ \ov{\cal H}e^{{\cal A}_{\cal H}} {\cal H} \Bigr]
+ {i\over\sqrt2} \Fint \widetilde{\cal H} X_{\cal H}{\cal H}  + {\rm h.c.}
\end{array}
\eeq
where ${\cal A}_{\cal H}$ and $X_{\cal H}$ are matrix-valued in the representation $r_{\cal H}$ of 
the hypermultiplets. The wave-function renormalization constants are then 
\beq
Z_X = {1\over g^2}\widehat Z_X, \qquad\qquad\qquad \widehat Z_X= Z_{\cal H} = Z_{\widetilde{\cal H}} = 1,
\eeq
where the last equalities are due to the non-renormalization theorem of ${\cal N}=2$ theories. 
With these choices, the NSVZ $\beta$ function becomes
\beq
\label{betaNSVZ3}
\beta_{\rm NSVZ} (g^2) = -{g^4\over8\pi^2} {b_0 + \widehat\gamma_XT(r_X) + 2\gamma_{\cal H}T(r_{\cal H})
\over 1 - {g^2\over8\pi^2}
[C(G)-T(r_X)]}
= - {g^4\over4\pi^2} [C(G) - T(r_{\cal H})] .
\eeq
In this expression, 
\beq
\widehat\gamma_X = - M{d\over dM} \ln\widehat Z_X =0, \qquad
\gamma_{\cal H} = - M{d\over dM} \ln Z_{\cal H} = 0, \qquad
T(r_X)= C(G).
\eeq
If the hypermultiplet is in the adjoint representation, $\beta=0$ and the theory has ${\cal N}=4$ supersymmetry.

\subsection{Gaugino condensates, nonperturbative superpotentials}

The effective action describing gaugino condensates $\langle \widetilde\Tr {\cal WW}\rangle 
= - \langle \widetilde\Tr\lambda\lambda \rangle$ is formally derived by coupling the operator
$\widetilde\Tr {\cal WW}$ to a classical source field $J$ in the path integral and taking the Legendre 
transformation exchanging $J$ with the condensate classical field $U$. In the supersymmetric context,
$J$ and $U$ are expected to be chiral superfields since $\widetilde\Tr {\cal WW}$ is chiral, 
but $U$ should 
also keep track of the relation $\widetilde\Tr {\cal WW} = \ov{DD}\Omega$. Consider again the 
microscopic theory (\ref{CSF4}):
$$
{\cal L} = \Dint {\cal H}(\hat L, Y) + \Fint W(\Phi) + \Fbarint \ov W(\ov \Phi).
$$
As explained in Section \ref{seclinear}, this is equivalent to
\beq
\label{cond1}
\begin{array}{rcl}
{\cal L} &=& \Dint \Bigl[ {\cal H}(V, Y)
 - {1\over2}(S+\ov S) (V+2\Omega)\Bigr] + \Fint W(\Phi) + \Fbarint \ov W(\ov \Phi)
\crbig
&=& \Dint \Bigl[ {\cal H}(V, Y) - {1\over2}(S+\ov S) V \Bigr] + {\rm derivative}
\crbig
&& + \Fint \Bigl[ W(\Phi) + {1\over4}S\,\widetilde\Tr{\cal WW}\Bigr] 
+ \Fbarint \Bigl[ \ov W(\ov \Phi) + {1\over4}\ov S\,\widetilde\Tr\ov{\cal WW} \Bigr] .
\end{array}
\eeq
The real, unconstrained gauge-invariant superfield $V$ has the same canonical dimension 
two as $\hat L$. We assume that $S$ has natural scale dimension and $U(1)_R$ charge 
$w=q=0$. The $S$--dependent terms in the Lagrangian are then scale invariant and do not 
modify the scale-breaking superfield $\Delta$.

The last equality (\ref{cond1}) firstly shows that $S$ is actually the source superfield $J$.\footnote{
Introducing the source is equivalent to replace $S$ by $J$. The condensate superfield $U$ is then
the Legendre dual of $J$ calculated at $J=S$ and the effective Lagrangian depends then on $U$ and $S$.}
Secondly, the integration over the gauge superfield is now confined in a universal 
(${\cal H}$--independent) term and in the matter dependence of ${\cal H}$. Finally, the Lagrangian has
an axionic shift symmetry $\delta S = ic$ which in the last line exists because 
$\widetilde\Tr {\cal WW} = \ov{DD}\Omega$. 

Consider first pure super-Yang-Mills theory:
\beq
\label{cond2}
{\cal L}_{SYM} = \Dint \Bigl[ {\cal H}(V) - {1\over2}(S+\ov S) V \Bigr] 
+ {1\over4}\Fint S\,\widetilde\Tr{\cal WW} + {1\over4}\Fbarint \ov S\,\widetilde\Tr\ov{\cal WW} .
\eeq
Using anomaly-matching, the effective Lagrangian is then of the form
\beq
\label{cond3}
\begin{array}{rcl}
{\cal L}_{SYM, eff.} &=& \displaystyle
\Dint \left[ {\cal H}(V) + {d(w)\over8\pi^2} \, V\left(\ln {V\over m^2}-1\right) + {\cal K}_{(U)}  \right] 
\crbig
&&\displaystyle + {1\over4}\Fint \left[ S\left( U + {1\over2}\ov{DD}V\right) 
+ {b(w)\over24\pi^2} \, U\left(\ln {U\over M^3} -1 \right) \right]
\crbig
&& \displaystyle
+ {1\over4}\Fbarint \left[\ov S\left( \ov U + {1\over2}DDV\right) + {b(w)\over24\pi^2} \, 
\ov U\left(\ln{\ov U\over M^3}-1\right)\right] ,
\end{array}
\eeq
with $b(w) = 3C(G) = 3d(w)$ in the absence of chiral superfields. In the first line, $m$ is the
irrelevant mass parameter already present, for instance, in the Wilson Lagrangian (\ref{Wils10}).
However, the chiral, $U$--dependent contributions match the one-loop anomaly induced by a 
rescaling of the physical scale $M$, as in the Wilson Lagrangian (\ref{Wils4}), and
corresponding to the identification (\ref{Wils8b}).
The field equation for $S$ is
\beq
\label{cond4}
U = -{1\over2} \ov{DD} \,V \qquad\Longleftrightarrow\qquad \langle \widetilde\Tr{\cal WW}\rangle
= -{1\over2}\ov{DD} \,\langle\hat L\rangle
\eeq
as required and the effective Lagrangian is then a function of $V$, $DD V$ and 
$\ov{DD}V$, 
\beq
\label{cond5}
\begin{array}{rcl}
{\cal L}_{SYM, eff.} &=& \displaystyle
\Dint \left[ {\cal H}(V) + {d(w)\over8\pi^2} \, \left( V \ln {V\over m^2} - V \right) + {\cal K}_{(U)} 
\right]_{U=-{1\over2}\ov{DD}\,V}
\crbig
&&\displaystyle + {b(w)\over96\pi^2} \, \Fint  \left[ U\ln {U\over M^3} - U 
\right]_{U=-{1\over2}\ov{DD}\,V} + {\rm h.c.}
\end{array}
\eeq
The real function ${\cal K}_{(U)}$ of $U$ is the induced K\"ahler potential which
generates kinetic terms for the 
components of the condensate superfield $U$.\footnote{While ${\cal H}$ generates the kinetic terms
of the gauge coupling field and supersymmetry partners as in the microscopic Lagrangian.} 

To derive the gaugino condensate, we need the bosonic component expansion of  $V$:
\beq
\label{cond6}
\begin{array}{rcl}
V &=& C - \theta\theta\,\ov F - \ov{\theta\theta}\, F + \theta\sigma^\mu\ov \theta v_\mu
+ \theta\theta\ov{\theta\theta}\, \left({1\over2}D + {1\over4}\Box C \right),
\crbig
U &=& u - \theta\theta f_u,
\crbig
u &=& -2F,
\qquad\qquad
\Re f_u \,\,=\,\, -D,
\qquad\qquad
\Im f_u \,\,=\,\, - \partial^\mu v_\mu.
\end{array}
\eeq
The gaugino condensate is then the value of the (classical) superfield $U$ at the minimum of the 
effective potential included in the effective Lagrangian: $\langle\widetilde\Tr\lambda\lambda\rangle
= - \langle u \rangle = 2\langle F\rangle$. The scalar potential is the sum of three squares induced 
by the field equations of the three real auxiliary fields $D$, $\Re F$ and $\Im F$ included in $V$.
As usual with a coupling field, gaugino condensation alone does not lead to a stabilized ground state: there 
is a runaway behaviour and further contributions would be needed to determine the ground state value of $C$,
{\it i.e.} to dynamically determine the value of the gauge coupling. But the gaugino condensate is 
determined as a function of $C$ by the cancellation of the terms linear in $D$, which
in ${\cal L}_{SYM, eff.}$ are 
\beq
\label{cond7}
D\Bigl[ {1\over2} {\cal H}_C + {d(w)\over16\pi^2} \ln {C\over m^2} +{b(w)\over48\pi^2}\ln {|u|\over M^3} \Bigr] 
\eeq
and a quadratic term is generated by $K_{(U)}$. The linear terms cancel at
the supersymmetric ground state:
\beq 
\label{cond8}
|u| = M^3 \left[{C\over m^2}\right]^{-3d(w)/b(w)}  \exp\left(-{24\pi^2\over b(w)}{\cal H}_C\right) 
\eeq
or, with the identification $C = m^2 g^2(M)$, 
\beq
\label{cond9}
|u| = M^3 [g^2(M)]^{-3d(w)/b(w)} \exp \left( -{24\pi^2\over b(w)}{\cal H}_C \right).
\eeq
This formula holds for super-Yang-Mills theory in which ${\cal H} = m^2 \ln(\hat L/m^2)$,
and the gaugino condensate is then
\beq
\label{cond10}
|u| = { M^3 \over g^2(M)} \exp \left( -{8\pi^2\over C(G)g^2(M)} \right).
\eeq

The gaugino condensate is a physical quantity which is then invariant under the 
renormalisation group. The condition $M{d\over dM}|u|=0$ applied on formula (\ref{cond8})
provides then a derivation of the beta function
$$
\beta(g^2) = - { g^4 \over 4\pi^2} { b(w) \over 1 - {g^2\over8\pi^2}d(w)},
$$
and it also provides a definition for the RG-invariant scale characterizing the strength of the gauge 
interaction,
\beq
\label{cond11}
|u|  = \Lambda^3,
\eeq
which is a derived quantity. 

The effective Lagrangian (\ref{cond3}) has a continous $\widetilde R$ symmetry with transformations
\beq
\label{cond12}
U \quad\longrightarrow\quad e^{3i \alpha}U, \qquad\qquad
S \quad\longrightarrow\quad S - i {b(w)\over8\pi^2} \alpha
\eeq
and $\theta\rightarrow e^{{3\over2}i\alpha}\theta$. The rotation of the Grassmann coordinates induces 
the rotation of $U$ once $S$ imposes $U=-{1\over2}\ov{DD} V$ and the shift of $S$ is induced by the 
chiral anomaly. In version (\ref{cond5}) of the theory, the $\widetilde R$ symmetry is manifest since 
$U=-{1\over2}\ov{DD}V$ and $\Re\int d^2\theta\, U\ln U$ transforms then with a derivative.
The $\widetilde R$ symmetry is spontaneously broken by the gaugino condensate and, since
the $D$ contribution to the scalar potential only specifies the modulus 
$|\langle\widetilde\Tr\lambda\lambda\rangle|$, the
condensate phase provides the expected ground state degeneracy.

The effective Lagrangian (\ref{cond5}) has been derived from perturbative anomaly arguments.
It is expected that non-perturbative contributions discretize the $\widetilde R$ transformations.  
Concentrating now
on $SU(N)$ super-Yang-Mills theory, $b(w) = 3C(G)=3N$, discretization to $Z_{2N}$ implies
that the parameter $\alpha$ in transformations (\ref{cond12}) has values
\beq
\label{cond13}
{3\over2}\alpha = {\pi k\over N}, 
\qquad\qquad \makebox{$k$ integer}.
\eeq
This follows from the shift in $S$ (\ref{cond12}) which effectively corresponds to an anomalous 
$\widetilde R$ transformation
\beq
\delta {\cal S}_{SYM} = - 3Nq\, \alpha, \qquad\qquad
q = {1\over32\pi^2} \int d^4x\,\widetilde\Tr F_{\mu\nu}\widetilde F^{\mu\nu}
\eeq
of the super-Yang-Mills action. Since for non-trivial gauge field configurations $q$ is an integer, 
condition (\ref{cond13}) follows.

Since $U^{kN}$ is now invariant, we may add non-perturbative contributions to the superpotential 
term of the effective Lagrangian 
(\ref{cond3}):
\beq
\label{cond14}
\begin{array}{rcl}
W(S,U) &=& W_{pert.}(S,U) + W_{np}(U) + {1\over8}S\ov{DD}V, 
\crbig
W_{pert.} (S,U) &=& \displaystyle
{1\over4} U \left[S + {N\over8\pi^2} \, \left(\ln {U\over M^3} -1 \right)\right], 
\qquad
W_{np} (U) \,\,=\,\,
{1\over4} U \sum_{n\ge1} {1\over kN} c_k \, U^{kN} ,
\end{array}
\eeq
with complex coefficients $c_k$. 

To verify that the non-perturbative contribution sums $k$--instanton terms, a standard approach is 
to neglect the K\"ahler potential ${\cal K}_{(U)}$, omit the last term in $W$ and eliminate $U$ as a 
function of $S$. Both $S$ and $U$ are then unconstrained chiral superfields
and the superpotential $W(S,U)$ generates the Legendre transformation 
\beq
\label{cond15}
0 = {\partial\over\partial U} [W_{pert.}(S,U) + W_{np}(U) ]
\eeq
which expresses $U$ as a function of $S$. 
Perturbatively, 
\beq
\label{cond16}
0 = {\partial\over\partial U}W_{pert.}(S,U)   \qquad\Longrightarrow\qquad
U = M^3 e^{-8\pi^2 S / N}.
\eeq
Replacing in $W_{pert.}(S,U) + W_{np}(U)$ leads to
\beq
\label{cond17}
W_{(S)}= {1\over4}M^3  e^{-8\pi^2 S / N} \left[ - {N\over8\pi^2}  
+  \sum_{k\ge1} {1\over kN}c_k \Bigl(M^3 e^{-8\pi^2 S}\Bigr)^k \right]
\eeq
as expected from $k$--instanton contributions expressed in terms of the Wilson holomorphic coupling 
field $S$. The complete Legendre transformation (\ref{cond15}) is of course much more complicated. 
In any case, this procedure is a crude approximation of the effective Lagrangian (\ref{cond3}) which in
particular turns background equations into overconstrained superfield equations. For instance, a term quadratic in 
$D$ is generated by ${\cal K}_{(U)}$. Without this term, the field equation for $D$ is the contraint 
${\partial\over\partial C} {\cal L}_{SYM,eff.}=0$.

\section{Discussion} \label{secdisc}

In this work we have studied effective actions obtained by replacing the Yang--Mills coupling constant of $\mathcal{N}=1$ SYM by a 
real field embedded in a linear superfield. As a consequence, holomorphicity is not relevant. This choice introduces a second real gauge-invariant 
super-Yang-Mills operator $\hat L$ to be used in $D$--terms of effective Lagrangians. We have then shown how this approach allows to correctly 
treat the quantum anomalies of $U(1)_R$ and dilatation transformations. In particular, we have shown that the one-loop running of the 
Wilsonian action and anomaly matching in this effective approach are sufficient to derive the NSVZ $\beta$ function, provided a specific 
$D$--term anomaly counterterm constructed with $\hat L$ is used to account for the discrepancy in the anomalous behaviours of $R$ and 
dilatation transformations. This counterterm is at the origin of the denominator of $\beta$. 
We have also shown that a similar approach leads to an effective Lagrangian for super-Yang-Mills condensates, in terms of
a real superfield $V$ (for $\langle\hat L\rangle$) and a chiral $U=-{1\over2}\ov{DD}V$ (for $\langle\widetilde\Tr{\cal WW}\rangle$), with two 
outcomes: a scalar potential predicting the value of the modulus of the gaugino condensate, as function of the physical gauge coupling
with correct all-order behaviour and another derivation of the NSVZ $\beta$ function.
Since the real countertem in terms of $\hat L$ cannot be (analytically) transformed into a $F$--term by chiral-linear duality, we conclude
then that embedding the coupling field in a linear superfield is not only useful but actually necessary for a correct description of super-Yang-Mills theory.

That $R$ and dilatation transformations must be considered in a super-Poincar\'e theory has a simple origin. These transformations are symmetries 
of the ${\cal N}=1$ superconformal algebra and the multiplets of Poincar\'e supersymmetry carry a representation of the full superconformal algebra 
(with same scale dimension and $R$ charge for chiral superfields, as unique restriction).
Hence, the $R$ and dilatation currents are well-defined but not conserved in a generic Poincar\'e theory. 
We have illustrated this point in our construction of the supercurrent structures of ${\cal N}=1$ gauge theories coupled to a linear superfield, 
with the occurrence of the superfields $\Delta$ and $\widetilde\Delta$ in the source superfields $X$ and $\chi_\alpha$. The use of the 
$16_B+16_F$ operators of the ${\cal S}$ supercurrent structure \cite{KS} is extremely useful in this respect, in contrast with the Ferrara-Zumino 
$12_B+12_F$ structure with $X$ only. This construction is also the main tool in our treatment of $R$ and dilatation anomalies, and then in the
construction of effective Lagrangians. 

Since we have considered anomalies for a generic simple gauge group with an arbitrary matter content, allowing also anomalous dimensions
for the chiral superfields, it is tempting to generalize the NSVZ $\beta$ function (\ref{Wils8c}), (\ref{Wils9}) derived for pure super-Yang-Mills theory  
to 
\beq
\label{discbeta}
\beta(C) = - { C^2 \over 8\pi^2} { b(w) \over 1 - {1\over8\pi^2}d(w) C}
\qquad\qquad
\beta(g^2) = - { g^4 \over 8\pi^2} { b(w) \over 1 - {g^2\over8\pi^2}d(w)},
\eeq
where $b(w)$ and $d(w)$ are given by expressions (\ref{an8b}) and (\ref{an11}). This equation would hold for background matter superfields and
anomalous dimensions. However, anomalous dimensions are related to wave-function renormalization, $\gamma=-M{d\over dM}\ln Z$, and the 
significance of eq.~(\ref{discbeta}) can only be established in relation with a dynamical Lagrangian for the matter superfields, a point illustrated
in our discussion of ${\cal N}=2$ theories. In ${\cal N}=1$ theory with chiral matter superfields, a treatment with the field coupling $C$ also 
requires an approriate formulation of Konishi anomalies, and the outcome should be an expression of the anomalous dimension as a function of $C$, 
in analogy with the case of $\beta$. This problem goes far beyond the present paper and our present knowledge, which some of us try to improve.

\section*{Acknowledgements}

The work of J.~H. is partially supported by a Marina Solvay fellowship and by the advanced 
ERC grant `Symmetries and Dualities in Gravity and $M$--theory' (SyDuGraM) 
of Marc Henneaux. The work of N.~A.,
D.~A. and J.-P.~D. is supported by the Albert Einstein Center for Fundamental Physics (AEC) at the 
University of Bern. J.-P.~D. has benefitted from extended stays at Ecole Normale Sup\'erieure, Paris and 
Ecole Polytechnique, Palaiseau. He wishes to thank his  colleagues for hospitality and intelligence.
J. H. wishes to thank the University of Bern for its hospitality and financial support.
Early stages of this research have been supported by the Swiss National Science Foundation. 

\renewcommand{\thesection}{\Alph{section}}
\setcounter{section}{0}
\setcounter{equation}{0}
\renewcommand{\theequation}{\thesection.\arabic{equation}}

\section{The supercurrent superfield equation}\label{AppA}
\setcounter{equation}{0}

This appendix presents the superfield and component formulas for the supercurrent 
structures used in the main text. 
One needs to solve the supercurrent superfield equations
\beq
\label{appD1}
\ov D^\dalpha\,J_{\alpha\dalpha} = D_\alpha X + \chi_\alpha, \qquad
\ov D_\dalpha X=0,\qquad
\chi_\alpha = -{1\over4}\ov{DD}D_\alpha\, U, \qquad U=U^\dagger
\eeq
for the components of the supercurrent superfield $J_{\alpha\dalpha}$, as a function of
the anomaly sources $X$ and $U$.
We use the following expansion of the chiral superfields $X$ and $\chi_\alpha$:
\beq
\label{appD2}
\begin{array}{rcl}
 X &=& x + \sqrt2\,\theta\psi_X - \theta\theta\, f_X-i\theta\sigma^\mu\bar\theta\partial_\mu x-\frac{i}{\sqrt{2}}\theta\theta\bar\theta\bar\sigma^\mu\partial_\mu\psi_X-\frac{1}{4}\theta\theta\ov{\theta\theta}\Box x ,
 \crbig
 \chi_\alpha &=& -i \lambda_\alpha + \theta_\alpha \, D + \frac{i}{2}(\theta\sigma^\mu\ov\sigma^\nu)_\alpha F_{\mu\nu} - \theta\sigma^\mu\bar\theta\partial_\mu\lambda_\alpha-\theta\theta(\sigma^\mu\partial_\mu\ov\lambda)_\alpha\nonumber
 \crbig
 &&- \frac{1}{2}\theta\theta(\sigma^\mu\bar\theta)_\alpha(\partial_\nu F^\nu{}_\mu-i\partial_\mu D)+\frac{i}{4}\theta\theta\ov{\theta\theta}\Box\lambda_\alpha,
\end{array}
\eeq
where
\beq
\label{appD3}
U = \theta\sigma^\mu\ov\theta\, U_\mu + i \,\theta\theta\ov{\theta\lambda} 
- i \, \ov{\theta\theta}\theta\lambda
+ {1\over2}\theta\theta\ov{\theta\theta}\, D + \ldots ,
\eeq
$F_{\mu\nu}=\partial_\mu U_\nu-\partial_\nu U_\mu$ and the dots denote contributions in $U$ which do not 
appear in $\chi_\alpha$.
The real field $D$ is defined by
\beq
\label{appD3b}
{1\over2}D = U|_{\theta\theta\ov{\theta\theta}} + {1\over4}\Box\, U|_{\theta=0}.
\eeq
With this definition, the lowest component $U|_{\theta=0}$ of $U$ does not
appear in the expansion (\ref{appD2}) of $\chi_\alpha$. Eq.~(\ref{appD3b}) is used
in theories with a linear superfied, and in the next Appendix.

Then, the supercurrent equation is solved 
by the component expansion\footnote{We do not choose a particularly suitable normalization for the 
supercurrent $S_\mu$, which we never use here. }
\beq
\label{appD4}
\begin{array}{rcl}
J_\mu (x,\theta,\ov\theta) &=&  (\ov\sigma_\mu)^{\dalpha\alpha} J_{\alpha\dalpha} 
\crbig
&=& {8\over3}\, j_\mu
+ \theta (S_\mu + 2\sqrt2\, \sigma_\mu\ov\psi_X )
+ \ov\theta (\ov S_\mu - 2\sqrt2\, \ov\sigma_\mu\psi_X )
\crbig
&& - 2i \,\theta\theta\,\partial_\mu\ov x + 2i \,\ov{\theta\theta}\, \partial_\mu x
\crbig
&& \displaystyle
+ \theta\sigma^\nu\ov\theta\left[ 8\,T_{\mu\nu} - 4\,\eta_{\mu\nu}\Re f_X
- {1\over2}\epsilon_{\mu\nu\rho\sigma}\left( {8\over3}\,\partial^\rho j^\sigma - F^{\rho\sigma} \right)\right]
\crbig
&& \displaystyle - {i\over2} \theta\theta\ov\theta ( \partial_\nu S_\mu \sigma^\nu 
+ 2 \sqrt2\, \ov\sigma_\mu \sigma^\nu \partial_\nu\ov\psi_X)
\crbig
&& \displaystyle + {i\over2} \ov{\theta\theta}\theta ( \sigma^\nu \partial_\nu \ov S_\mu
+ 2\sqrt2\, \sigma_\mu \ov\sigma^\nu \partial_\nu\psi_X)
\crbig
&& \displaystyle
- {2\over3}\theta\theta\ov{\theta\theta}\, \Bigl(  2 \, \partial_\mu \partial^\nu j_\nu
- \Box j_\mu \Bigr),
\end{array}
\eeq
with $T_{\mu\nu} = T_{\nu\mu}$. Eq.~(\ref{appD1}) implies that $T_{\mu\nu}$ and $S_\mu$ are 
conserved. They are identified with the energy-momentum tensor and the supercurrent of the 
super-Poincar\'e theory with supercurrent superfield $J_{\alpha\dalpha}$. Eq.~(\ref{appD1}) also 
imposes the relations
\beq
\label{appD5}
\begin{array}{c}
4 \, {T^\mu}_\mu = D + 6 \Re f_X, \qquad\qquad
\partial^\mu\,j_\mu = - {3\over2} \,\Im f_X, 
\crbig
(\sigma^\mu \ov S_\mu)_\alpha = 6\sqrt2\,\psi_{X\,\alpha} + 2i\,\lambda_\alpha
\end{array}
\eeq
between components of $J_{\alpha\dalpha}$ and the anomaly superfields $X$ and $\chi_\alpha$.
The first equation is useful when discussing the behaviour of the theory under dilatations: the trace
of the energy-momentum tensor is related, but in general not equal, to the divergence of the dilatation 
current. The second equation controls the behaviour of the theory under a $U(1)_R$ transformation 
with current $j_\mu$.
Note that these relations can be used to modify the component expansion (\ref{appD4}), which is then 
not unique. Our expansion is as in ref.~\cite{KS}.

The canonical scale dimension $w$ and chiral $R$ charge $q$ of the supercurrent superfield
$J_{\alpha\dalpha}$ are $w=3$ and $q=0$. In the superconformal case where 
$\ov D^\dalpha J_{\alpha\dalpha}=0$, these dimensions are as required for a conserved
dimension three $R$--current and a conserved dimension four symmetric tensor.
The natural weights $(w,q)$ of the source superfields $X$, $\chi_\alpha$ and $U$ are 
then respectively $( 3,3 )$, $( 7/2,3/2 )$ and $( 2,0 )$.

The supercurrent superfield equation (\ref{appD1}) is sufficient for all theories considered in this article.  
The superfields $X$ and $\chi_\alpha$ are usually called {\it chiral} and {\it linear} sources or anomalies. 
Their existence has been known for a long time \cite{FZ, SWest} but an unfortunate claim that their 
simultaneous presence in the supercurrent equation is not compatible with a conserved 
energy-momentum tensor \cite{CPS} soon propagated in the literature. 
It is the merit of Komargodski and Seiberg \cite{KS} to have eliminated this mistake.\footnote{See also
for instance ref.~\cite{MSW}.} Relations 
(\ref{appD5}) indicate that the linear source leads to a conserved $R$ current while the chiral source
correlates ${T^\mu}_\mu$ and the divergence of the $R$ current. Hence, different order parameters
for $\partial^\mu j_\mu$ and ${T^\mu}_\mu$ require both sources. Notice also that the dilatation
current $j_\mu^D$ is not present in the supercurrent structure. It is defined (up
to the addition of identically conserved currents) as the current for which the variation $\delta$
of the Lagrangian under scale transformations equals $\partial^\mu j_\mu^D$ on-shell.

Improvement transformations of the energy-momentum tensor and the supercurrent can be 
induced by observing that the superfield identity
\beq
\label{appD6}
2\ov{D}^\dalpha [D_\alpha,\ov{D}_\dalpha]{\cal G}=D_\alpha\ov{DD}{\cal G}+3\ov{DD}D_\alpha{\cal G},
\eeq
which holds for any superfield ${\cal G}$, is a solution of the supercurrent equation (\ref{appD1}). 
It can thus be used to transform the supercurrent structure as
\beq
\label{appD7}
\begin{array}{rcl}
J_{\alpha\dalpha} \qquad&\longrightarrow&\qquad \widetilde J_{\alpha\dalpha} = J_{\alpha\dalpha} 
+ 2\,[ D_\alpha , \ov D_\dalpha ] \, {\cal G} , 
\crbig
X \qquad&\longrightarrow&\qquad \widetilde X = X + \ov{DD}\,{\cal G} ,
\crbig
\chi_\alpha \qquad&\longrightarrow&\qquad \widetilde\chi_\alpha = 
\chi_\alpha + 3\, \ov{DD} \, D_\alpha \,{\cal G},
\end{array}
\eeq
with any real ${\cal G}$.
If $\cal G$ has the expansion
\beq
\label{appD8}
\begin{array}{rcl}
 \cal G &=& C_g+i\theta\chi_g-i\bar\theta\bar\chi_g+\theta\sigma^\mu\bar\theta v_{g\mu}+\frac{i}{2}\theta\theta(M_g+iN_g)-\frac{i}{2}\ov{\theta\theta}(M_g-iN_g)
\crbig
&& + i\theta\theta\bar\theta(\bar\lambda_g+\frac{i}{2}\partial_\mu\chi_g\sigma^\mu)-i\ov{\theta\theta}\theta(\lambda_g-\frac{i}{2}\sigma^\mu\partial_\mu\bar\chi_g)+\frac{1}{2}\theta\theta\ov{\theta\theta}(D_g-\frac{1}{2}\Box C_g),
\end{array}
\eeq
then the components of the transformed superfields $\widetilde J_\mu$, $\widetilde X$ 
and $\widetilde\chi_\alpha$ read
\beq
\label{appD9}
\begin{array}{rcl}
\widetilde j_\mu &=& j_\mu-3v_{g\mu},
\crbig
\widetilde S_\mu &=& S_\mu+8\sigma_{[\mu}\bar\sigma_{\nu]}\partial^\nu\chi_g,
\crbig
\widetilde\psi_X &=& \psi_X+2\sqrt{2}i\lambda_g+2\sqrt{2}\sigma^\mu\partial_\mu\bar\chi_g,
\crbig
\widetilde x &=& x+2i(M_g-iN_g),
\crbig
\widetilde T_{\mu\nu} &=& T_{\mu\nu}+(\partial_\mu\partial_\nu C_g-\eta_{\mu\nu}\Box C_g),
\crbig
\widetilde f_X &=& f_X+2D_g-2\Box C_g+2i\partial_\mu v^\mu_g,
\crbig
\widetilde F_{\mu\nu} &=& F_{\mu\nu}-24\partial_{[\mu}v_{g\nu]},
\crbig
\widetilde\lambda &=& \lambda-12\lambda_g,
\crbig
\widetilde D &=& D-12D_g.
\end{array}
\eeq
Hence, the scalar quantity $C_g$, the lowest component of ${\cal G}$ which defines the whole superfield,
induces an improvement of the energy-momentum tensor. It also modifies $\Re f_X$ to verify the first 
equation (\ref{appD5}). Similarly, the fermionic quantity $\chi_g$ improves the supercurrent $S_\mu$ 
and changes $\psi_X$ to maintain the validity of the third eq.~(\ref{appD5}). The vector field $v^\mu_g$
modifies the nature of the $U(1)$ current $j_\mu$, which becomes in general the current of another 
$U(1)$ transformation. The other components of ${\cal G}$ only exchange quantities in the anomaly 
superfields $X$ and $\chi_\alpha$.

In practice, we use the superfield transformation (\ref{appD7}) to improve the energy-momentum
tensor and then to modify the relation between its trace and the divergence of the dilatation current,
which does not appear in the supercurrent structure. This is useful to have a firm control of
scale invariance anomalies. We are, of course, particularly interested in the improvement in which
the trace of the energy-momentum tensor {\it equals} the divergence of the dilatation current, if it exists.
The object to consider is the virial current ${\cal V}_\mu$ which, under the improvement (\ref{appD9})
of the energy-momentum tensor, transforms according to
\beq
\label{appD10}
\widetilde{\cal V}_\mu = {\cal V}_\mu + 3\,\partial_\mu C_g,
\eeq
where we used that the dilatation current satisfies $j_\mu^D = {\cal V}_\mu + x^\nu T_{\mu\nu}
= \widetilde{\cal V}_\mu + x^\nu \widetilde T_{\mu\nu}$ up to identically conserved currents.
This is the subject of Subsection \ref{secscale} and Appendix \ref{AppC}.
It would also be interesting to see what the inclusion of the virial current superfield introduced in ref.~\cite{Nak}
would bring to this analysis.


\setcounter{equation}{0}

\section{On the superfield \boldmath{$\Delta( L, \Phi, \ov\Phi)$} and omitted derivatives}
\label{AppB}

This appendix applies to any real function of $L$, $\Phi$ and $\ov\Phi$, like the Lagrangian function
${\cal H}(L,\Phi, \ov\Phi)$  (omitting gauge superfields)
but more specifically to $\Delta$ and to the discussion in subsection \ref{secscale}.

A chiral superfield is usually expanded as
$$
\Phi = z + \sqrt2\theta\psi - \theta\theta f 
- i\theta\sigma^\mu\ov\theta \, \partial_\mu z
+ {i\over\sqrt2}\theta\theta \, \partial_\mu\psi\sigma^\mu\ov\theta
- {1\over4}\theta\theta\ov{\theta\theta}\, \Box z
$$
to solve $\ov D_\dalpha\Phi = 0$. 
To solve $DDL = 0$, a real linear superfield writes
$$
L = C + i \theta\chi - i \ov{\theta\chi}
+ \theta\sigma^\mu\ov\theta \, v_\mu 
+ {1\over2}\theta\theta \, \partial_\mu\chi\sigma^\mu\ov\theta
+ {1\over2}\ov{\theta\theta} \, \theta\sigma^\mu\partial_\mu\ov\chi
+ {1\over4}\theta\theta\ov{\theta\theta}\, \Box C,
$$
with $v_\mu = {1\over6}\epsilon_{\mu\nu\rho\sigma}h^{\nu\rho\sigma}
= {1\over2}\epsilon_{\mu\nu\rho\sigma}\partial^\nu B^{\rho\sigma}$.
The opposite sign in the highest component introduces a subtle novelty
in the highest component of a function $\Delta( L, \Phi, \ov\Phi)$ of
both chiral and linear superfields:
\begin{eqnarray}
\label{appH1}
\Delta( L, \Phi, \ov\Phi)|_{\theta\theta\ov{\theta\theta}} &=&
{\cal L}_\Delta
- {1\over4}\Box \Delta( C, z, \ov z) 
+ {1\over2} \partial^\mu [ \Delta_C\partial_\mu C] , \nonumber
\crbig
{\cal L}_\Delta &=& -{1\over4}\Delta_{CC} [(\partial^\mu C)(\partial_\mu C)
+ {1\over6} h^{\mu\nu\rho} h_{\mu\nu\rho} ]
+ \Delta_{z\ov z}[(\partial^\mu\ov z)(\partial_\mu z) + \ov ff ] \nonumber
\crbig &&
- {i\over2} v^\mu [ \Delta_{Cz}\partial_\mu z - \Delta_{C\ov z}\partial_\mu\ov z ]
+ \makebox{fermion terms}.
\end{eqnarray}
Since total derivatives are irrelevant in a Lagrangian, we for instance use in Section \ref{secsupercurrent}
\beq
\label{appH2}
\int d^4x \Dint {\cal H}(L,\Phi,\ov\Phi) = \int d^4x\,{\cal L}_{\cal H}.
\eeq
From this Lagrangian, we derive canonical (Noether) energy-momentum tensor and dilatation current
which are not gauge invariant. The symmetric gauge-invariant Belinfante tensor\footnote{Omitting 
fermions and gauge fields, its expression is given in eq.~(\ref{CSF13b}).} is then obtained by improving the terms 
involving the antisymmetric tensor and the gauge fields, using field equations.

It is however important to realize that ${\cal L}_\Delta$ {\it differs} from the $D$ 
component of the superfield $\Delta$, as defined in the expansion (\ref{appD3}), (\ref{appD3b}) of
a real superfield. Instead, 
\beq
\label{appH4}
D = 2 {\cal L}_\Delta + \partial^\mu [ \Delta_C\partial_\mu C ]
\eeq
and the derivative term due to the linear superfield must be included when using $D$.
This is in particular the case when evaluating ${T^\mu}_\mu$ in any supercurrent structure 
using the first eq.~(\ref{appD5}).

The derivatives present in the expansion (\ref{appH1}) and neglected in Lagrangians would however, if retained,
contribute to the naive form of Noether currents. Since these derivatives do not break translation symmetry, 
they would affect the energy-momentum tensor by an irrelevant improvement term. For instance, 
applying standard Noether methods to $-{1\over4}\Box\Delta$ leads to the contribution
$-{1\over4}(\partial_\mu\partial_\nu - \eta_{\mu\nu}\Box)\Delta$ to the energy-momentum tensor.
But if scale or chiral $U(1)$ transformations are broken by $\Delta$, the corresponding dilatation 
and $U(1)$ currents could receive new derivative contributions which can always be safely omitted.
But the point is that superfield expressions in general include some of these derivatives, as displayed
for instance in eq.~(\ref{appH1}).

To illustrate this remark, consider a single chiral superfield $\Phi$ with K\"ahler potential ${\cal K}(\Phi,\ov\Phi)$
and a non--$R$ chiral $U(1)$ variation $\delta\Phi=iq\Phi$, $\delta\ov\Phi=-iq\ov\Phi$. The canonical Noether 
current derived from the standard ${\cal N}=1$ sigma-model Lagrangian is
\beq
V_\mu = iq \, {\cal K}_{z\ov z}\, \Bigl(z\partial_\mu \ov z - \ov z\partial_\mu z \Bigr) 
+ q \Bigl( {\cal K}_{z\ov z} + {1\over2}z{\cal K}_{zz\ov z} + {1\over2}\ov z{\cal K}_{z\ov z\ov z}
\Bigr) \psi\sigma_\mu\ov\psi.
\eeq
The vector current in the $\theta\sigma_\mu\ov\theta$ component of the current superfield 
${\cal Z} = {q\over2}({\cal K}_\Phi\Phi + {\cal K}_{\ov\Phi}\ov\Phi)$ is however different:
\beq
\begin{array}{rcl}
{\cal V}_\mu &=& \displaystyle
{iq\over2} \, \Bigl( \ov z{\cal K}_{\ov z\ov z} + {\cal K}_{\ov z} + z{\cal K}_{z\ov z} \Bigr)
\partial_\mu \ov z
- {iq\over2} \, \Bigl( z {\cal K}_{zz} + {\cal K}_z + \ov z {\cal K}_{z\ov z} \Bigr) \partial_\mu z
\crbig
&& \displaystyle
+ {q\over2} \Bigl( 2\,{\cal K}_{z\ov z} + z{\cal K}_{zz\ov z} + \ov z{\cal K}_{z\ov z\ov z}\Bigr) 
\psi\sigma_\mu\ov\psi.
\end{array}
\eeq
The difference
\beq
{\cal V}_\mu - V_\mu = -{iq\over2} \, \partial_\mu \Bigl( \ov z{\cal K}_{\ov z} - z{\cal K}_z \Bigr)
\eeq
is not an improvement term and vanishes if the K\"ahler potential has $U(1)$ symmetry.
Neglecting fermions since these derivatives affect scalar contributions only, 
\beq
\delta_{U(1)}\, \Bigl[{\cal K}_{z\ov z}(\partial_\mu z)(\partial^\mu\ov z) \Bigr] =
iq {\partial^2\over\partial z \partial\ov z} ( z{\cal K}_z - \ov z{\cal K}_{\ov z}  )
= \partial^\mu V_\mu,
\eeq
the second equality holding on-shell, while
\beq
\partial^\mu {\cal V}_\mu = iq \left[{\partial^2\over\partial z \partial\ov z} + {1\over2}\Box \right]
( z{\cal K}_z - \ov z{\cal K}_{\ov z}  ).
\eeq
The new contribution does not use any field equations. What actually matters is that the quantity
$z{\cal K}_z - \ov z{\cal K}_{\ov z}$ measures the violation of the $U(1)$ symmetry in both cases,
and one can safely use either the standard Noether current $V_\mu$ or the superfield current 
${\cal V}_\mu$.

\section{Improving the energy-momentum tensor}\label{AppC}
\setcounter{equation}{0}

This appendix is mostly concerned with scale (non-)invariance and also to 
its relation to conformal symmetry in the
context of classical theories and at the Lagrangian level. 

It should be familiar that, in general, for a given field theory, an energy-momentum tensor 
$T_{\mu\nu}$ verifies
\beq
\label{B1}
j_\mu^{(dilatations)} \ne x^\nu T_{\mu\nu}, \qquad\qquad
\partial^\mu j_\mu^{(dilatations)} \ne {T^\mu}_\mu,
\eeq
where $j_\mu^{(dilatations)}$ is the current for scale transformations. An improvement of $T_{\mu\nu}$ 
may turn these relations into equalities, modifying the dilatation current while keeping the (on-shell) value
of its divergence unchanged. But this improvement transformation does not always exist.  

For instance, in the canonical formalism, in a Poincar\'e-invariant Lagrangian depending on fields
\footnote{Which are not necessarily scalars only.} $\varphi_i$ with scaling dimension 
$w_i$ and their first derivatives, ${\cal L}(\varphi_i,\partial_\mu\varphi_i)$, 
the Noether current for dilatations is
\beq
\label{B2}
j^{(dilatations)}_\mu = \sum_i w_i\, {\partial{\cal L}\over\partial\partial^\mu\varphi_i}\varphi_i + x^\nu 
\, T^{(can.)}_{\mu\nu},
\eeq
where
\beq
\label{B3}
T^{(can.)}_{\mu\nu} = \sum_i{\partial{\cal L}\over\partial\partial^\mu\varphi_i}\partial_\nu\varphi_i
- \eta_{\mu\nu}{\cal L}
\eeq
is the canonical (Noether) energy-momentum tensor. The first term is induced by the transformation 
of the fields at fixed $x$, the second by the transformation of the coordinates.
The field
\beq
\label{B4}
\Delta \equiv \delta{\cal L}-4{\cal L}= \sum_iw_i{\partial{\cal L}\over\partial\varphi_i}\varphi_i 
+ \sum_i(w_i+1) {\partial{\cal L}\over\partial\partial^\mu\varphi_i}\partial^\mu\varphi_i - 4{\cal L} \,
\eeq
is a measure for the violation of scale invariance.\footnote{We assume that the assignment of scale dimensions $w_i$ has some justification even if $\Delta$ does not vanish with any assignment, as in 
a generic theory without scale invariance.} The currents $T_{\mu\nu}^{(can.)}$ and $j_\mu^{(dilatations)}$
and the quantity $\Delta$ are in general calculated in terms of off-shell fields, but the conservation
equations
\beq
\label{B5}
\partial^\mu\,T^{(can.)}_{\mu\nu} =0, \qquad\qquad
\partial^\mu j_\mu^{(dilatations)} = \Delta
\eeq
are verified on shell.
From eq.~(\ref{B2}), the trace of the canonical energy-momentum tensor satisfies on shell
\beq
\label{B6}
T^{(can.)\mu}{}_\mu=\Delta-\partial^\mu\left(\sum_iw_i\, {\partial{\cal L}\over\partial\partial^\mu\varphi_i}\varphi_i\right),
\eeq
and it is in particular not traceless in a scale-invariant theory.

Except in general for the contribution of scalar fields, 
the canonical energy-mo\-men\-tum tensor is not symmetric (and not gauge
invariant). Lorentz invariance of the theory can be used to improve $T_{\mu\nu}^{(can.)}$ to
a symmetric Belinfante tensor\footnote{For detail, see for instance refs.~\cite{CCJ, CJ, C, O}.
Field equations are used. In this sense, the transformation from the canonical to the symmetric
tensor is not truly an improvement.} 
${\cal T}_{\mu\nu}$, which also turns out to be gauge invariant. 
The improvement procedure uses a tensor ${\cal X}^{(Bel.)}_{\mu\rho\nu} = 
-{\cal X}^{(Bel.)}_{\rho\mu\nu}$, and  
\beq
\label{B7}
{\cal T}_{\mu\nu} = T_{\mu\nu}^{(can.)} + \partial^\rho {\cal X}_{\mu\rho\nu}^{(Bel.)}.
\eeq
In view of eq.~(\ref{B2}), the canonical dilatation current improves to a Belinfante current
according to 
\beq
\label{B9}
\begin{array}{l}
j_\mu^{(dilatations)} = V_\mu + x^\nu T_{\mu\nu}^{(can.)} \,\longrightarrow\,
{\cal J}_\mu^{(dilatations)} = {\cal V}_\mu + x^\nu {\cal T}_{\mu\nu}, 
\crbig
{\cal V}_\mu = V_\mu + {{\cal X}_{\quad\mu\rho}^{(Bel.)}}^\rho,
\end{array}
\eeq
omitting in ${\cal J}_\mu$ the improvement term $-\partial^\rho({\cal X}_{\mu\rho\nu}x^\nu)$.
The vector field ${\cal V}_\mu$ is called the {\it virial current}. 

The possibility to improve the energy-momentum tensor suggests that there may exist another
symmetric energy-momentum tensor $\Theta_{\mu\nu}$ verifying
\beq
\label{CCJ1}
\partial^\mu J_\mu^{(dilatations)} = {\Theta^\mu}_\mu, \qquad\qquad
\partial^\mu\Theta_{\mu\nu}=0.
\eeq
Its existence is linked to the interplay of scale and conformal transformations in Poin\-ca\-r\'e 
theories: with any symmetric energy-momentum tensor $T_{\mu\nu}$, one can 
define four additional currents
\beq
\label{CCJ2}
{\cal K}_{\mu\nu} = x^2 \, T_{\mu\nu} - 2 \, x_\nu x^\rho \, T_{\mu\rho} 
= - (2x_\nu x_\rho - \eta_{\nu\rho}x^2) {T_\mu}^\rho
\eeq
verifying
\beq
\label{CCJ3}
\partial^\mu {\cal K}_{\mu\nu} = -2 \, x_\nu \, {T^\mu}_\mu.
\eeq
Hence, if $\Theta_{\mu\nu}$ exists, $\partial^\mu {\cal K}_{\mu\nu} = -2 x_\nu \,\partial^\mu
J_\mu^{(dilatations)}$ (on shell) and the four currents ${\cal K}_{\mu\nu}$ constructed with 
$\Theta_{\mu\nu}$  are always conserved in a scale-invariant theory.
Since the ${\cal K}_{\mu\nu}$ appear to be the currrents for conformal transformations (conformal boosts), 
a scale-invariant theory is then also conformal.
The non-existence of such an energy-momentum tensor is then a feature of field 
theories where scale invariance does not imply conformal invariance. These Lagrangians are not 
renormalizable and scale invariance, if present, is in general spontaneously broken.

We then wish to construct a symmetric tensor such that $J_\mu^{(dilatations)} $$= 
x^\nu \Theta_{\mu\nu}$ (off shell), or equivalently such that the improved virial current vanishes up to
an improvement or a conserved current: ${\Theta^\mu}_\mu=\Delta$ on shell.
A first method would be to improve ${\cal T}_{\mu\nu}$ to
\beq
\label{CCJ4}
\Theta_{\mu\nu} = {\cal T}_{\mu\nu} - {1\over3} \Bigl( \partial_\nu {\cal V}_\mu 
- \eta_{\mu\nu}\partial^\rho{\cal V}_\rho \Bigr).
\eeq
In terms of the improved tensor, the dilatation current is 
\beq
\label{CCJ5}
{\cal J}_\mu^{(dilatations)} = x^\nu \Theta_{\mu\nu} 
- {1\over3}\partial^\nu (x_\mu {\cal V}_\nu - x_\nu {\cal V}_\mu),
\eeq
the second term is an improvement with zero divergence which can be omitted to obtain
\beq
\label{CCJ6}
J_\mu^{(dilatations)} = x^\nu \Theta_{\mu\nu},
\qquad\qquad
\partial^\mu J_\mu^{(dilatations)} = {\Theta^\mu}_\mu, \qquad\qquad
\partial^\mu\Theta_{\mu\nu}=0.
\eeq
However, both energy-momentum tensors are symmetric only if $\partial_{[\mu} {\cal V}_{\nu]} = 0$, up maybe to an improvement term. It is clearly solved if
${\cal V}_\mu = \partial_\mu {\cal G}$, for some function ${\cal G}$ of the off-shell fields.
In this case,
\beq
\label{CCJ7}
\Theta_{\mu\nu} = T_{\mu\nu} - {1\over3} (\partial_\mu\partial_\nu - \eta_{\mu\nu}\Box) {\cal G},
\eeq
but the existence of ${\cal G}$ in terms of off-shell fields
is a non-trivial conditions on ${\cal V}_\mu$ and then on the Lagrangian.\footnote{
With a single field $\varphi$ in a two-derivative Lagrangian, 
${\cal V}_\mu = f(\varphi)\partial_\mu\varphi$,
which can be written as $\partial_\mu g(\varphi)$ with $g^\prime = f$. The condition is already
nontrivial with two real scalar fields, as in supersymmetric theories with chiral superfields.}

How to improve the energy-momentum tensor and the dilatation current to obtain equalities
(\ref{CCJ6}) has been discussed in more general terms long ago and in particular 
by Callan, Coleman and Jackiw (CCJ)
\cite{CCJ, CJ}.\footnote{See also Coleman \cite{C}, Polchinski \cite{P} or Ortin's book \cite{O}, section 2.4.} 
They first observe that the tensor $\Theta_{\mu\nu}$ differs by an improvement  from the Belinfante tensor 
only for spin (or helicity) zero fields.
To summarize the improvement procedure, it is assumed that there exists a tensor $\sigma_{\mu\nu}$ such 
that (off shell)
\beq
\label{CCJ8}
{\cal V}_\mu = \partial^\nu \sigma_{\mu\nu} = \partial^\nu \sigma_{[\mu\nu]} + \partial^\nu \sigma_{(\mu\nu)},
\qquad\qquad
\sigma_{[\mu\nu]} = -\sigma_{[\nu\mu]}, \qquad
\sigma_{(\mu\nu)} = \sigma_{(\nu\mu)}.
\eeq
The first term $\partial^\nu \sigma_{[\mu\nu]}$ is an improvement which can be omitted in the dilatation current,
and the second term can be written
\beq
\label{CCJ8b}
\partial^\nu \sigma_{(\mu\nu)} = {\widehat{\cal X}_{\mu\nu}}\,^\nu
\eeq
with
\beq
\label{CCJ9}
\widehat{\cal X}_{\mu\rho\nu} = {1\over2} \Bigl[ \partial_\mu\sigma_{(\rho\nu)}
- \partial_\rho\sigma_{(\mu\nu)} -  \eta_{\mu\nu} \partial^\lambda\sigma_{(\rho\lambda)} 
+\eta_{\rho\nu} \partial^\lambda\sigma_{(\mu\lambda)} \Bigr]
+ {1\over6} \Bigl[ \eta_{\mu\nu}\partial_\rho {\sigma^\lambda}_\lambda
-  \eta_{\rho\nu}\partial_\mu {\sigma^\lambda}_\lambda \Bigr] , 
\eeq
verifying also
\beq
\label{CCJ10}
\widehat{\cal X}_{\mu\rho\nu} = -  \widehat{\cal X}_{\rho\mu\nu}
 \qquad\qquad
\partial^\rho \widehat{\cal X}_{\mu\rho\nu} = \partial^\rho \widehat{\cal X}_{\nu\rho\mu}.
\eeq
Then, the improvement formula (\ref{CCJ4}) can be extended to
\beq
\label{CCJ10b}
\Theta_{\mu\nu} = {\cal T}_{\mu\nu} - \partial^\rho\widehat{\cal X}_{\mu\rho\nu}
\eeq
which relates two symmetric energy-momentum tensors.
The corresponding improvement of the  dilatation current is then
\beq
\label{CCJ11}
{\cal J}_\mu^{(dilatations)} = \partial^\nu\sigma_{\mu\nu} + x^\nu{\cal T}_{\mu\nu} 
\qquad\Longrightarrow\qquad
J_\mu^{(dilatations)} = x^\nu\Theta_{\mu\nu}
\eeq
omitting improvement terms. 

Hence, if the condition (\ref{CCJ8}) on the virial current is verified, there exists an improved energy-momentum 
tensor and a dilatation current verifying conservation equations (\ref{CCJ6}) and then scale invariance implies
conformal symmetry.

In a two-derivative theory with scalar fields only, the virial current is linear in the field derivatives
\beq
\label{CCJ14}
{\cal V}_\mu = \sum_i {\cal F}_i(\varphi_j) \, \partial_\mu\varphi_i
\eeq
and the only available tensor is then $\sigma_{\mu\nu} = \eta_{\mu\nu}{\cal F}$, leading to
\beq
\label{CCJ15}
\begin{array}{rcl}
\widehat\chi_{\mu\rho\nu} &=& {1\over3}( \eta_{\nu\rho}\partial_\mu{\cal F}
- \eta_{\mu\nu}\partial_\rho{\cal F} ),
\crbig
\partial^\rho\widehat\chi_{\mu\rho\nu} &=& {1\over3}( \partial_\mu\partial_\nu{\cal F}
- \eta_{\mu\nu}\Box {\cal F}), 
\end{array}
\eeq
as in eq.~(\ref{CCJ7}). 
If ${\cal V}_\mu= \partial_\mu{\cal F}$, or
\beq
\label{CCJ16}
\partial_{[\mu}{\cal V}_{\nu]}=0,
\eeq
the tensor $\Theta_{\mu\nu}$ exists and scale invariance implies conformal symmetry.
There are however many scalar Lagrangians for which this condition is not verified.

Consider for instance the scalar sector of a Wess-Zumino model with a single chiral superfield and K\"ahler 
potential ${\cal K}(z, \ov z)$.
Since
\beq
\begin{array}{rcl}
{\cal V}_\mu &=& wz{\cal K}_{z\ov z}\,\partial_\mu\ov z + w\ov z{\cal K}_{z\ov z}\,\partial_\mu z ,
\crbig
\partial_{[\mu}{\cal V}_{\nu]} &=& w ( z{\cal K}_{zz\ov z} - \ov z{\cal K}_{z\ov z\ov z}) 
(\partial_{[\mu} z)(\partial_{\nu]}\ov z),
\end{array}
\eeq
the condition (\ref{CCJ16}) is $z{\cal K}_{zz\ov z} = \ov z{\cal K}_{z\ov z\ov z}$ which 
integrates\footnote{Up to an irrelevant K\"ahler transformation of ${\cal K}$.} into
\beq
z{\cal K}_z = \ov z {\cal K}_{\ov z}.
\eeq
Then, 
\beq
{\cal V}_\mu = w\,\partial_\mu (z{\cal K}_z) = w\,\partial_\mu (\ov z{\cal K}_{\ov z})
\eeq
and the improved energy momentum tensor is
\beq
\Theta_{\mu\nu} = T_{\mu\nu} - {1\over3}(\partial_\mu\partial_\nu - \eta_{\mu\nu}\Box )
wz{\cal K}_z.
\eeq
The outcome is that the scale dimension $w$ must be chosen to correspond to a $U(1)$ symmetry
of the K\"ahler potential acting with charge $w$ on $z$, $\delta z = iw z$. 
If such a $U(1)$ symmetry does not exist,
one can of course assign $w=0$, 
in which case the condition $j_\mu^{(dilatations)}= x^\nu T_{\mu\nu}$
is trivially true already for canonical (Noether) currents. But this choice does not lead to scale invariance.

The simple K\"ahler potential ${\cal K} = {1\over2}(z^2\ov z + \ov z^2z)$ does not have a $U(1)$ symmetry.
The  Lagrangian is
\beq
\label{ex1}
{\cal L} = (z+\ov z) (\partial^\mu z)(\partial_\mu\ov z).
\eeq
It is scale-invariant with dimension $w=2/3$,
\beq
\label{ex2}
\Delta = (3w-2){\cal L},
\eeq
but since the point $z=0$ is excluded, scale invariance is spontaneously 
broken by $\langle\Re z\rangle \ne 0$. Under the conformal transformation
\beq
\label{ex3}
\delta_\alpha z = (2x_\alpha x_\rho - \eta_{\alpha\rho}x^2) \partial^\rho z + 2wzx_\alpha,
\eeq
the variation of the Lagrangian is
\beq
\label{ex4}
\delta_\alpha{\cal L} = \partial^\mu[(2x_\alpha x_\mu - \eta_{\alpha\mu}x^2){\cal L}]
+ 2{\cal V}_\alpha + 2x_ha\Delta,
\eeq
where
\beq
\label{ex5}
{\cal V}_\alpha = w(z+\ov z) (z\partial_\alpha \ov z + \ov z \partial_\alpha z)
= j_\alpha^{(dilatations)} - x^\nu T_{\mu\nu}^{(can.)},
\eeq
using the field equation. Scale invariance ($\Delta=0$) does not imply 
conformal invariance: $\delta_\alpha{\cal L}$ is not a derivative since ${\cal V}_\alpha$ is not a derivative. 

In this paper, we are mostly interested in theories without scale invariance, but we need to have control
of the relation between the divergence of the dilatation current and the trace of the energy-momentum tensor.
The introduction of the linear superfield leads to more subtleties, discussed in Subsection \ref{secscale}.


\section{The Ferrara-Zumino supercurrent structure}\label{AppFZ}

The supercurrent structure originally found by Ferrara and Zumino (FZ) \cite{FZ} has $\chi_\alpha=0$. It
is obtained by improving the supercurrent structure (\ref{Imp4}) using identity (\ref{appD6}) with
${\cal G} = \Delta_{(w)}/6$ to eliminate $\chi_\alpha$. The resulting $J_{(FZ)\alpha\dalpha}$  
does not depend on $w$:
\beq
\label{SCFZ2}
\begin{array}{rcl}
\ov D^\dalpha  J_{(FZ)\alpha\dalpha} &=& D_\alpha X_{(FZ)} ,
\crbig
J_{(FZ)\alpha\dalpha} &=&
- 2 \Bigl[ (\ov{\cal D}_\dalpha\ov\Phi) {\cal H}_{\Phi\ov\Phi}  ({\cal D}_\alpha\Phi)
- {\cal H}_{LL}(\ov D_\dalpha\hat L)(D_\alpha\hat L) 
+ 2 \, {\cal H}_L \widetilde\Tr ({\cal W}_\alpha e^{-{\cal A}} \ov {\cal W}_\dalpha e^{\cal A}) \Bigr] 
\crbig
&& - {2\over3} [ D_\alpha , \ov D_\dalpha ] ({\cal H} - \hat L{\cal H}_L) , 
\crbig
X_{(FZ)} &=& - {4\over3}\widetilde\Delta_{(w)} + {1\over6} \ov{DD} \Delta_{(w)}
+ {1\over6}\ov{DD} ( w{\cal H}_\Phi\Phi - w\ov\Phi{\cal H}_{\ov\Phi} )
\crbig
&=& 4\, W - {1\over3}\ov {DD}({\cal H} - \hat L{\cal H}_L) .
\end{array}
\eeq
In the second equality for $X_{(FZ)}$, the superfield equation for $\Phi$ has been used but the first expression is
actually more significant since it depends on the three {\it off-shell} superfields
$$
\widetilde\Delta_{(w)} , \qquad\qquad \Delta_{(w)}, \qquad\qquad
w{\cal H}_\Phi\Phi - w\ov\Phi{\cal H}_{\ov\Phi}
$$
which control the scale and $R$ symmetries of the theory.
With chiral and gauge multiplets only, the function ${\cal H}$ is replaced by the gauge invariant K\"ahler 
potential ${\cal K}$.
In the scale-invariant case, $\Delta=\widetilde\Delta=0$, ${\cal H} - \hat L{\cal H}_L = 
{1\over2}(w\Phi{\cal H}_\Phi + w\ov\Phi{\cal H}_{\ov\Phi})$, the FZ structure coincides with our
improved supercurrent structure (\ref{Imp4}). But both structures significantly differ if scale
transformations are not symmetries.\footnote{Without a linear superfield, this has been observed 
in ref.~\cite{ADH}.}

This is of minor importance for the energy-momentum tensor: the structures differ by improvements.
In the FZ supercurrent superfield $J_{\alpha\dalpha}$,
\beq
\label{SCFZ4}
\widetilde T_{\mu\nu} = T_{\mu\nu} 
- {1\over3} (\partial_\mu\partial_\nu-\eta_{\mu\nu}\Box)({\cal H}-C{\cal H}_C) ,
\eeq
where $T_{\mu\nu}$ is the Belinfante tensor present in the natural structure (\ref{CSF12}), to be compared
with expression (\ref{Imp7}) for the improved supercurrent. Accordingly, the dilatation current becomes
\beq
\label{SCFZ5}
j_\mu^D = \widetilde{\cal V}_\mu + x^\nu\widetilde T_{\mu\nu}
\eeq
with virial current\footnote{See eqs.~(\ref{CSF20}) and (\ref{appD10}).}
\beq
\label{SCFZ6}
\widetilde{\cal V}_\mu = 
{1\over2}\,\partial_z\Delta_{(w)}D_\mu z + {1\over2}\,\partial_{\ov z}\Delta_{(w)}D_\mu\ov z, 
\eeq
omitting an improvement term. 

More significant is the chiral $U(1)$ current present in the lowest component of the supercurrent 
superfield $\widetilde J_{\alpha\dalpha}$. Compared with the $U(1)_{\widetilde R}$ current 
$j_\mu^{\widetilde R}$ (\ref{CSF13}) present in the natural, Belinfante structure (\ref{CSF12}), we now find
\beq
\label{SCFZ7}
\begin{array}{rcl}
\widetilde j_\mu - j_\mu^{\widetilde R} &=&
- C{\cal H}_{CC} {1\over6} \epsilon_{\mu\nu\rho\sigma} H^{\nu\rho\sigma} 
- i ({\cal H}-C{\cal H}_C)_z \, D_\mu z
+ i ({\cal H}-C{\cal H}_C)_{\ov z} \, D_\mu \ov z
\crbig
&& -C{\cal H}_{CC} \widetilde\Tr \lambda\sigma_\mu\ov\lambda 
+ [ {\cal H}_{z\ov z} -C{\cal H}_{Cz\ov z} ]\, \psi\sigma_\mu\ov\psi 
- {1\over2}\,[{\cal H}_{CC} + C{\cal H}_{CCC}]\, \chi\sigma_\mu\ov\chi 
\crbig
&& 
- {i\over\sqrt2}\,C{\cal H}_{CC\ov z} \, \chi\sigma_\mu\ov\psi
+ {i\over\sqrt2}\,C{\cal H}_{CCz} \, \psi\sigma_\mu\ov\chi.
\end{array}
\eeq
Alternatively, in terms of derivatives of $\Delta = 2(C{\cal H}_C - {\cal H})$,  
\beq
\label{SCFZ8}
\begin{array}{rcl}
\widetilde j_\mu - j_\mu^{\widetilde R} &=&
- {1\over12}\Delta_C \epsilon_{\mu\nu\rho\sigma} H^{\nu\rho\sigma} 
+ {i\over2}\Delta_z \, D_\mu z
- {i\over2} \Delta_{\ov z} \, D_\mu \ov z
\crbig
&& - {1\over2}\Delta_C \widetilde\Tr \lambda\sigma_\mu\ov\lambda 
- {1\over2}\Delta_{z\ov z}\, \psi\sigma_\mu\ov\psi 
- {1\over2}\,\Delta_{CC}\, \chi\sigma_\mu\ov\chi 
\crbig
&& 
- {i\over2\sqrt2}\,\Delta_{C\ov z}\, \chi\sigma_\mu\ov\psi
+ {i\over2\sqrt2}\,\Delta_{Cz}\, \psi\sigma_\mu\ov\chi.
\end{array}
\eeq
With chiral multiplets only, ${\cal H}_C=0$, 
\beq
\label{SCFZ9}
\widetilde j_\mu - j_\mu^{\widetilde R} = - i{\cal H}_z \, \partial_\mu z
+ i{\cal H}_{\ov z} \, \partial_\mu \ov z
+ {\cal H}_{z\ov z}\, \psi\sigma_\mu\ov\psi 
\eeq
and ${\cal H}$ is the K\"ahler potential. Hence, the Ferrara-Zumino structure includes 
the current $\widetilde j_\mu$ which is actually the K\"ahler connection derived from 
K\"ahler potential ${\cal H}$. 

The conclusion is that while our natural, Belinfante (\ref{CSF12}) or improved (\ref{Imp4}) supercurrent 
structures include the currents naturally related to $U(1)_R$ transformations rotating chiral superfields
with angle zero or $w$, the Ferrara-Zumino structure includes a K\"ahler current which is
not the Noether current of $U(1)$ transformations acting on superfields. Of course, if
the theory is scale invariant with scale dimensions $w$, 
$2({\cal H} - C{\cal H}_C) = -w[{\cal H}_\Phi\Phi + \ov\Phi{\cal H}_{\ov\Phi}]$ and the Ferrara-Zumino and improved structures coincide.

\section{Legendre identities} \label{appLeg}
\setcounter{equation}{0}

This Appendix collects some useful formula induced by the Legendre transformation \beq
\label{Leg1}
{\cal K}(X, Y) = {\cal F}(L,Y) - {1\over2} XL,
\eeq
which generates the chiral-linear duality (Section \ref{seclinear}).
It implies in particular:
\begin{eqnarray}
&& \displaystyle {\cal K}_X = -{1\over2}L, \qquad\qquad
{\cal F}_L = {1\over2} X, \qquad\qquad
{\cal K}_Y={\cal F}_Y,
\crbig
&& \displaystyle {\partial L\over\partial X} = -2\,{\cal K}_{XX}, \qquad \qquad
{\partial X\over\partial L} = 2\,{\cal F}_{LL}, \qquad\qquad
-4 \, {\cal K}_{XX}{\cal F}_{LL} = 1, 
\crbig
&& \displaystyle {\partial X\over\partial Y} = 2{\cal F}_{LY}, \qquad\qquad
{\partial L\over\partial Y} = -2{\cal K}_{XY}, 
\crbig
&& \displaystyle {\cal K}_{XY} = {{\cal F}_{LY} \over 2{\cal F}_{LL}}, \qquad\qquad
{\cal F}_{LY} = -{{\cal K}_{XY} \over 2{\cal K}_{XX}}, 
\crbig
&& \displaystyle {\cal K}_{YY} = {\cal F}_{YY} - {{\cal F}_{LY}^2 \over {\cal F}_{LL}},
\qquad\qquad
{\cal F}_{YY} = {\cal K}_{YY} - {{\cal K}_{XY}^2\over {\cal K}_{XX}}.
\end{eqnarray}


\newpage
\begin{thebibliography}{99}

\bibitem{Seiberg}
N.~Seiberg,
Phys.\ Lett.\ B {\bf 318} (1993) 469 [hep-ph/9309335];
Phys.\ Rev.\ D {\bf 49} (1994) 6857 [hep-th/9402044].

\bibitem{Seiberg2}
N.~Seiberg,
Phys.\ Lett.\ B {\bf 206} (1988) 75.
  
\bibitem{SW}
N.~Seiberg and E.~Witten,
Nucl.\ Phys.\ B {\bf 426} (1994) 19 [Erratum-ibid.\ B {\bf 430} (1994) 485]
[hep-th/9407087]; 
Nucl.\ Phys.\ B {\bf 431} (1994) 484 [hep-th/9408099].

\bibitem{BDFS}
F.~Buccella, J.-P.~Derendinger, S.~Ferrara and C.~A.~Savoy,
Phys.\ Lett.\ B {\bf 115} (1982) 375; \\
C.~Procesi and G.~W.~Schwarz,
Phys.\ Lett.\ B {\bf 161} (1985) 117.

\bibitem{SV86}
M.~A.~Shifman and A.~I.~Vainshtein,
Nucl.\ Phys.\ B {\bf 277} (1986) 456.

\bibitem{SV}
M.~A.~Shifman and A.~I.~Vainshtein,
Nucl.\ Phys.\ B {\bf 359} (1991) 571.

\bibitem{AHM}
N.~Arkani-Hamed and H.~Murayama,
JHEP {\bf 0006} (2000) 030
[hep-th/9707133].

\bibitem{KS}
Z.~Komargodski and N.~Seiberg,
JHEP {\bf 1007}, 017 (2010)
[arXiv:1002.2228 [hep-th]].

\bibitem{ADH}
D.~Arnold, J.-P.~Derendinger and J.~Hartong,
Nucl.\ Phys.\ B {\bf 867}, 370 (2013)
[arXiv:1208.1648 [hep-th]].

\bibitem{CCJ}
C.~G.~Callan, Jr., S.~R.~Coleman and R.~Jackiw,
Annals Phys.\  {\bf 59} (1970) 42.

\bibitem{CJ}
S.~R.~Coleman and R.~Jackiw,
Annals Phys.\  {\bf 67} (1971) 552.

\bibitem{NSVZ}
V.~A.~Novikov, M.~A.~Shifman, A.~I.~Vainshtein and V.~I.~Zakharov,
Nucl.\ Phys.\ B {\bf 229} (1983) 381;
B {\bf 229} (1983) 407; 
Phys.\ Lett.\ B {\bf 166} (1986) 329.

\bibitem{DFKZ2}
J.-P.~Derendinger, S.~Ferrara, C.~Kounnas and F.~Zwirner,
Phys.\ Lett.\ B {\bf 271} (1991) 307.

\bibitem{BDQQ}
C.~P.~Burgess, J.-P.~Derendinger, F.~Quevedo and M.~Quiros,
Phys.\ Lett.\ B {\bf 348} (1995) 428
[hep-th/9501065].

\bibitem{Siegel}
W.~Siegel,
Phys.\ Lett.\ B {\bf 85} (1979) 333.

\bibitem{FZ}
S.~Ferrara and B.~Zumino,
Nucl.\ Phys.\ B {\bf 87} (1975) 207.

\bibitem{CFV}
S.~Cecotti, S.~Ferrara and M.~Villasante,
Int.\ J.\ Mod.\ Phys.\ A {\bf 2} (1987) 1839.

\bibitem{CO}
G.~Lopes Cardoso and B.~A.~Ovrut,
Nucl.\ Phys.\ B {\bf 369}, 351 (1992); 
B {\bf 392}, 315 (1993) [hep-th/9205009].

\bibitem{DFKZ1}
J.-P.~Derendinger, S.~Ferrara, C.~Kounnas and F.~Zwirner,
Nucl.\ Phys.\ B {\bf 372} (1992) 145.

\bibitem{DKL}
L.~J.~Dixon, V.~Kaplunovsky and J.~Louis,
Nucl.\ Phys.\ B {\bf 355} (1991) 649.

\bibitem{FWZ}
S.~Ferrara, J.~Wess and B.~Zumino,
Phys.\ Lett.\ B {\bf 51} (1974) 239.

\bibitem{DQQ}
J.~P.~Derendinger, F.~Quevedo and M.~Quiros,
Nucl.\ Phys.\ B {\bf 428} (1994) 282 [hep-th/9402007].

\bibitem{Mack}
G.~Mack,
Commun.\ Math.\ Phys.\  {\bf 55} (1977) 1.

\bibitem{MSW}
M.~Magro, I.~Sachs and S.~Wolf,
Annals Phys.\  {\bf 298} (2002) 123
[hep-th/0110131].

\bibitem{Pons}
J.~M.~Pons,
J.\ Math.\ Phys.\  {\bf 52} (2011) 012904
[arXiv:0902.4871 [hep-th]].

\bibitem{AB}
I.~Antoniadis and M.~Buican,
Phys.\ Rev.\ D {\bf 83} (2011) 105011
[arXiv:1102.2294 [hep-th]].

\bibitem{VY}
G.~Veneziano and S.~Yankielowicz,
Phys.\ Lett.\ B {\bf 113} (1982) 231.

\bibitem{SV88}
M.~A.~Shifman and A.~I.~Vainshtein,
Nucl.\ Phys.\ B {\bf 296} (1988) 445 [Sov.\ Phys.\ JETP {\bf 66} (1987) 1100].

\bibitem{Gross}
D.~J.~Gross, {\it Applications of the Renormalization Group to High-Energy Physics},
in {\it Methods in Field Theory, Les Houches 1975}, North Holland, Amsterdam, 1976.

\bibitem{Jones} 
D.~R.~T.~Jones,
Phys.\ Lett.\ B {\bf 123}, 45 (1983).

\bibitem{JJN}
I.~Jack, D.~R.~T.~Jones and A.~Pickering,
Phys.\ Lett.\ B {\bf 435} (1998) 61
doi:10.1016/S0370-2693(98)00769-2
[hep-ph/9805482];\\
I.~Jack, D.~R.~T.~Jones and C.~G.~North,
Nucl.\ Phys.\ B {\bf 486} (1997) 479
doi:10.1016/S0550-3213(96)00637-2
[hep-ph/9609325].

\bibitem{KataevS}
A.~L.~Kataev and K.~V.~Stepanyantz,
Nucl.\ Phys.\ B {\bf 875} (2013) 459
doi:10.1016/j.nuclphysb.2013.07.010
[arXiv:1305.7094 [hep-th]] ;
Theor.\ Math.\ Phys.\  {\bf 181} (2014) 1531
doi:10.1007/s11232-014-0233-3
[arXiv:1405.7598 [hep-th]].

\bibitem{Trace}
S.~L.~Adler, J.~C.~Collins and A.~Duncan,
Phys.\ Rev.\ D {\bf 15} (1977) 1712; \\
P.~Minkowski,
unpublished Bern report 76-0813 (September 1976); \\
N.~K.~Nielsen,
Nucl.\ Phys.\ B {\bf 120} (1977) 212; \\
J.~C.~Collins, A.~Duncan and S.~D.~Joglekar,
Phys.\ Rev.\ D {\bf 16} (1977) 438.
  
\bibitem{LS}
R.~G.~Leigh and M.~J.~Strassler,
Nucl.\ Phys.\ B {\bf 447} (1995) 95
[hep-th/9503121].

\bibitem{SWest}
M.~F.~Sohnius and P.~C.~West,
Phys.\ Lett.\ B {\bf 105} (1981) 353.

\bibitem{CPS}
T.~E.~Clark, O.~Piguet and K.~Sibold,
Nucl.\ Phys.\ B {\bf 143} (1978) 445.

\bibitem{Nak}
Y.~Nakayama,
Phys.\ Rev.\ D {\bf 87} (2013) 8,  085005
[arXiv:1208.4726 [hep-th]].

\bibitem{C}
S.~Coleman, {\it Aspects of Symmetry}, Cambridge University Press, Cambridge (1988), 402 pages. 
Chapter 3.

\bibitem{O}
T.~Ortin, {\it Gravity and strings}, Cambridge University Press, Cambridge (2004), 684 pages.

\bibitem{P}
J.~Polchinski,
Nucl.\ Phys.\ B {\bf 303} (1988) 226.

\end{thebibliography}
\end{document}